\documentclass[letterpaper, aps, tightenlines, nofootinbib, 11pt]{article}
\pdfoutput = 1

\usepackage{setspace}
\usepackage[bookmarks = false, colorlinks = true, linkcolor = blue, citecolor = purple]{hyperref}
\usepackage{shortcuts}
\usepackage{titlesec}
\usepackage{cite}
\usepackage{tocloft}
\usepackage[vcentermath]{youngtab}
\usepackage{tensor}

\usepackage{tikz}
\usetikzlibrary{arrows, calc, decorations.pathmorphing, backgrounds, positioning, fit, petri}

\usepackage[margin = 2.5cm]{geometry}
\begin{document}

\begin{center}

\thispagestyle{empty}

\vspace*{5em}

{\LARGE \bf The Ergodicity Landscape of Quantum Theories}

\vspace{1cm}

{\large Wen Wei Ho$^\spadesuit$ and \DJ or\dj e Radi\v cevi\'c$^\clubsuit$}
\vspace{1em}

$^\spadesuit${\it Department of Theoretical Physics, University of Geneva, 1211 Geneva, Switzerland}\footnote{Present address: Department of Physics, Harvard University, Cambridge, Massachusetts 02138, USA.}\\
\texttt{wenwei.ho@unige.ch}
\\
\vspace{1.5em}

$^\clubsuit${\it Perimeter Institute for Theoretical Physics, Waterloo, Ontario, Canada N2L 2Y5}\\
\texttt{djordje@pitp.ca}\\

\vspace{0.08\textheight}
\begin{abstract}
This paper is a physicist's review of the major conceptual issues concerning the problem of spectral universality in quantum systems. Here we present a unified, graph-based view of all archetypical models of such universality (billiards, particles in random media, interacting spin or fermion systems). We find phenomenological relations between the onset of ergodicity (Gaussian-random delocalization of eigenstates) and the structure of the appropriate graphs, and we construct a heuristic picture of summing trajectories on graphs that describes why a generic interacting system should be ergodic. We also provide an operator-based discussion of quantum chaos and propose criteria to distinguish bases that can usefully diagnose ergodicity. The result of this analysis is a rough but systematic outline of how ergodicity changes across the space of all theories with a given Hilbert space dimension. As a particular example, we study the SYK model and report on the transition from maximal to partial ergodicity as the disorder strength is decreased.
\end{abstract}
\end{center}

\newpage

\tableofcontents

\newpage

\section{Introduction and summary}

\subsection{Taxonomy by ergodicity}

The classification of quantum field theories (QFTs) is an old and important problem. This subject is riddled with subtleties: not only is there no satisfying rigorous definition of many quantum theories, but even the more physical approach --- specifying field content and an action or a Hamiltonian --- is not free of shortcomings. On the one hand, there seem to exist perfectly acceptable QFTs without known actions or field content; on the other hand, theories with different actions and fields can turn out to be dual to each other, thereby really representing the same theory.

Wilson's \emph{renormalization group (RG)} procedure is an extremely important conceptual beacon that has shed much light upon the structure of the space of possible theories. RG systematically zooms into the long-distance degrees of freedom of a given theory, and so it defines a monotonic flow on the space of local QFTs. Fixed points of this flow possess conformal symmetry, and this provides a great degree of control over them. One can typically understand many relevant questions about a given QFT by identifying the conformal theory (CFT) it flows to under RG. In this sense, CFTs represent a useful set of \emph{universality classes} that chart the set of QFTs on which RG can be defined.

Unfortunately, not every quantum theory admits a useful Wilsonian RG procedure. The key assumptions behind the RG paradigm include a tensor product structure of the Hilbert space and a sufficient locality of the Hamiltonian. In the absence of a product structure, e.g.~in a $(0+1)$-dimensional spacetime, the notion of coarse-graining the short-distance degrees of freedom no longer applies. Moreover, even in the presence of such a product structure, a sufficiently nonlocal Hamiltonian can render the procedure meaningless. An appropriate generalization of RG may be formulated in terms of decimating general operator algebras\cite{Radicevic:2016kpf}, as we review below.

These issues have been brought to the fore by the recent surge of interest in connections between \emph{quantum chaos} and \emph{quantum gravity}\cite{Shenker:2013pqa, Maldacena:2015waa}. This has in turn caused the rise to prominence of the SYK model, a (mostly) tractable, (mostly) chaotic theory of many fermions without a concept of spatial locality but with an alluring similarity to quantum gravity \cite{Kitaev:2015, Maldacena:2016hyu, Polchinski:2016xgd}.

We will have nothing to say about quantum gravity, and will instead fully focus on quantum chaos. Quantum chaos deals with typical long-time behavior of a quantum system, mirroring the study of sensitive dependence on initial conditions that underpins the field of classical chaos \cite{Gutzwiller:1990}. Alternatively, quantum chaos can be seen as a field that studies general properties of entire spectra of quantum theories. As such, it can be contrasted to the Wilsonian philosophy, which is oriented toward studying long-wavelength states only.

Intriguingly, quantum chaos brings forth a different kind of universality. Quantizing most classically chaotic systems gives spectra much like those of random matrices drawn from one of the standard Gaussian ensembles, with eigenvalues exhibiting \emph{Wigner-Dyson statistics} \cite{Wigner:1951, Dyson:1970tza, Bohigas:1983er}. Conversely, quantizing many classically integrable systems produces Hamiltonians whose eigenvalue gaps have Poisson statistics \cite{Berry:1977}.\footnote{Chaotic systems typically exhibit \emph{eigenvalue repulsion}: the difference between neighboring eigenvalues is statistically unlikely to be small compared to the mean eigenvalue spacing. The Poisson distribution of gaps in typical integrable systems means that in those cases there is no such eigenvalue repulsion.}  There exists a rich mathematical literature on this subject, and the origin of this universality is in some cases well understood --- particularly in systems with semiclassical limits \cite{Shnirelman:1974, Zelditch:1987, CdV:1985} (see Refs.~\citen{Muller:2005, Muller:2009} for recent physics developments) and in wide classes of non-Gaussian random matrices (see e.g.~Ref.~\citen{Erdos:2011}).

Wilsonian RG cannot access universality properties of the entire spectrum --- but its algebraic generalization\cite{Radicevic:2016kpf} can, as it allows for a much more general set of coarse-grainings. Using this set of ideas, we will see that the ``landscape'' of spectral universality turns out to be relatively simple compared to the landscape of CFTs (i.e.~of universality classes under Wilsonian RG). All evidence shows that there exist two extremal regimes --- quantum ergodicity (random-matrix-like behavior) and localization (integrable-system-like behavior). Other regimes appear to be mixes or crossovers between the two, where one part of the spectrum is ergodic while the other one is integrable. In general, we can expect ten different classes of quantum ergodicity, depending on which discrete symmetries (time reversal, orientation reversal, and charge conjugation) are present or broken in the theory \cite{Gnutzmann:2003, Altland:1997}; the three classes that are most frequently encountered are the Gaussian ensembles, GOE, GUE, and GSE. On the other hand, each localized regime is governed by the particular set of symmetries present in the system. Crossovers between these regimes may also be marked by a third, critical regime with power-law localized, self-similar wavefunctions \cite{Mirlin:2000, Kravtsov:1997, Kaplan:2009kr, Kachiga:1986, Geisel:1991}, but we currently know very little about when this phenomenon generally appears.

The \textbf{goal} of this paper is to overview a large number of examples of different chaotic behaviors, and to utilize the algebraic version of RG in order to chart this \emph{ergodicity landscape} of quantum theories (Fig.~\ref{fig landscape medium}) and explain the findings mentioned in the previous paragraph. The methods we use are only slight generalizations of what is already a staple in the quantum chaos literature, but the systematic algebraic development and comparison of different theories is new, to the best of our knowledge. While we do not possess a prescription as powerful as Wilsonian RG, we use numerical results and analytic arguments to describe what traits cause different theories to end up in different ergodic universality classes. Along the way, we stress that \emph{any} regulated quantum theory can be mapped to motion on a so-called \emph{state graph}, and that this idea is particularly useful for analyzing ergodicity; it applies whether the theory is a spin system, a single quantum particle, or a suitably discretized QFT. Concretely, we argue that the ergodicity of a theory depends on a certain distribution of weighted trajectories on its state graph --- and we make this a little more precise by proposing and justifying specific definitions of ergodicity and of the regularity criteria that separate ergodic Hamiltonians from the nonergodic ones.

\subsection{How to read this paper}

The ideas we present here stem from various fields --- high energy physics, condensed matter physics, and random matrix theory --- and we have attempted to make the presentation accessible to members of all of these communities, even though the review is primarily meant for high energy physicists. In particular, we have strived to make this paper as self-contained as possible, and all models are presented from a unified point of view. Section \ref{sec rgqc} covers physical ideas about operators and coarse-graining that provide a universal way to view all theories, while section \ref{sec review} synthesizes known rigorous and numerical results on ergodic properties of many theories. In the rest of the current section we present three major ideas needed to understand our results, so readers may choose to skim either (or both) of sections \ref{sec rgqc} and \ref{sec review} and to focus on results in sections \ref{sec chart} and \ref{sec syk}.

\begin{enumerate}
\item We take \emph{quantum ergodicity} to mean that, in an admissible basis (defined more precisely in sections \ref{subsec qc} and \ref{subsec ops}), eigenstates of the Hamiltonian have the same statistics as a set of uniformly random vectors with unit norm. A set like this is almost surely provided by eigenvectors of random self-adjoint matrices drawn from Gaussian ensembles. To measure this randomness of eigenstates, we coarse-grain the theory from a full quantum algebra down to a classical algebra of observables and calculate the von Neumann entropies of the reduced density matrices that eigenstates reduce to. When the classical observables at the end of our coarse-graining are position operators, we recover entropic delocalization measures present in much of the quantum chaos literature, e.g.~Refs.~\citen{Chirikov:1988, Izrailev:1990, Santos:2011, Santos:2010}.
\item We do \emph{not} rely on eigenvalue statistics as a proxy for quantum ergodicity. This is because eigenvalues can diagnose quantum chaos when no quantum ergodicity is present in the eigenstates, as in the case of certain metric graphs \cite{Kaplan:2001, Berkolaiko:2003}. We will see a few more examples of this dissonance between eigenvalue statistics and ergodicity in models of spin chains or fermions.
\item  The structure that controls ergodicity is most clearly seen by viewing Hamiltonians as adjacency matrices of graphs. Each Hamiltonian then describes motion between the vertices of an appropriate graph; the graph edges correspond to hoppings enacted by the operators in the Hamiltonian. The symmetries of this \emph{state graph} determine the ergodicity properties of the theory.\footnote{A similar connection between chaos and symmetries of the Hamiltonian matrix for spin chains was observed in Ref.\ \citen{Santos:2011}.} Sufficiently small perturbations of the graph in question are found not to strongly influence ergodicity, meaning that ergodicity is a smooth function of the structure of the state graph; $O(1)$ perturbations that are not highly symmetric always increase the ergodicity. Sufficiently strong perturbations, on the other hand, always lead to localized theories. The critical perturbation strength needed to transition between regimes is a function of the system size and its analysis will generally be beyond our abilities.
\end{enumerate}

In more detail, this paper is organized as follows. In section \ref{sec rgqc} we discuss operator algebras and their basis choices in general quantum theories. These algebraic methods provide a useful conceptual way to think of all theories on the same footing --- as matrices in a fixed basis. By studying natural reductions of operator algebras, we clarify why the usual Wilsonian RG is not a good tool to use in order to understand quantum chaos, and we define the entropy of reduced algebras as a tool that can be used to quantify the information lost in a given state upon algebraic coarse-graining.

In section \ref{sec review} we report on the current understanding of ergodic universality across many examples. We are necessarily restricted to low-dimensional theories in which the entire spectrum is known. Within this class, we examine many one-particle quantum mechanics problems, including the typical examples of particles moving in flat and hyperbolic billiards. We also review particles moving on discrete spaces, random Hamiltonians of the Wigner type, and one-dimensional spin chains, including the SYK model as a special case. We diagnose ergodicity in all these models by numerically obtaining the entire spectrum and computing the distribution of coarse-grained entropies of states in the spectrum.

In section \ref{sec chart} we bring together all these special cases and a provide more in-depth analysis of their features. Using our knowledge of how ergodicity changes with perturbations of these models, we draw a sketch of the \emph{ergodic landscape} in the space of all quantum theories with a fixed Hilbert space dimension. Our findings can be understood by invoking ordinary perturbation theory of quantum mechanics, which allows us to express the eigenstate coefficients as a sum over trajectories on the appropriate graphs; we conjecture how the lack of structure or overly strong interactions can lead to ergodicity via a central limit theorem.

As an application of these ideas, in section \ref{sec syk} we focus on SYK and related models. We show that, in the standard spin basis and for up to $N = 22$ Majorana fermions, the eigenstates of the standard $q = 4$ SYK are maximally ergodic within their appropriate parity sectors, in keeping with the theoretical expectations in the $N \sim q^2$ regime. We also present a way to dial the strength of disorder in SYK, show how ergodicity is lost as disorder is weakened, and we compare this behavior to ergodic Ising chains and disordered random graphs.

\section{Renormalization group and quantum chaos} \label{sec rgqc}

\subsection{Preliminaries: generalized position and momentum} \label{subsec prelim}

As our goal is to compare ergodicity properties of various models, it is important to develop a way to express different theories in the same basis. Essentially, our goal is to view each Hamiltonian as a large matrix and to make sure we compare energy eigenstates expressed in a position basis that is held fixed as Hamiltonians are changed. We will comment more on this in section \ref{subsec ops}, but now we establish conventions for operator algebras that will apply to any regularized theory with a finite-dimensional Hilbert space. More details can be found in Ref.~\citen{Ohya:2004}.

Consider a quantum theory with a $D$-dimensional Hilbert space. The largest algebra of quantum operators on this Hilbert space is $\C^{D\times D}$, the vector space of $D\times D$ complex matrices.\footnote{Discrete antilinear transformations such as time- and parity reversal do not belong to this algebra, and we do not include them in our algebraic considerations.} Whatever algebra defines our quantum theory, it will be a subalgebra of $\C^{D\times D}$, and any operator statement (e.g.~a definition of algebraic coarse-graining) can be recast as a statement about complex matrices. It is very covenient to specify a complete orthonormal basis of $\C^{D \times D}$ via a \emph{group} of operators. We choose this group such that for any two of its elements $\O_i$ and $\O_j$ we have $\Tr(\O_i^{-1} \O_j) = D\, \delta_{i,j}$. A simple and always definable group with such properties is generated by any two operators satisfying the \emph{clock algebra}
\bel{\label{def U L}
  U^D = L^D = \1, \quad UL = e^{2\pi i/D} LU.
}
We will call $U$ the \emph{position} and $L$ the \emph{shift (momentum)}. A concrete example are matrices $U_{nm} = e^{2\pi i n/D} \delta_{n,\,m}$ and $L_{nm} = \delta_{n,\,(m + 1)\, \trm{mod}\, D}$. We typically assume that $U$ and $L$ have been fixed and a basis has been chosen to achieve the above representation; different theories will have Hamiltonians built out of different combinations of $U$ and $L$.

Whenever $D$ is not prime, the Hilbert space is isomorphic to a tensor product of spaces of different dimensions $D_i$, with $D = \prod_{i = 1}^N D_i$. This is the case in lattice field theories, where one has $N$ identical Hilbert spaces of dimension $D_0$ such that $D = D_0^N$, with $N$ being the number of lattice sites. Then it is possible to consider generators $U_i$ and $L_i$ on individual lattice sites. These satisfy
\bel{\label{def Ui Li}
  U_i^{D_0} = L_i^{D_0} = \1,\quad U_i L_i = e^{2\pi i/D_0} L_i U_i, \quad U_i L_j = L_j U_i\ \trm{for} \ i \neq j.
}
These are analogs of quantum fields and their conjugate momenta. As they generate the same algebra, the generators $\{U_i, L_i\}$ and $\{U, L\}$ are \emph{dual} to each other. When dealing with lattice field theories we will express all operators in terms of the basis generated by $\{U_i, L_i\}$; for more general theories all operators will be expressed in terms of $U$ and $L$.

A typical example to keep in mind is the Ising model with $N$ spins. Here we have $D_0 = 2$, and the generators $U_i$, $L_i$ at each site $i$ can be taken to be the Pauli matrices $\sigma^z_i$, $\sigma_i^x$. Another familiar example is the limit $D_0 \gg 1$, where one can write $U_i = e^{i \theta_i}$ and $L_i \sim \pder{}{\theta_i}$, thus obtaining the algebra of the $O(2)$ model.\footnote{We will make the connection between $L$ and $\pder{}{\theta}$ more precise in section \ref{subsec qm1}.} In a non-linear sigma model (e.g.~an $O(N)$ quantum rotor with $N \geq 3$), one usually takes the generating operators on each site to be the field operators $\phi_i$ and $\pi_i$ that do not satisfy the clock algebra, but rather that measure the position and momentum of a particle on the target space of the sigma model. Such models are reduced to our paradigm by triangulating the target space into a lattice with $D_0$ sites. The discrete analogues of $\phi_i$ and $\pi_i$ can then be expressed as (possibly very complicated) linear combinations of products of clock operators $U_i$ and $L_i$. Appropriate generalizations of this procedure exist for fermionic and gauge theories, but we will not discuss them in this paper.

One reason why the operator approach is useful is that any state in the Hilbert space can be represented as a density matrix belonging to the algebra of operators on that space. As a result, any systematic reduction in the operator algebra induces a systematic reduction of density matrices. For any basis $\B = \{\O_i\}$ whose elements form a group, the density matrix is
\bel{\label{def rho}
  \rho = \frac1D \sum_{i = 1}^{D^2} \avg{\O^{-1}_i} \O_i.
}
Here $\avg \O$ denotes the expectation value of $\O$ in the given state, and we emphasize that this formula follows from the condition $\Tr(\O_i \O_j) = D\, \delta_{ij}$ introduced above. This compact formula is one reason why it is convenient to use a group as a basis for the quantum algebra.

Finally, the Hamiltonian $H$ is also an element of the operator algebra. Once it is given, one can meaningfully discuss the dimensionality of the space on which the theory lives.\footnote{Note that this is not the same as the dimension of the Hilbert space, but rather the dimension of the spatial manifold on which our dynamics happens.} If $N = \ell^d$, we can label each site $i$ with $d$ separate indices $i_1, \ldots, i_d$, each running from 1 to $\ell$.\footnote{This assumes that degrees of freedom live on sites of a hypercubic lattice. Other arrangements are of course also possible, but the simple flat-space setup is enough to illustrate all our points.} This labeling is reasonable only if the Hamiltonian is appropriately local, i.e.~if it is a linear combination of basis operators $\O_i \O_j \O_k \ldots$, where for each term the mutual distance of indices $i$, $j$, $k$, etc.~on the $d$-dimensional lattice is of order one. (It goes without saying that we will always be imagining large systems, with $D \gg N  \gg 1$.) Dualities between operator algebras are really only useful when they preserve the locality of the Hamiltonian. Nevertheless, we emphasize that \emph{any} theory can be dualized to a single particle hopping between $D$ sites in some possibly very irregular way, with an algebra generated by a single pair of operators $U$ and $L$.

The point of view presented here is a slightly unusual reformulation of a very traditional story. There is reason to believe that not all ingredients specified above are necessary to define a satisfactory quantum theory. For instance, we needn't have started with a Hilbert space --- specifying one state, or one set of expectation values $\avg{\O_i}$, is enough to reconstruct the rest through the GNS construction, modulo some subtleties \cite{Ohya:2004, Balachandran:2013}.  Similarly, the Hamiltonian need not be explicitly known --- it may be sufficient to specify a symmetry principle which, along with a few parameters, fixes the whole time evolution, like it is done in many conformal theories without a known action and with a Hamiltonian defined formally via a bootstrap equation \cite{Polyakov:1974gs}. Nevertheless, in this paper we will assume that the Hamiltonian and Hilbert space are explicitly known.

\subsection{Reducing operator algebras} \label{subsec rg}

As we view the operator algebra as the central object in a quantum theory, we should recall that not all $D \times D$ matrices are necessary to define a theory --- it is sometimes possible to consistently project down to a subset of the operator algebra, as done in gauge or orbifold theories. Moreover, as $D \rar \infty$ this question leads to the vast field of subfactor theory \cite{Kawahigashi, Jones}. We will not discuss such issues here. Rather, we will tackle a different but related question: what can be learned about a theory from a subalgebra of its operators?

In general, given a state $\rho$ and a subalgebra $\A_1$ of the original algebra $\A$, one can construct the reduced density matrix
\bel{
  \rho_1 = \frac1D \sum_{\O \in \B_1} \avg{\O^{-1}} \O,
}
where $\B_1$ is the group whose elements form the basis of $\A_1$. This is merely a projection of the original density matrix \eqref{def rho} to the chosen set of basis operators. It can be checked that this is the unique matrix that satisfies the condition $\Tr(\rho_1 \O) = \avg{\O}$ for all $\O \in \B_1$ (and, by linearity, for all $\O \in \A_1$). Two major areas of research analyze these reduced density matrices in different contexts, though neither were initially defined from this operator perspective:
\begin{enumerate}
  \item Algebras supported on spatial subregions lead to many useful characterizations of a quantum state, such as the entanglement entropy and its cousins (Renyi entropies, relative entropy with respect to a fiducial state, etc)\cite{Ohya:2004}. Most of these quantities can be computed for an arbitrary subalgebra and state, but they are of greatest utility in local theories at $d \geq 1$, as there they can be used to detect the structure of spatial entanglement between different degrees of freedom in the given state. A related question is whether typical eigenstates in a theory reduce to thermal density matrices when restricted to a sufficiently small region. This behavior, \emph{eigenstate thermalization}, is a consequence of quantum ergodicity \cite{Srednicki:1994, Deutsch:1991}.
  \item Renormalization group techniques provide complementary insight: instead of zooming into a spatial subregion, RG uniformly coarse-grains the operator algebra throughout the entire space. (Said another way, RG traces out regions in momentum space.) For a typical state in a generic theory, doing this just gives another reduced density matrix for which the various entropies could be computed. For a local theory and in the lowest states in the energy spectrum, however, RG usually does much more: it leads to a pure reduced density matrix. This is because typical low-lying states will also have very low spatial momentum, so an effective Hamiltonian description may hold in the coarse-grained picture. The resulting, Wilsonian RG flow on the space of Hamiltonians leads to the great benefits outlined in the Introduction.
\end{enumerate}

It is important to stress that RG does not lead to an effective quantum theory without the slow spatial variation in the given state. If this assumption is violated, any state will become mixed after coarse-graining; moreover, the purity of this reduced state will vary with time, meaning that its evolution will not be unitary. Also, note that the product structure of the Hilbert space is crucial here: in a general quantum problem without this structure, there is no reason to expect a coarse-grained effective theory for \emph{any} state.

Of course, a decimation of a general quantum algebra can always be performed. For instance, if $D$ is even, the set of generators $\{U, L\}$ can be replaced with $\{U^2, L\}$, and this will yield a new, smaller algebra.\footnote{If $D$ is even, we can choose a basis generated by two operators, $U_1$ and $U_2$, such that $U_1^2 = U_2^{D/2} = \1$. Then the coarse-graining described above would correspond to removing all basis operators that contain $U_1$. This is not equivalent to a ``tracing out'' of degrees of freedom associated to the index 1; to do this we would have to remove the generator $L_1$, too.} Of particular interest are Abelian algebras of quantum operators, e.g.~the ones generated only by $L$ or by $U$ \cite{Radicevic:2016kpf}. The entropies associated to these Abelian algebras play a very useful role in diagnosing quantum ergodicity, as we will discuss next.

\subsection{Probing quantum chaos via IR diversities} \label{subsec qc}

As mentioned in the Introduction, studying quantum chaos requires analyzing the entire spectrum, not just the eigenstates for which RG works --- if any even exist. The signature of chaos we will look for is the randomness of energy eigenstates. More precisely, we want to know whether for $D \gg 1$ every eigenstate looks like a set of $D$ random variables drawn independently from a distribution that, as $D \rar \infty$, tends to a Gaussian centered at $1/\sqrt D$ and with variance that vanishes with $D$ \cite{Berry:1977b}. (At finite $D$, the overall constraint on the coefficients to have unit norm causes deviations from Gaussianity \cite{Izrailev:1990}.) Said another way, we wish to know if the eigenstates behave like random matrix eigenvectors \cite{Brody:1981, Tao:2010, ORourke:2016}. This behavior is to be contrasted to eigenstates being sharply localized (as found in Anderson- or many-body-localized systems) or to having uniformly delocalized states with all entries equal to $1/\sqrt D$ with exponentially small corrections (as found in free theories).

This criterion is only meaningful if there exists a preferred basis of states with respect to which the eigenstate entries are supposed to look random. The existing literature has almost exclusively used the position basis for this purpose, and all the results stated in the previous paragraph were assuming this choice.  The point of view of this paper is that we can use the eigenbasis of any operator with nondegenerate, uniformly distributed eigenvalues, subject to some additional criteria. The position basis, furnished by a single operator $U$ represented as a diagonal matrix with entries $e^{2\pi i n/D}$, is an example of such a tame basis, and it will be used for most of this paper. We will discuss other admissible bases in section \ref{subsec ops}.

Given a reference eigenbasis $\{\qvec{e_m}\}$ of an appropriate operator $\O$, what remains is to find a useful diagnostic of randomness for each state. Two particularly natural ones are the distributions of inverse participation ratios (IPR) and infrared (IR) entropies across the spectrum. Given an energy eigenstate $\qvec n =\sum_m \psi_{mn} \qvec{e_m}$, the IPR is defined as
\bel{
  \xi^{(n)} = \frac1{\sum_m |\psi_{mn}|^4},
}
and the IR entropy is
\bel{
  S\_{IR}^{(n)} = - \sum_m |\psi_{mn}|^2 \log |\psi_{mn}|^2.
}
The latter quantity is called the infrared entropy following Ref.~\citen{Radicevic:2016kpf}, where it was shown to be the natural characterization of the eigenstate's reduced density matrix that is obtained by coarse-graining the full operator algebra down to the Abelian algebra generated by $\O$. As a successful diagnostic of quantum ergodicity, however, it has appeared in multiple previous studies, e.g.~Refs.~\citen{Santos:2010, Izrailev:1990}. In this paper we will typically use its normalized exponential, which we will call the \emph{IR diversity},
\bel{\label{def Omega}
  \Omega^{(n)} = \frac{\exp S\_{IR}^{(n)}}D, \quad 0 \leq \Omega^{(n)} \leq 1.
}
Note that the IPR can be understood as the exponential of the second R\'enyi entropy. There exist many other possible diagnostics. For instance, one may ask for the distribution of $L_p$ norms or nodal or level sets of eigenstates, but we will not use these in this work.

Studying the statistics of eigenstate coefficients is sensible if the dimension of the Hilbert space is large enough. Perhaps less obviously, though, prematurely taking the thermodynamic limit $D \rar \infty$ may make quantum ergodicity much harder to study. If $D$ is strictly infinite, the system will not explore its whole phase space in finite time, and many hallmarks of quantum chaos will be much harder (or impossible) to detect.\footnote{This is why continuum QFTs do not exhibit the semicircle law even when they are chaotic. For example, in two spacetime dimensions, chaotic CFTs have the Cardy density of states, $\varrho(E) \sim e^{\sqrt E}$, which can only agree with the Wigner-Dyson formula $\varrho(E) \sim \sqrt{E_0^2 - E^2}$ in a narrow band of low energies where the continuum limit applies.} Typically, if we consider a particle moving on a smooth manifold and ask for eigenstates $\psi_n(x)$ of the corresponding Laplacian, studying quantum ergodicity amounts to proving that $\{|\psi^2_n(x)|\}$ weakly converges to the volume form on the manifold or that matrix elements of observables converge to their thermodynamic averages, and then one needs to appeal to results from classical ergodic theory to show that in these cases delocalization is random as opposed to uniform. There are several other diagnostics; for a review of mathematical approaches see Refs.~\citen{Zelditch:2005, Nonnenmacher:2010}, and for related work on the physical side --- defining and computing quantum Lyapunov exponents --- see Ref.~\citen{Shenker:2013pqa}. We will work at finite $D$ and will be able to clearly see the characteristic $D$-dependence of the IR diversity in the ergodic regime as $D$ is increased.

In this paper we focus on Hamiltonians with real entries, with the SYK model being the only exception. Fully ergodic real Hamiltonians have the same spectral properties as matrices drawn from the Gaussian orthogonal ensemble (GOE) \cite{Brody:1981}. Recall that for GOE matrices each off-diagonal entry is independently drawn from a Gaussian of zero mean and unit variance, and each on-diagonal entry is independently drawn from a Gaussian of zero mean and variance two.

\begin{figure}
  \centering
  \includegraphics[width=0.6\textwidth]{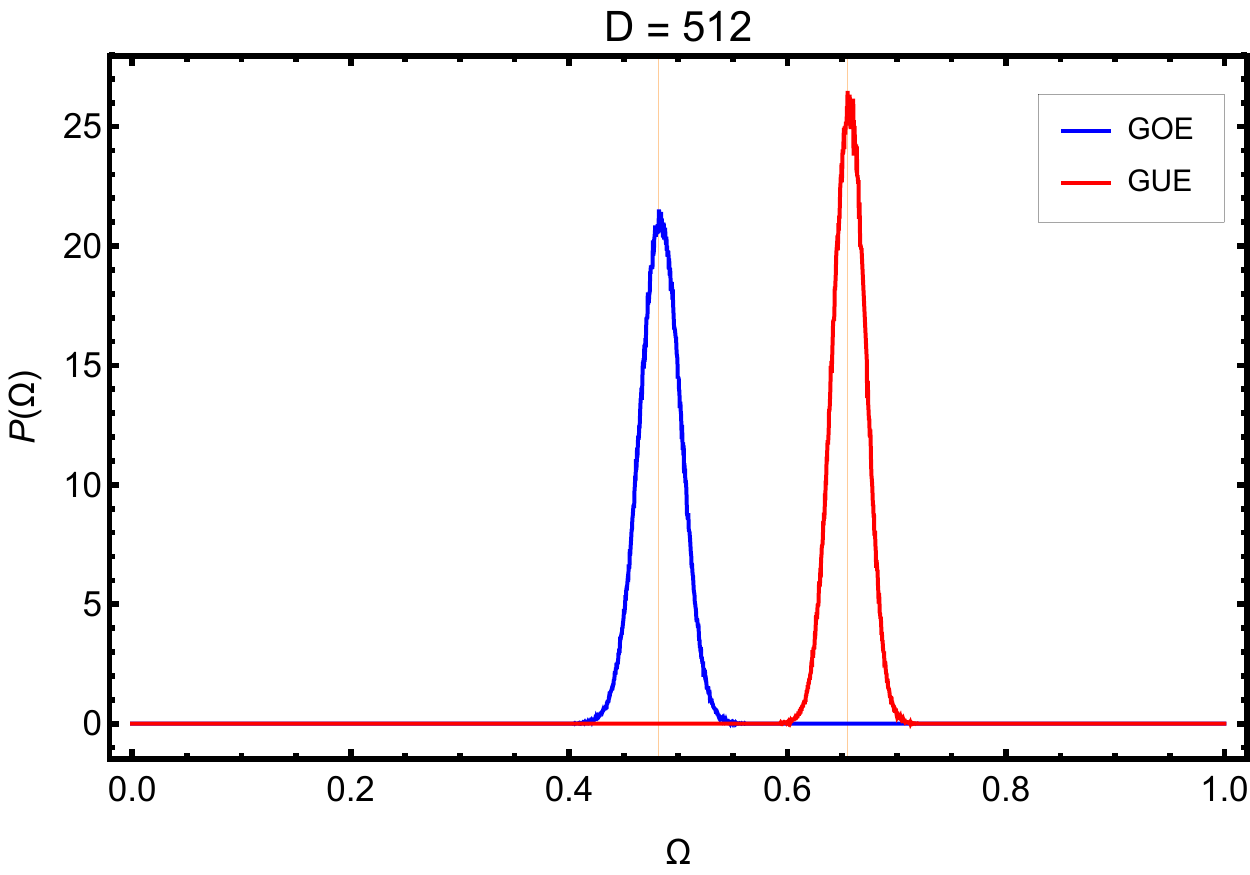}
  \caption{\footnotesize Distributions of IR diversities $\Omega$ of eigenvectors of random matrices from the GOE and GUE ensembles. The characteristic Gaussian distribution around $\Omega \approx 0.48$ and $\Omega \approx 0.65$ is clearly visible. We present aggregated results for 2000 instances of $512 \times 512$ matrices drawn from each ensemble.}\label{fig rmt}
\end{figure}

In all systems with real eigenvectors that we have studied, a fully nonergodic theory will produce a sharp peak in the distribution of position basis IR diversities near $\Omega = 0$ (for localized theories) or near $\Omega = 1$ (for free theories, which are localized in momentum space). This is to be contrasted with ergodic theories, where in position basis we find that, as $D$ tends to infinity, the distribution of IR diversities asymptotes to a Gaussian centered around $\Omega = \frac{2}{e^{2 - \gamma}} \approx 0.48$ with variance $\sim 1/D$. (Here $\gamma$ is the Euler-Mascheroni constant.) The latter is, unsurprisingly, also the kind of distribution expected for eigenvectors of GOE matrices, see Fig.~\ref{fig rmt}\cite{Izrailev:1990}. All other real-valued Hamiltonians can be seen as interpolating between these two extremes.

Systems with complex eigenvectors in a (real) position basis come with an additional quirk: the real and imaginary parts of each component are effectively independent random variables. The norm $|\psi_{mn}^2|$ of each eigenvector coefficient is now a sum of squares of two Gaussian random variables, and as a result the IR diversity distribution is centered around $\Omega \approx 0.65$. This result is found for eigenvectors of Gaussian symplectic and unitary matrices. It can also be found for eigenvectors of GOE matrices when entropies are not computed a basis in which the Hamiltonian is real.

Symmetries can constrain eigenvector coefficients and affect the IR entropies/diversities. In many examples the presence of a $\Z_2$ parity or reflection symmetry will restrict the IR diversity to $0 \leq \Omega \leq 1/2$. We remind the reader that this restriction is the property of the basis (or of the IR algebra), which is chosen on physical grounds as explained in section \ref{subsec ops}. In the presence of degeneracies, the choice of basis must be supplemented with a choice of \emph{polarization} in the degenerate subspace, i.e.~by a specific choice of eigenvectors that span this subspace.

Finally, while we do not diagnose ergodicity using eigenvalue gap statistics, we still compute it in all models we study, and we find good agreement with previous studies. Instead of looking directly at eigenvalue gaps, we look at the statistics of the $r$-parameter \cite{Oganesyan:2006}. Let $E_n$ be an ordered set of energy levels of a Hamiltonian, and $s_n = E_{n+1} - E_n$ the nearest-neighbor spacings. The $r$-parameter is defined as
\bel{
  r_n = \frac{\min(s_n, s_{n-1} )}{\max(s_n, s_{n-1})} \in [0,1].
}

For Hamiltonians with Poisson-distributed gaps, the $r$-distribution is $P(r) = 2/(1+r)^2$. For Hamiltonians with Wigner-Dyson gap statistics, a Wigner-like surmise for $P(r)$ \cite{Atas:2012} is
\bel{\label{eq surmise}
  P(r) = \frac 1{Z_\beta}  \frac{(r+r^2)^\beta}{(1+r+r^2)^{1+3\beta/2}},
}
where $\beta$ is the Dyson index equal to 1 (GOE), 2 (GUE), or 4 (GSE). The normalization is $Z_\beta = \frac 4{27}$, $\frac{ 2\pi}{81 \sqrt{3} }$, $\frac{ 2\pi }{729 \sqrt{3} }$ for $\beta = 1,2,4$, respectively. Eigenvalue repulsion is captured in the fact that the $P(r)$ vanishes for small $r$ in a distinctive fashion, $P(r) \sim r^\beta$, depending on the three Wigner-Dyson classes.

\section{Localization vs.~ergodicity: a review} \label{sec review}

\subsection{Quantum mechanics in one dimension} \label{subsec qm1}

We start from the simplest possible setting: a single particle moving on a one-dimensional target space, say $\Z_D$. No product structure of the Hilbert space is assumed, and $D$ can also be prime. The operator algebra is the clock algebra with position and shift operators $U$ and $L$ defined in Eq.~\eqref{def U L}.

The free Hamiltonian describes hopping between nearest neighbors and in our basis is
\bel{
  H = L + L^{-1} \Longleftrightarrow H_{nm} = \delta_{n,\, (m + 1)\, \trm{mod}\, D} + \delta_{n,\, (m - 1)\, \trm{mod}\, D}.
}
This theory is easily solvable by transforming to the momentum ($L$) eigenbasis, and thus the eigenstates all have zero IR entropy in the momentum basis.\footnote{Each energy level is doubly degenerate in this theory. The momentum basis provides a polarization that lifts all of these degeneracies. The position basis does not, and the IR entropies are not well-defined in this basis.} They are completely localized in momentum space. The same holds for less local hoppings described by higher powers of $L$ (without any $U$'s).

The next simplest theory is a perturbed free particle of the form
\bel{
  H = L + L^{-1} + \lambda V(U).
}
We consider potential to be a function with $O(1)$ coefficients, for instance $V = U^n + U^{-n}$ for some $n$. We wish to highlight the following facts:
\begin{enumerate}
  \item The standard continuum limit is reached when $\lambda = -2\pi^2 \~\lambda/D^2$ for $\~\lambda = O(D^0)$. In that regime, one can expand $L = \exp\{-\d \theta \der{}\theta\}$ in powers of $\d\theta = 2\pi /D$ and, after rescaling the Hamiltonian, get $\~H = -\frac12 \dder{}\theta + \~\lambda V(e^{i\theta})$. This approximation fails for highly excited states.
  \item If $\lambda \rar 0$, the ergodicity properties will be the same as in the free theory. When $\lambda \gtrsim 1/D$, at least some states will localize in position space. The nature of this crossover depends on the potential. For very smooth potentials such as $V = U + U^{-1}$, full localization happens only when $\lambda \sim D$. These effects are mostly invisible in the continuum limit, where $\lambda\sim 1/D^2$; however, dialing $\~\lambda \gg 1$ does show signs of localization in the lowest-lying states.
  \item Conversely, for very quickly varying potentials, full position localization may happen at small couplings. Consider the disordered background $V(U) = \sum \zeta_n \delta(U - e^{2\pi i n/D})$ with random numbers $\zeta_n$ of unit variance. This model does not admit a continuum limit, so we will restrict ourselves to the discrete case. When the coupling is large enough, all states are localized in position space. This is the familiar Anderson localization in one dimension \cite{Anderson:1958vr}. The lack of localization at small couplings is a finite size effect, as high-energy states may be localized at finite lengths much greater than $D$. The rigorous theory of Anderson localization is reviewed in e.g.~Ref.~\citen{Stolz:2011}.
\end{enumerate}

\begin{figure}
  \centering
  \includegraphics[width=0.32\textwidth]{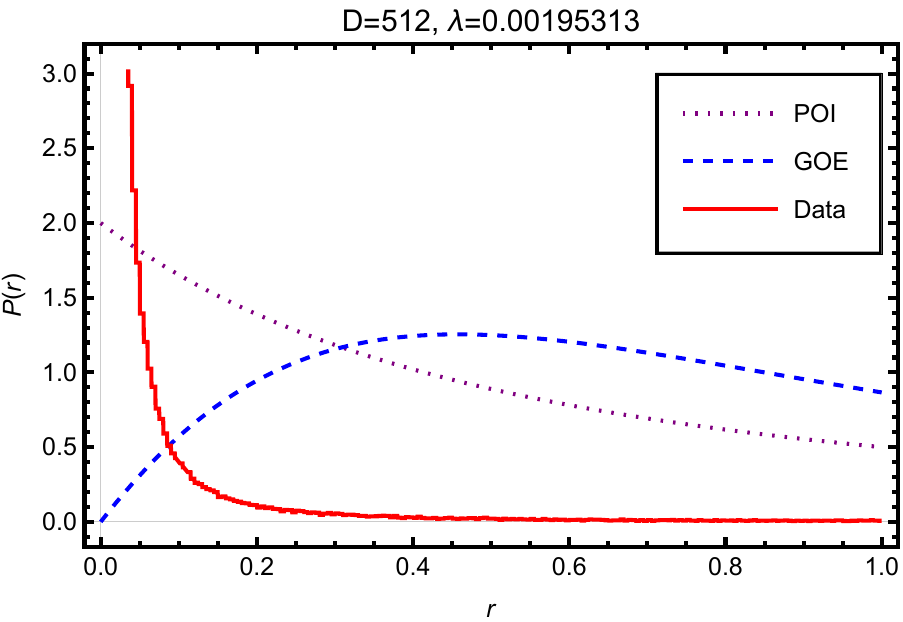} \includegraphics[width=0.32\textwidth]{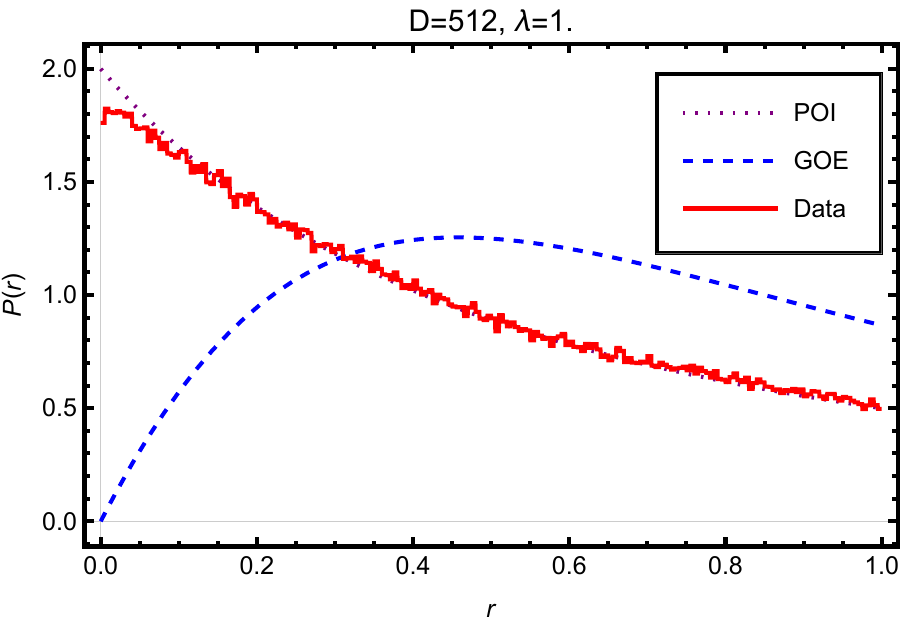} \includegraphics[width=0.32\textwidth]{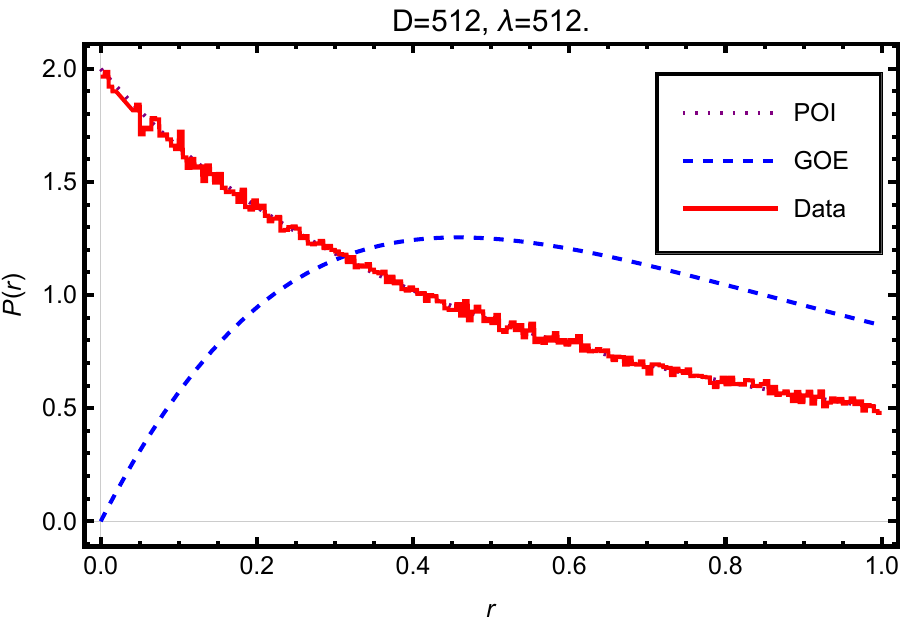}
  \caption{\footnotesize Eigenvalue statistics for a particle moving on a circle in the presence of a random potential, $H = L + L^{-1} + \lambda \sum_n \zeta_n \delta(U - e^{2\pi i n/D})$, aggregated over 2000 instances of disorder. The random strengths $\zeta_n$ are independently drawn from the Gaussian distribution $\mathcal N(0, 1)$. Eigenvalue statistics are shown for $\lambda = 1/D \approx 0.002$, $\lambda = 1$, and $\lambda = D = 512$, and they show how the system transitions from a free theory with uniform eigenvalue spacing to an Anderson-localized one with Poisson-distributed eigenvalue gaps.} \label{fig AL evals}
\end{figure}

\begin{figure}
  \centering
  \includegraphics[width=0.48\textwidth]{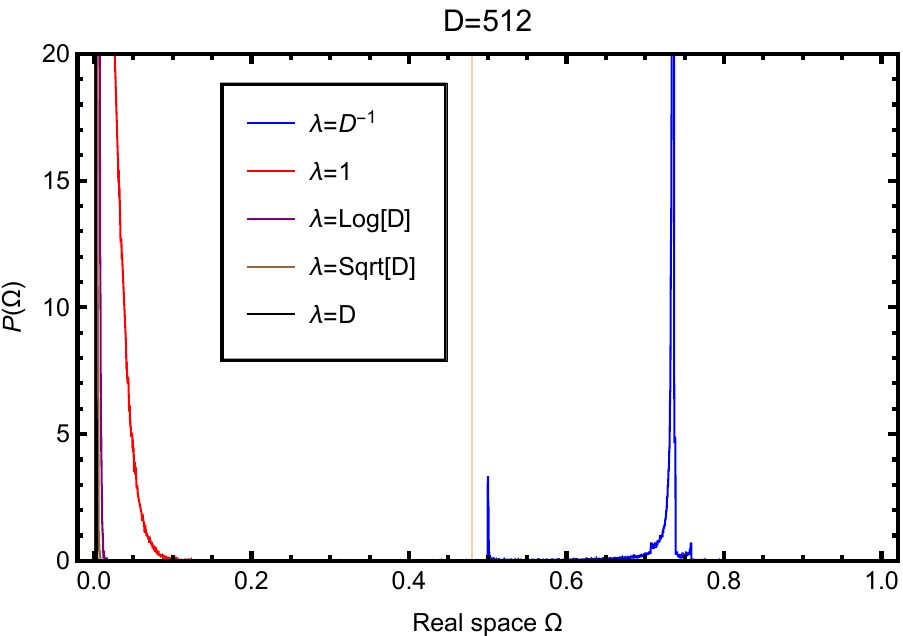} \includegraphics[width=0.48\textwidth]{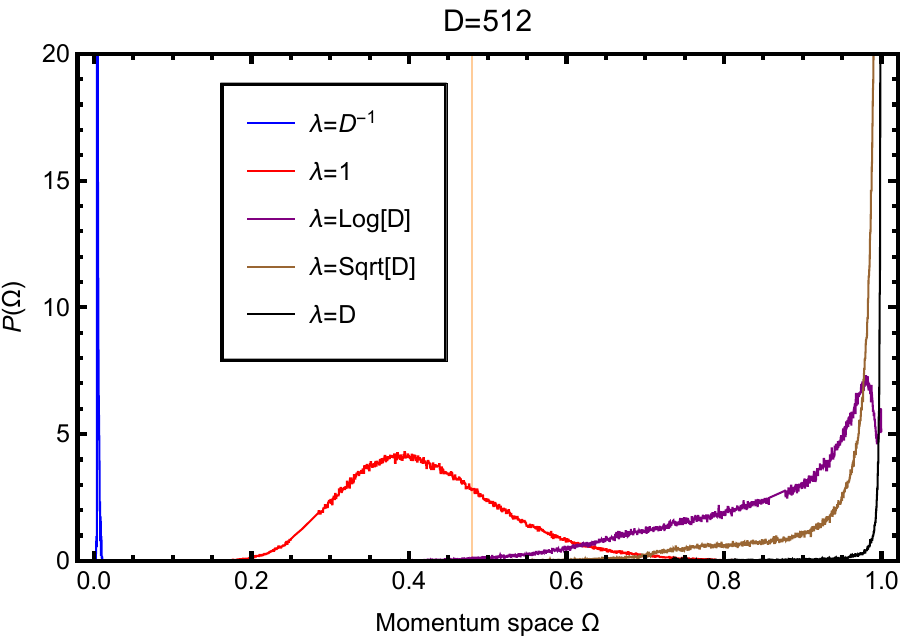}

  \caption{\footnotesize Distributions of IR diversities $\Omega$ for the particle in a random potential, with the coupling $\lambda$ is varied from $1/D$ to $D$. The left plot shows IR diversities calculated in the position basis, and sharp localization at both strong and weak coupling is clearly visible. The localization at $\Omega < 1$ in the free theory limit is due to the arbitrary choice of polarization between degenerate basis vectors. The right plot shows the distribution of diversities calculated in the momentum basis, in which a different polarization of degenerate states ensures that $\Omega = 0$ for the free theory states.} \label{fig AL ent}
\end{figure}

It is important to note that quantum ergodicity criteria are never fulfilled in these examples, even if $\lambda$ is allowed to take values that scale with $D$ so that a crossover is observed; at no point during the crossover do the eigenvalues  follow the GOE surmise \eqref{eq surmise}, as seen on Figs.~\ref{fig AL evals} and \ref{fig AL ent}. The effect we see should be understood as a smooth transformation between position and momentum space localization without emerging ergodicity. The behavior in the regime where the continuum limit applies is in agreement with the classical expectation that there exists no chaos in one spatial dimension.

\begin{figure}
  \centering
  \includegraphics[width=0.48\textwidth]{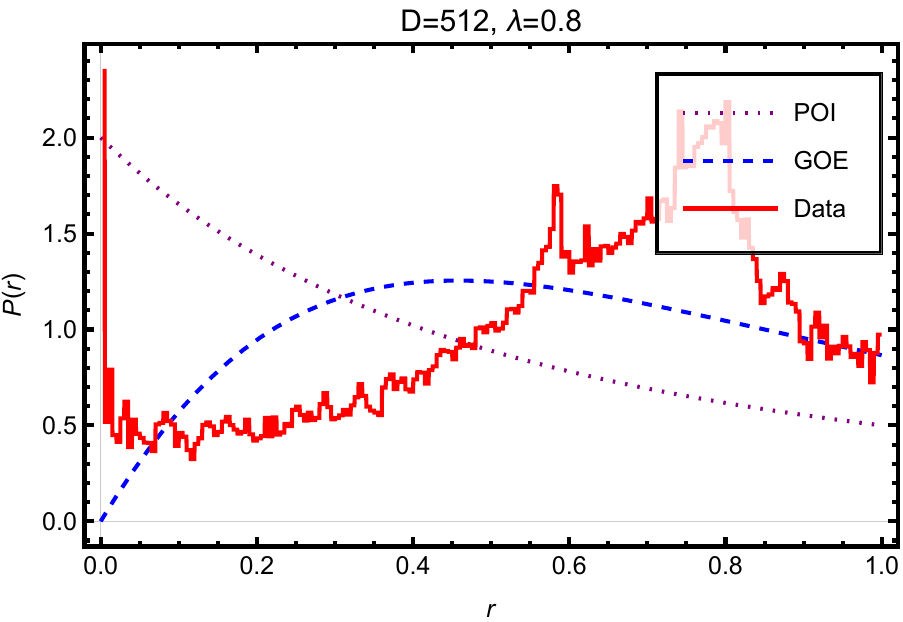} \includegraphics[width=0.48\textwidth]{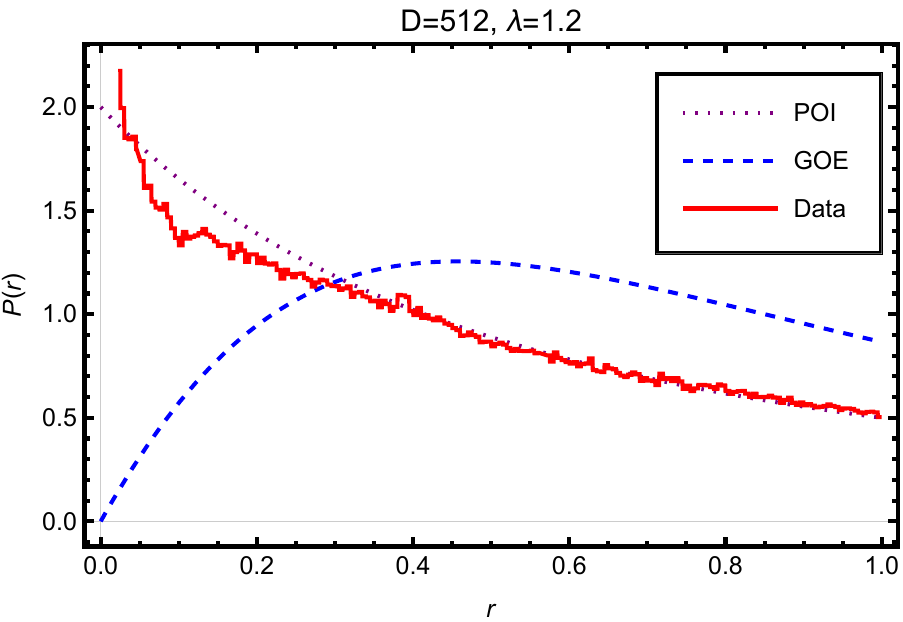}

  \includegraphics[width=0.48\textwidth]{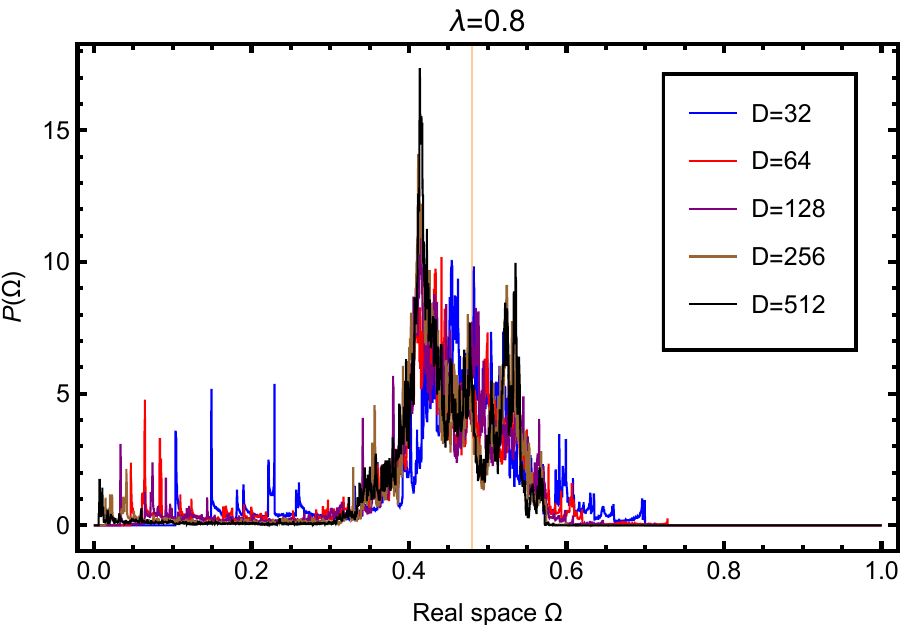} \includegraphics[width=0.48\textwidth]{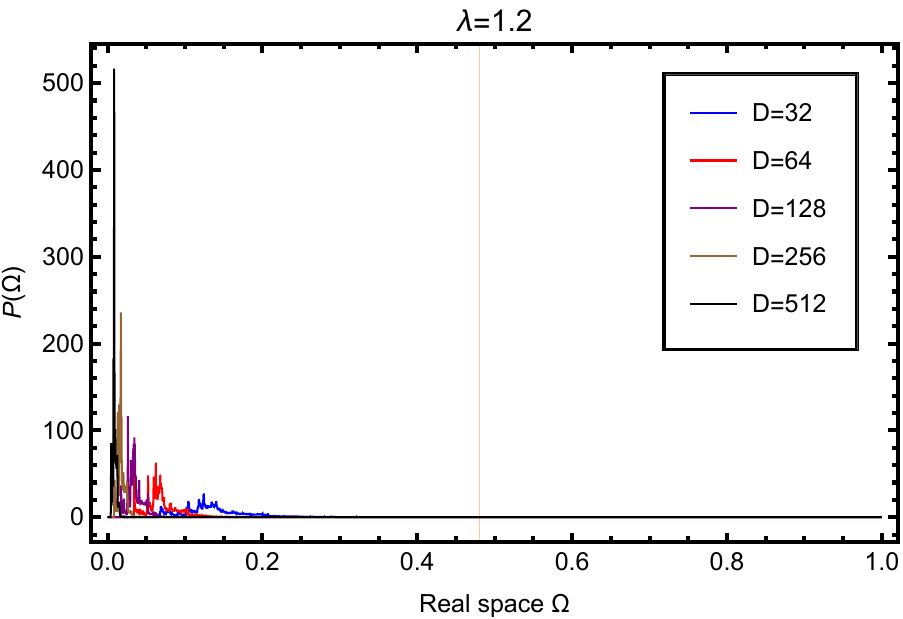}
  \caption{\footnotesize Eigenvalue statistics and IR diversities below and above the critical coupling in the Aubry-Andr\'e model with potential $V(e^{2\pi i n/D}) = 2\lambda \cos (2\pi \omega n + \phi)$, with $\omega = \frac{\sqrt 5 - 1}2$ and $\phi$ chosen uniformly from $[0, 2\pi]$. The plots show aggregate eigenvectors of $1,024,000/D$ realizations of $\phi$. The theory transitions from free (momentum-localized) at $\lambda \rar 0$ to position-localized with Poisson statistics at $\lambda \rar \infty$. The regime at finite $\lambda < 1$ is not universal, and does not appear to smooth out as $D$ is increased; this agrees with the known self-similarity of the eigenstates.}\label{fig AA}
\end{figure}

\begin{figure}
  \centering
  \includegraphics[width=0.48\textwidth]{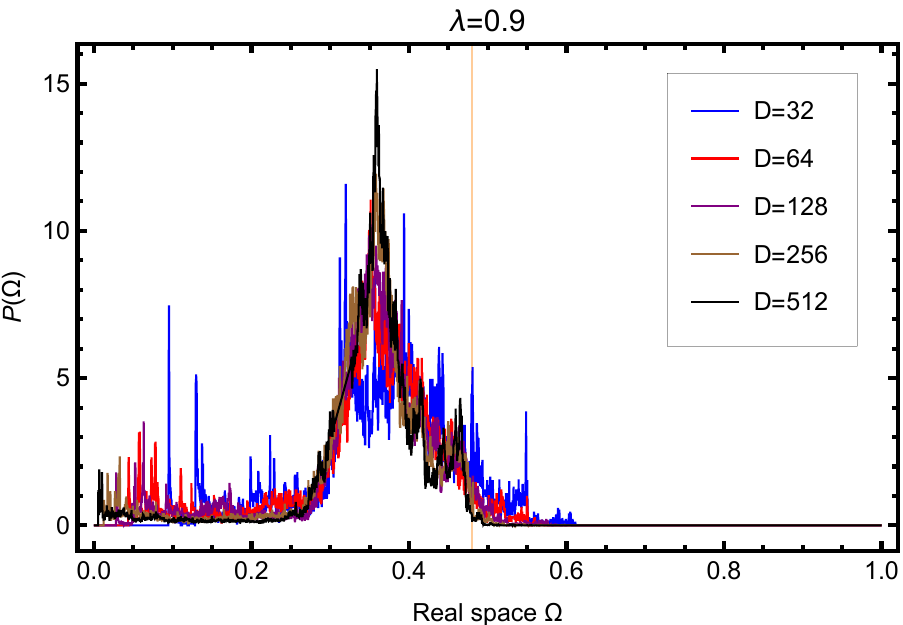}
  \includegraphics[width=0.48\textwidth]{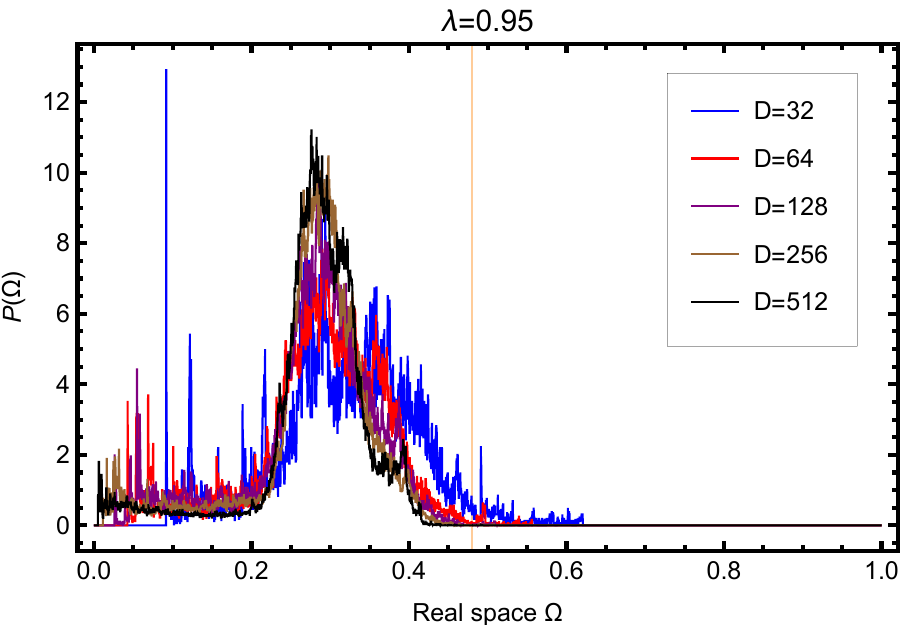}

  \includegraphics[width=0.48\textwidth]{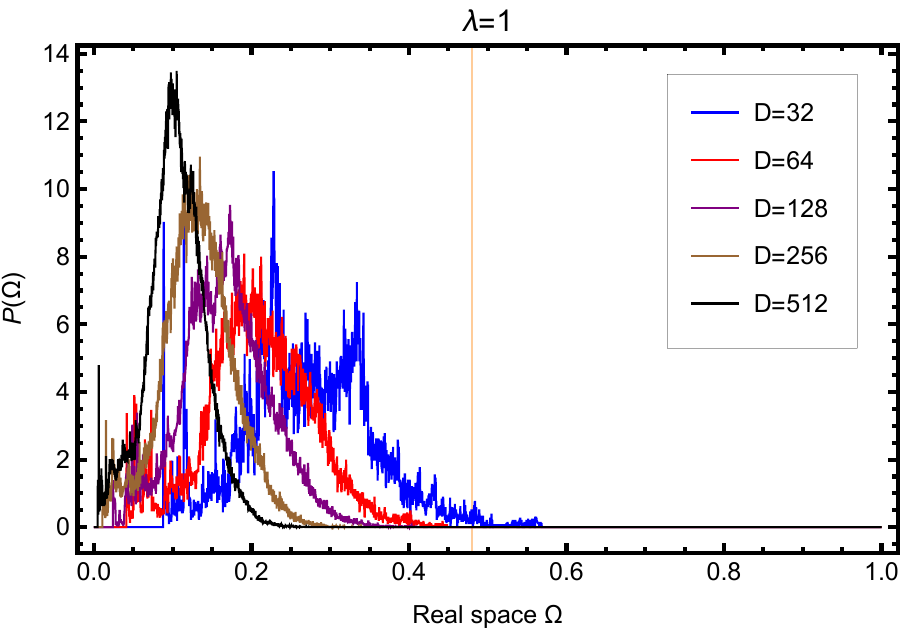}
  \includegraphics[width=0.48\textwidth]{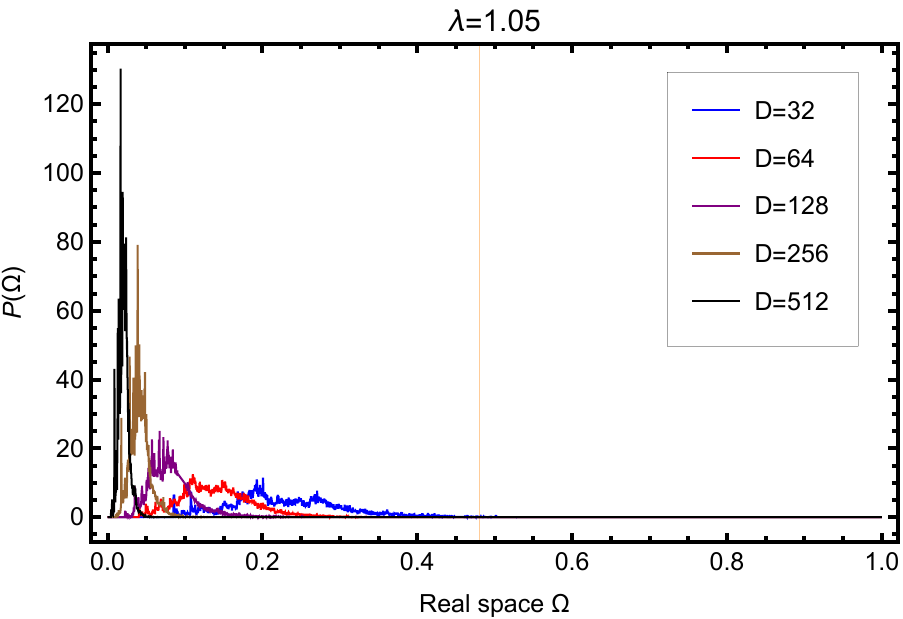}

  \caption{\footnotesize IR diversities of the Aubry-Andr\'e model near the transition location ($\lambda_c = 1$), calculated for the same model as on Fig.~\ref{fig AA} and shown for an aggregate of 1000 runs at each $D$. A quick crossover from momentum to position space localization is observed. At $\lambda = 0.90$ and $\lambda = 0.95$, the $\Omega$-distributions appear to keep roughly the same shape as the system size is increased. This may be a sign of an unusual kind of criticality.}\label{fig AA transition}
\end{figure}

A more subtle situation arises in models where the above crossover turns into a sharp transition at large $D$. An example exists already among theories of a single particle in one dimension: the Aubry-Andr\'e model \cite{Aubry:1980}, defined for an irrational $\omega$ as
\bel{\label{eq AA}
  H = L + L^{-1} + \lambda (U^\omega + U^{-\omega}).
}
This model has a crossover concentrated in a small neighborhood of $\lambda_c = 1$, and at $D \rar \infty$ this becomes a sharp transition, see Figs.~\ref{fig AA} and \ref{fig AA transition}. The detailed nature of this transition depends strongly on the algebraic properties of $\omega$, but in general all the eigenstates become self-similar and delocalized in a very special way \cite{Wilkinson:1984, Kachiga:1986, Geisel:1991}. A potentially related, unusual kind of criticality that is achieved as $D \rar \infty$ has also been studied in Ref.~\citen{Kaplan:2009kr} in the context of a particle in the $1/r^2$ potential, which is famous in the literature on conformal quantum mechanics \cite{deAlfaro:1976vlx, Chamon:2011xk}.

Instead of adding a potential, it is also possible to modify the theory by weighting the kinetic terms in various ways. Doing so in a position-dependent way amounts to including terms of the form $U^n L^m$ into the Hamiltonian. Let us consider one such Hamiltonian,
\bel{
  H = \sum_n \zeta_n \left(\delta(U - e^{2\pi i n/D}) L + \trm{h.c.}\right).
}
This describes a free particle in one dimension whose motion on each link is modulated by some random, link-dependent noise. We choose each $\zeta_n$ to be a Gaussian with mean one and variance $\lambda$. Just as with the particle in a random potential, the eigenstates start becoming delocalized when $\lambda \gtrsim 1/D$.  Unlike that theory, however, here dialing $\lambda$ until it is $O(1)$ does something novel: it effectively prohibits hopping between some randomly selected pairs of sites. This splits the target space into multiple disconnected components. Further increasing $\lambda$ causes more splittings to happen, as any $|\zeta_n|$ much smaller than the maximal one becomes a negligible term in the Hamiltonian. We thus reach a theory where, in effect, a sizeable portion of links was chosen at random and  diluted.

This kind of modulation of hopping elements is very different for quantum mechanics in higher dimensions. Instead of trivially localizing the theory, this dilution of links gives rise to ergodicity. Indeed, we will see in section \ref{sec syk} that this effect is necessary in order to understand the ergodic behavior of the SYK model.

Another fact of interest is that if $\zeta_n$ are drawn from a Rademacher distribution, with each $\zeta_n$ taking values $\pm 1$ with probability $1/2$, there is essentially no effect on quantum ergodicity. Thus, instead of having taken $\zeta_n$ to be Gaussian random, we could have taken them to be absolute values of Gaussian random variables, and our conclusions would have been unaltered.

We have now seen that restricting hopping between sites ruins the momentum space localization of free theory eigenstates. One may then ask whether it is possible to add more hopping terms so that:
 \begin{enumerate}
   \item the problem remains that of a particle moving in one dimension (i.e.~the number of nonzero terms above the diagonal in the Hamiltonian never deviates from $D$ by more than $O(1)$);
   \item momentum space localization stays ruined;
   \item the region remains connected, and thereby position space localization is prevented.
 \end{enumerate}   This leads us to the study of \emph{metric graphs} \cite{Kottos:1997}. As an example, consider a particle moving on the edges of a tetrahedron. This problem is essentially the same as a free particle moving on a circle: the only modification is that at a few points (the tetrahedron vertices) the incoming wavepacket can now split into multiple wavepackets.  There exists a wealth of graphs and boundary conditions that can be studied, and this subject is well-researched. In particular, it was found that graphs lacking any accidental symmetries --- such as tetrahedron or star graphs with incommensurate edges --- show random matrix eigenvalue statistics but \emph{not} quantum ergodicity \cite{Berkolaiko:2003}. When the semiclassical limit is applicable, this phenomenon can be understood in terms of summing short classical orbits \cite{Kaplan:2001}, and a similar (though fully quantum and more schematic) understanding will be offered in this paper. The conclusion is that satisfying all three criteria is essentially impossible: to achieve full quantum ergodicity, one needs to consider Hamiltonians with significantly more that $D$ hopping elements (e.g.~$2D$). These correspond to particles moving in more than one dimension. This is the quantum analog of the statement that there is no classical chaos in one dimension.

\subsection{Quantum mechanics in two dimensions} \label{subsec qm2}

Let us now inject the smallest amount of product structure into the Hilbert space: we consider the quantum particle moving in two dimensions. The algebra is generated by two position and momentum operators, and an appropriate local Hamiltonian is
\bel{
  H = L_1 + L_1^{-1} + L_2 + L_2^{-1} + \lambda V(U_1, U_2).
}
This is a particle moving on a torus. The situation is similar to the one in one dimension. For smooth potentials, $\lambda$ may need to be dialed as far as $\lambda \sim D$ for the entire spectrum to get localized. On the other hand, for random potentials, we again have Anderson localization for any $\lambda$, with the caveat that as the coupling goes to zero the localization length diverges as $\exp(1/\lambda^2)$, rendering all states delocalized at any finite system size \cite{GangOfFour}. Localization at arbitrary coupling is no longer possible in higher dimensions.

The problem of random hoppings between lattice sites is much more interesting now than in one dimension. It is also well-studied and goes by the name of \emph{quantum percolation} or \emph{off-diagonal disorder} \cite{deGennes:1959, Shapir:1982, Chayes:1986}. Off-diagonal disorder effectively dilutes (or even removes) some links from a regular $d$-dimensional grid. A quantum percolation transition is observed as the disorder strength varies. Once a critical percentage of links is removed, all states become localized in position space; below this value, states are extended, though not necessarily ergodic --- depending on the potential, they can be localized in momentum space. The critical link density is dimension-dependent. For $d = 2$ it is equal to unity, which means that for a particle moving on a two-dimensional manifold, any nonzero density of hopping defects leads to complete position space localization of the spectrum. For higher dimensions $d$, the critical link density decreases, so for $d \gg 1$ the spectrum is in general delocalized and possibly ergodic.

\begin{figure}[b]
  \centering
  \includegraphics[width=0.4425\textwidth, keepaspectratio = true]{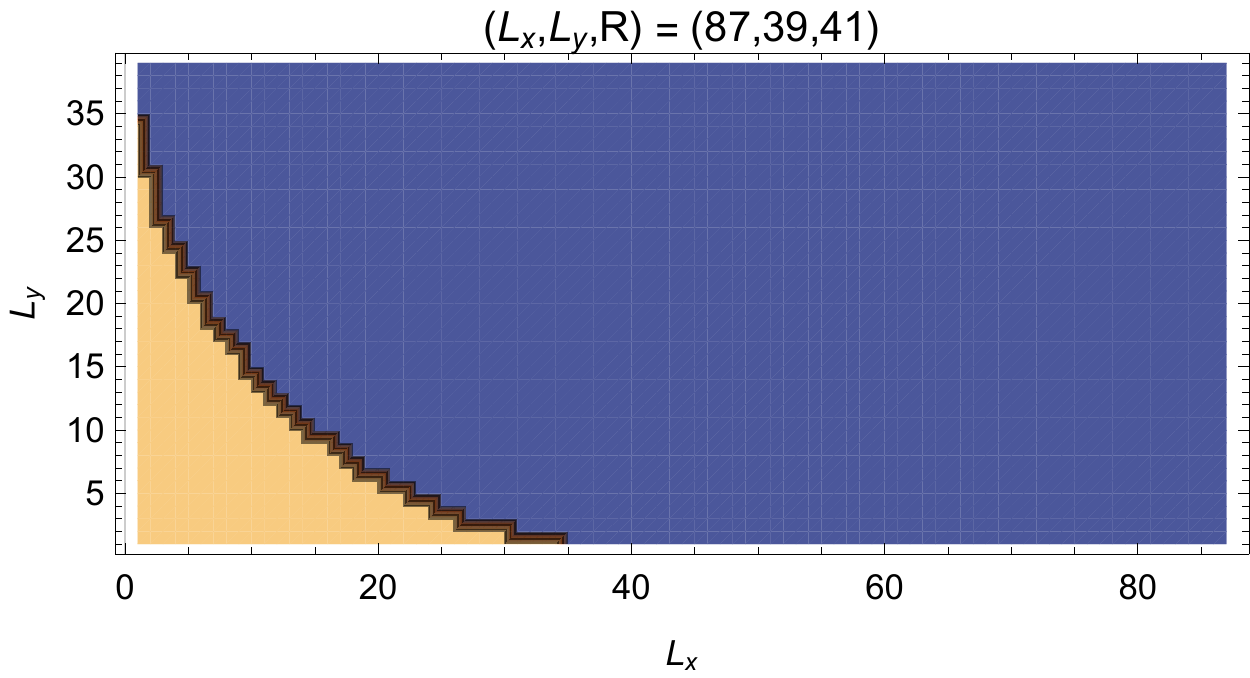}
  \includegraphics[width=0.52\textwidth, keepaspectratio = true]{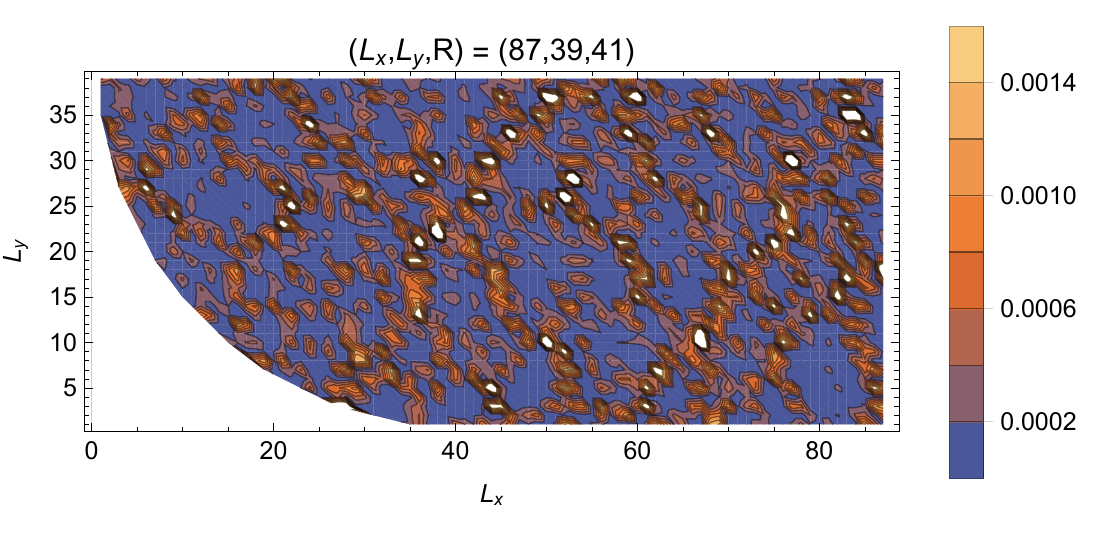}
  \caption{\footnotesize Left: the Bunimovich stadium \cite{Bunimovich:1970}, realized as a circular obstacle of radius $R$ on an $L_x \times L_y$ torus. Right: a typical eigenstate of the free particle on the Bunimovich stadium.}\label{fig stadium}
\end{figure}

\begin{figure}
  \centering
  \includegraphics[width=0.48\textwidth, keepaspectratio = true]{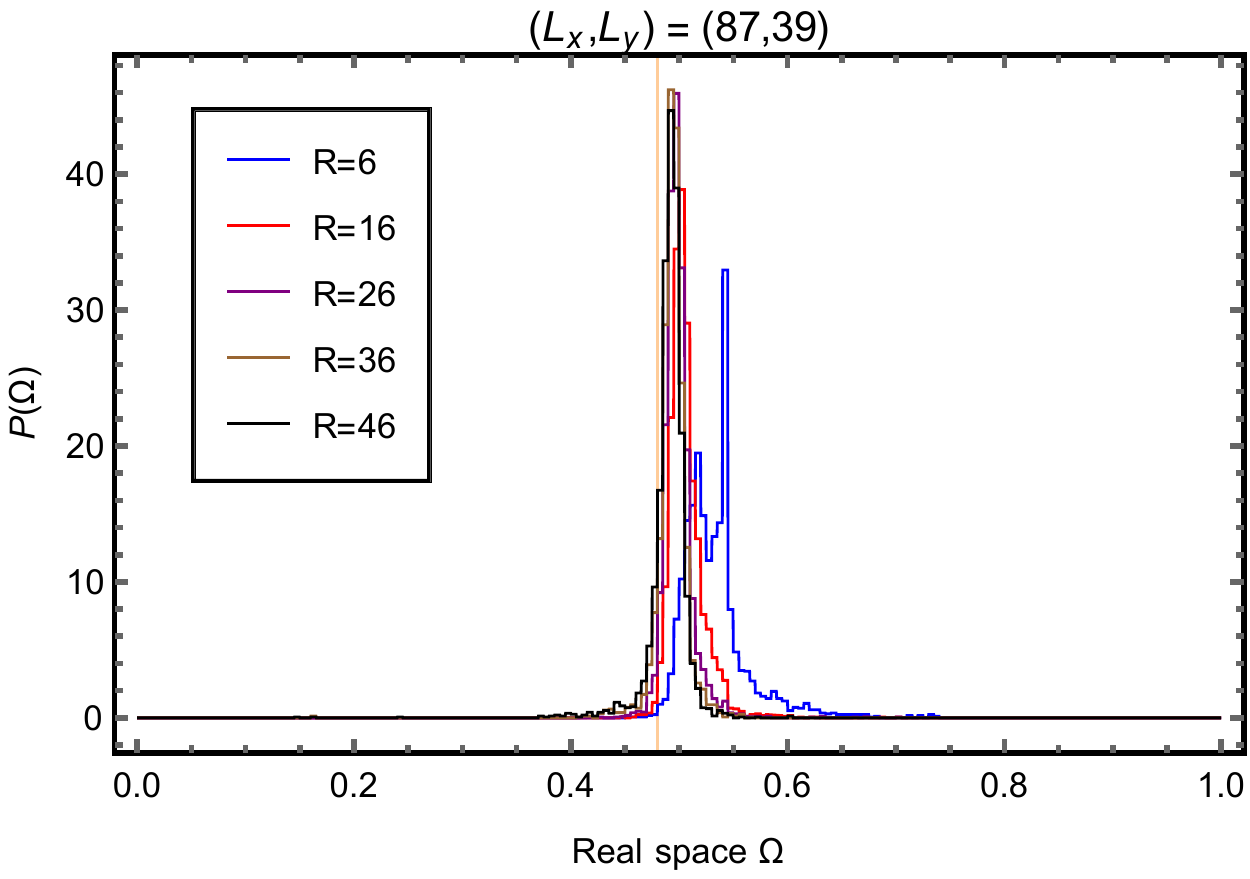}
  \includegraphics[width=0.48\textwidth, keepaspectratio = true]{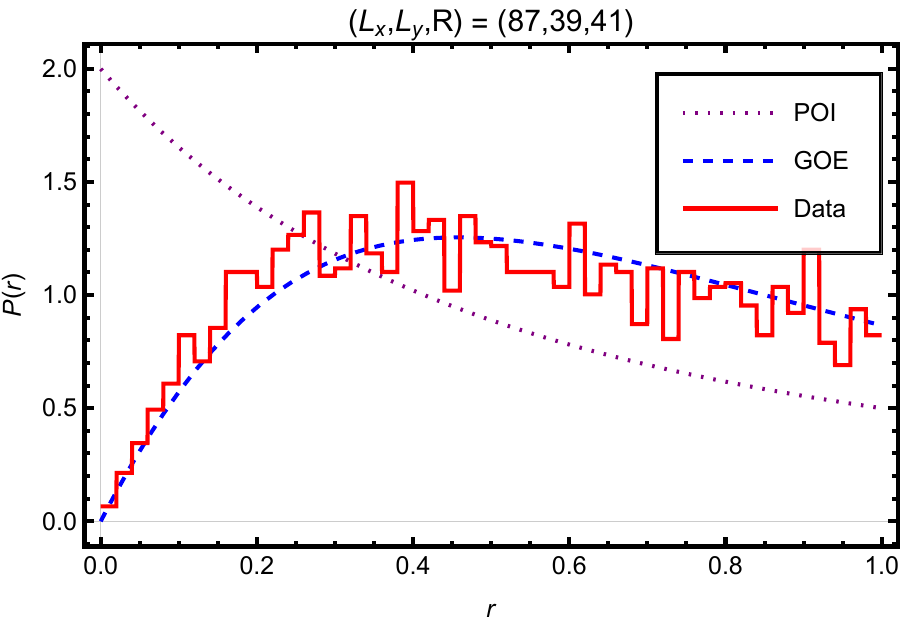}
  \caption{\footnotesize Left: the distribution of IR diversities for the Bunimovich stadium, with varying sizes of the obstacle. For all macroscopic obstacle sizes, the spectrum is almost fully ergodic. Right: the corresponding eigenvalue statistics, indicating good agreement with the expected GOE behavior.}\label{fig stadium ent}
\end{figure}

The famous stadium problems \cite{Sinai:1963, Bunimovich:1970, Berry:1981} that have long been archetypes of quantum chaos (see Fig.~\ref{fig stadium}) can also be accessed as special cases of weighted hoppings on a torus. Here, instead of randomly distributing the hopping weights, we choose to purposefully remove links (assign them weight zero) so that the particle starts feeling an impenetrable obstacle whose macroscopic shape corresponds to the classical boundaries of the billiard. A similar approach was taken in Ref.~\citen{Radicevic:2016kpf}, where $\delta$-function obstacles were placed on sites to mimic boundaries. Even without any randomness and with very few links removed, these link deletions profoundly alter the spectrum and bring it very close to that of a random matrix (Fig.~\ref{fig stadium ent}). It is a feature of many flat stadii that a sparse set of eigenstates remains localized; these are the famous quantum scars \cite{Heller:1984, Hassell:2008}.

We now move from flat manifolds with obstacles to curved manifolds. Most relevant results here are formulated directly in the continuum. Positively curved manifolds have received comparatively little attention in the quantum chaos literature. The motion of a particle on a two-sphere, while integrable, has a large number of degeneracies that make the usage of eigenvector statistics ill-defined unless a lot of other data (polarizations) is chosen; a randomly chosen polarization is known to be ergodic \cite{Zelditch:1992, Brooks:2015}. This shows that the notion of quantum ergodicity in systems with many symmetries can be quite subtle.

Particles moving freely on negatively curved two-manifolds are poster children of quantum chaos --- perhaps even more so than billiards. Laplacians on generic hyperbolic manifolds are expected to have  strongly ergodic spectra. This statement is made precise by the Quantum Unique Ergodicity (QUE) conjecture, which posits that hyperbolic surfaces have no quantum scars and that \emph{all} eigenstates are ergodic \cite{Rudnick:1994}. Perhaps ironically, the one set of manifolds for which QUE has been rigorously established is the set of \emph{arithmetic billiards} \cite{Bourgain:2003, Anantharaman:2008}, for which eigenvalue statistics is known to be that of integrable systems \cite{Bogomolny:2003, Bogomolny:2003b}. This happens because the manifolds in question are quotients of the hyperbolic plane by subgroups of the modular group, leaving a large number of symmetries present in the system --- without making the theory integrable.

Finally, \emph{good expander graphs} can be seen as discrete versions of hyperbolic manifolds. Examples of special importance here are regular graphs, which have an equal number of links emanating from each vertex. Note that these examples all involve a particle hopping from one vertex of a graph to another, with links representing hopping terms in the Hamiltonian. This is not the same situation as with metric graphs in the previous section, where the majority of possible positions were \emph{within} the links. Quantum mechanics on random regular graphs is known to be ergodic \cite{ORourke:2016, Gnutzmann:2006}. On the other hand, for highly symmetric regular graphs, e.g.~for the edges of a hypercubic lattice, the dynamics is integrable.

\subsection{A semiclassical interlude: trace formulae and periodic orbits} \label{subsec tr}

The results presented so far were based on exact diagonalization of various Hamiltonians and on model-specific analytic arguments. It is natural to ask whether an all-purpose perturbative approach may help us understand these results. Certainly, one can consider an exactly solvable Hamiltonian and develop perturbation theory in $\lambda$ around it. However, small perturbations of integrable systems typically do not lead to a fully ergodic phase. The alternative is to study chaos by expanding in $1/D$ while leaving $\lambda$ arbitrary (even of the order of $D$). This is the semiclassical limit.

For a classical limit to apply, it is not enough to have a particle hopping on a large graph; some regularity is necessary, such as translation invariance. A free particle hopping on a circle or a torus naturally admits a classical limit, as do slow-varying perturbations thereof. A particle hopping on a metric graph \emph{almost} admits a classical limit: it moves classically on each link, but encounters a singularity at a vertex. This is why the classical limit of a metric graph problem is always a random walk between the vertices of that graph \cite{Gnutzmann:2006}. A particle hopping on a torus with a singularity in the form of a $\delta$-function admits a classical limit as a particle moving on a torus with an obstacle with Dirichlet boundary conditions. A particle hopping on a graph that uniformly expands in all directions admits a classical limit of a particle on a hyperbolic manifold. In all these cases, semiclassical methods (systematically including long-time/finite-size effects in the form of $1/D$ corrections) provide powerful tools for understanding chaos.

The essence of semiclassicality is the use of \emph{trace formulae}, which relate the density of states to path integral problems that admit a saddle-point approach \cite{Gutzwiller:1971, Berry:1989}. To briefly summarize this philosophy, one expresses the density of states as
\bel{\label{eq trace}
  \varrho(E) = \frac1D \sum_n \delta(E - E_n) = -\frac1{\pi D} \trm{Im} \Tr \frac1{H - E + i\epsilon} =  \frac1{\pi D} \trm{Re} \int_0^\infty \d T\,\Tr\, e^{i(H - E + i\epsilon) T}.
}
The trace in this formula can now be expressed as a path integral over all periodic trajectories of period $T$. Classically, this path integral is dominated by the sum over classical orbits (with the appropriate boundary conditions at obstacles or graph vertices, as required). Thus the density $\varrho(E)$ is given by a the Fourier transform in $T$ of the sum over all periodic orbits of length $T$.

For many chaotic systems, the number of states in a window $\d E$ is known to look like the one found in Gaussian ensembles even at relatively small scales (e.g.~at $\d E \sim \frac{\log D}D$) \cite{Bauerschmidt:2016}. In order to access local eigenvalue statistics at microscopic scales $\eps \sim 1/D$, it is necessary to study correlations of the type $\varrho(E) \varrho(E + \eps)$ averaged over a large window of energies $E$ (see e.g.~Ref.~\citen{Berry:1989}). A careful semiclassical analysis (on an arbitrary classically chaotic manifold) can then account for finite $D$ effects due to nearby periodic orbits and give spectral correlators that agree with those from random matrix theory \cite{Muller:2005, Muller:2009}. Moreover, the sums of periodic orbits can be shown to be generated by an effective matrix model, thereby providing a link to random matrix theory \cite{Efetov:1997}. A similar story exists for metric graphs, where random matrix theory is accessed through counting graph cycles of a given length \cite{Gnutzmann:2006}.

These semiclassical methods show that spectral statistics no longer agrees with the random matrix one once there is a lot of structure in the sum over periodic orbits.  The simplest example, of course, comes from applying the trace formula to the free theory of a particle on a torus \cite{Bogomolny:2003}. More intricate examples come from the arithmetic billiards mentioned above. Similarly, spectral universality fails for metric graph problems when graphs have commensurate edges, as this leads to extra structure in the sum over cycles. What happens in all these examples is simple to state but difficult to prove in a robust way: the proliferation of incommensurate periodic orbits means that the phases that enter trace formulae are independent random phases, and a form of central limit theorem ensures that their sum has a universal distribution. If there is a lot of structure to the problem, the phases entering the trace formulae are no longer independent, and the central limit theorem fails. This is the thought that motivates us to investigate how having a lot of structure hidden in a Hamiltonian leads to the failure of ergodicity, even when semiclassical methods do not apply.

\subsection{Random matrix theory} \label{subsec rmt}

Having exhausted semiclassical methods in simple quantum mechanical problems, let us now step back and examine what is known about quantum ergodicity in random matrix theory. A good survey of recent results is Ref.~\citen{ORourke:2016}. We will pay extra attention to results obtained for sparse random matrices, as these are the ones that most closely correspond to physical Hamiltonians. As before, we focus on real-valued matrices that are expected to be in the universality class of Gaussian orthogonal ensembles (GOE).

There are many classical probability results about Gaussian ensembles. Gaussian random matrices are known to have eigenvectors drawn uniformly from the set of $D$-dimensional vectors with unit norm, and the largest entry of a random matrix eigenvector will be bounded by a number scaling as $\sqrt{\frac{\log D}D}$ with overwhelming probability. The density of eigenvalues of a properly normalized random matrix will obey the \emph{Wigner semicircle law} at scales greater than the average level spacing, i.e.~at $\d E \gg 1/D$,
\bel{
  \varrho(E) = \frac1{2\pi} \sqrt{4 - E^2}.
}

The concept of universality starts playing an interesting r$\hat{\trm o}$le once we shift our attention to Wigner matrices, whose each entry is an independent, not necessarily Gaussian random variable. Typically one takes off-diagonal entries to have one distribution and on-diagonal elements to have a different distribution. The statement of interest is that the spectra of Wigner matrices display the same statistics as Gaussian random matrices as long as the first four moments of the off-diagonal distribution and the first two moments of the on-diagonal distribution match the appropriate moments of a Gaussian distribution (for a review, see Ref.~\citen{Tao:2011}). Note that universality may fail at the edges of the spectrum; this means that some low-energy states may appear localized even if the rest of the spectrum is ergodic.

\begin{figure}
  \centering
  \includegraphics[width=0.48\textwidth,keepaspectratio = true]{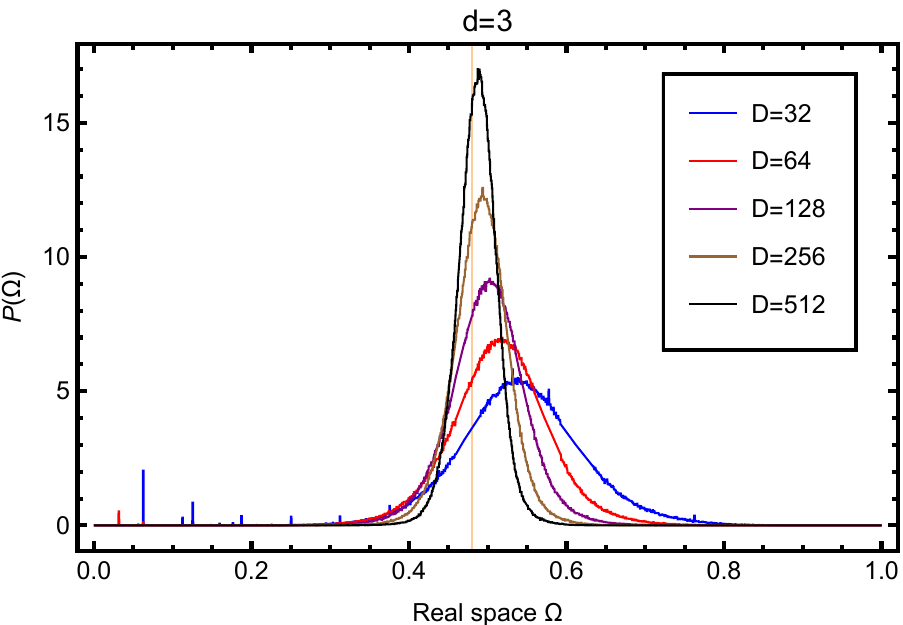}
  \includegraphics[width=0.48\textwidth,keepaspectratio = true]{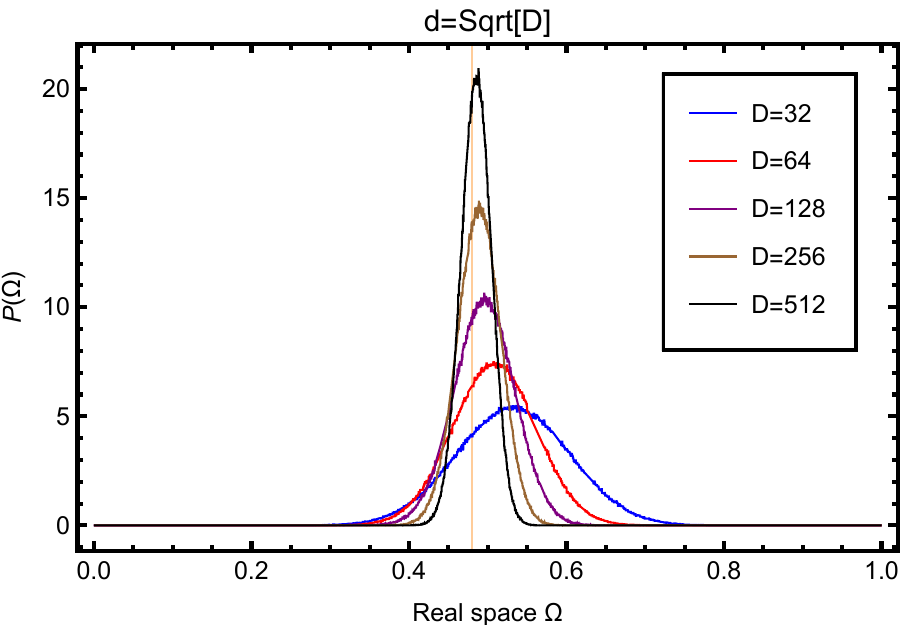}
  \caption{\footnotesize IR diversities for random regular graphs with fixed and varying connectivities $d$. In both cases the theory quickly approaches the GOE benchmark as $D$ grows. For each $D$, the presented diversities are aggregates of $1,024,000/D$ graph realizations, chosen uniformly from the space of all random regular graphs with a given connectivity.}\label{fig RR graph entropies}
\end{figure}

Both GOE and Wigner matrices are almost surely dense. Many sparse random matrices are also known to fall into the Gaussian universality class.  In particular, adjacency matrices of random graphs with connectivity much lower than $D$ are sparse and typically ergodic, as seen on Fig.~\ref{fig RR graph entropies} \cite{Erdos:2016, Bauerschmidt:2016}.\footnote{Recall that $H$ is an adjacency matrix of an unweighted graph with sites $i$ if $H_{ij} = H_{ji} = 1$ whenever sites $i$ and $j$ are connected by an edge, and $H_{ij} = H_{ji} = 0$ otherwise.} For a graph with $d$ links emerging from each site, the density of states of the properly normalized adjacency matrix follows the \emph{Kesten-McKay distribution},
\bel{
  \varrho(E) = \frac{d(d - 1)}{2\pi}\frac{\sqrt{4 - E^2}}{d^2 - (d - 1) E^2},
}
with the Wigner distribution reached when the connectivity $d$ is taken to infinity together with $D$, for instance if $d \sim \log^q D$ as is the case in typical local QFTs. In addition, in this case it is known that the bulk of the spectrum is ergodic --- though there may be individual localized states, such as the highest energy state which is localized in momentum space (uniformly delocalized across the entire graph in position space). 

\begin{figure}
  \centering
  \includegraphics[width=0.48\textwidth,keepaspectratio = true]{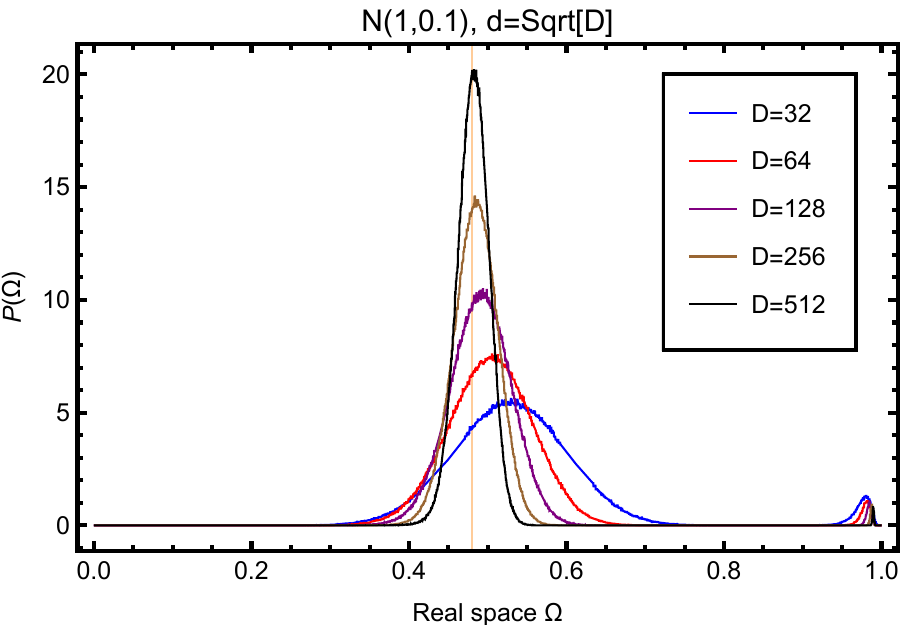}
  \includegraphics[width=0.48\textwidth,keepaspectratio = true]{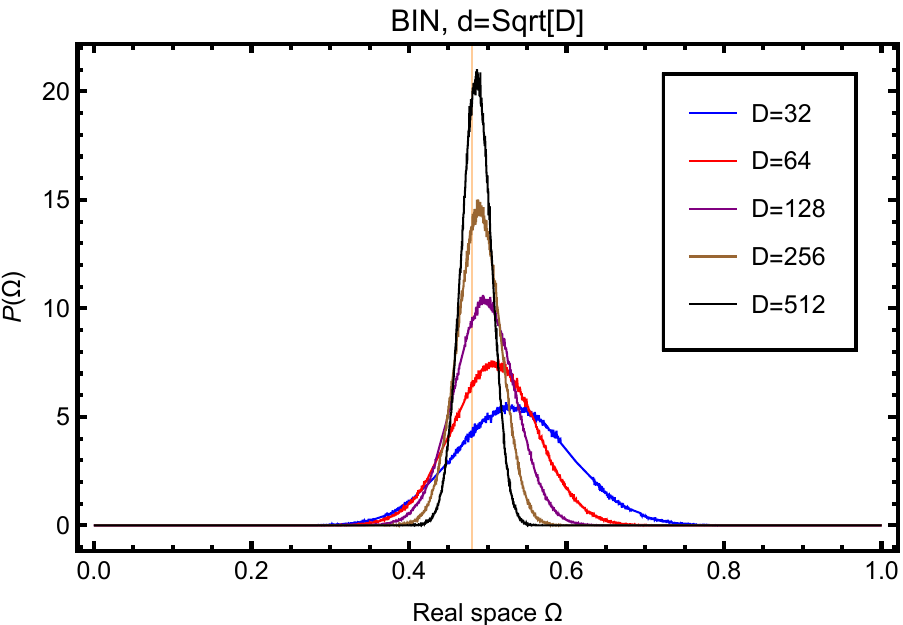}
  \caption{\footnotesize IR diversities of random regular graphs with disorder that does not affect their ergodicity. Left: each link is weighted with a random variable drawn from $\mathcal N(1, 0.1)$. Right: each link weight is randomly chosen from the set $\{+1, -1\}$.}\label{fig small disorder RR}
\end{figure}

\begin{figure}
  \centering
  \includegraphics[width=0.48\textwidth,keepaspectratio = true]{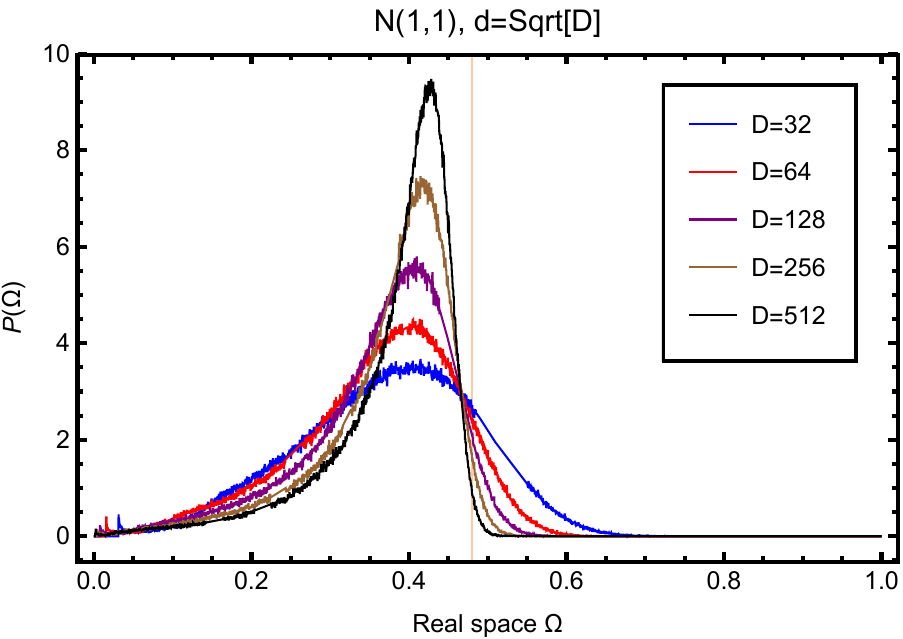}
  \includegraphics[width=0.48\textwidth,keepaspectratio = true]{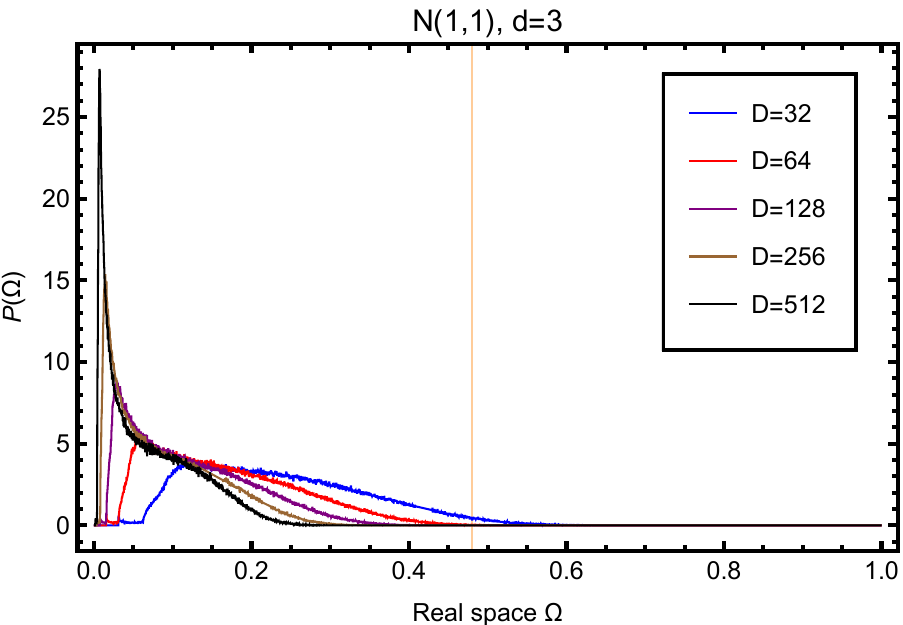}
  \caption{\footnotesize IR diversities of random regular graphs with disorder that observably affects their ergodicity. In both cases, each link is weighted with a random variable drawn from $\mathcal N(1, 1)$. Some weights are now very close to zero and can effectively be erased from the graph. If the connectivity $d$ of the graph is large enough, this makes the states relatively more localized, but this effect disappears as $D$ increases (left plot). If the connectivity is small, the graph becomes disconnected and most eigenstates are localized (right plot). }\label{fig large disorder RR}
\end{figure}

Adding relatively small disorder to the random graph Hamiltonian does not significantly affect its ergodicity properties, as seen on Fig.~\ref{fig small disorder RR}. Even at strong $O(1)$ disorder, the theory tends towards ergodicity in the large $N$ limit as long as the graph is sufficiently well connected (Fig.~\ref{fig large disorder RR}, left). If the disorder is strong enough to effectively separate the graph into disconnected pieces, however, ergodicity is lost; the situation is analogous to that in percolation theories (Fig.~\ref{fig large disorder RR}, right). The study of phases reached as disorder is increased has a rich history; see Ref.~\citen{Kravtsov:2015} and references therein.

Finally, let us comment on one case where random matrix universality fails:  Wigner matrices whose entries have very large tails or infinite variance. In this case it can be proven that a large number of excited states are localized, with the maximal entry of each eigenvector being greater than $1/\sqrt 2$. It was argued in Ref.\ \citen{Soshnikov:2006} that this is so because the matrix is dominated by its largest entries --- in terms of our previous discussions, these correspond to $\delta$-function bumps. Each bump serves to localize a state (or more, depending on the symmetries), and due to the heavy tails of the distribution we may expect a hierarchy of isolated bumps, each weaker than the previous but much stronger than the next one, each localizing a few states, until finally the gaps in the hierarchy become too small and the remaining states (if any) are ergodic.

\subsection{Spin chains} \label{subsec spin}

Finally, let us consider systems in which the operator algebra admits a large amount of product structure.  All quantum field theories/many-body systems fall into this category, but we will focus our attention on Ising spin chains, as they are the most amenable to numerical methods. Here we have $D = 2^N$, and the algebra of observables is generated by Pauli matrix pairs $\sigma_i^x$ and $\sigma^z_i$ for $i = 1, \ldots, N$. These will play the role of position and momentum generators $U_i$ and $L_i$ introduced in Eq.~\eqref{def Ui Li}.\footnote{As mentioned in section \ref{subsec prelim}, it is always possible to dualize a spin chain to a particle hopping between $D$ sites (the Fock space representation), but this does not preserve the locality of Hamiltonians. Just to illustrate this, the explicit duality is
\bel{
  U = \prod_{i = 1}^N (\sigma_i^z)^{1/2^{i - 1}}, \quad L = \sum_{i = 0}^{N - 1} \frac 1 {2^{i + 1}} (\sigma^x_{N - i} - \trm i \sigma^y_{N - i} )\prod_{j = 1}^{i} (\sigma^x_{N - i + j} + \trm i \sigma^y_{N - i + j}),
}
with the principal branch chosen in all non-integer powers involved, and where the imaginary unit is written as i to differentiate it from the counter $i$.} In what follows we will always refer to the $\sigma^z_i$'s as position space operators and the $\sigma^x_i$'s as momentum space operators.

\begin{figure}
  \centering
  \includegraphics[width=0.48\textwidth]{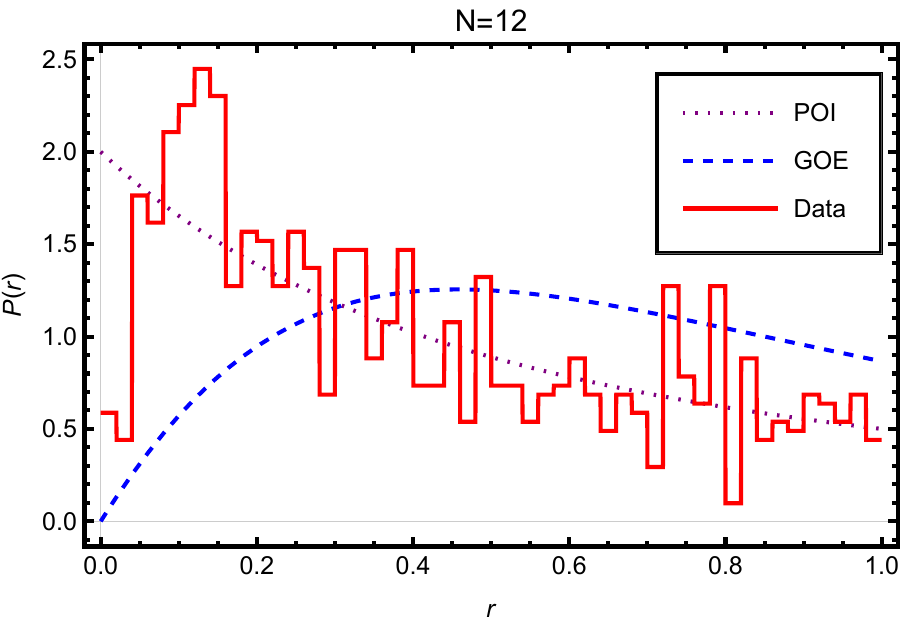}
  \includegraphics[width=0.4625\textwidth]{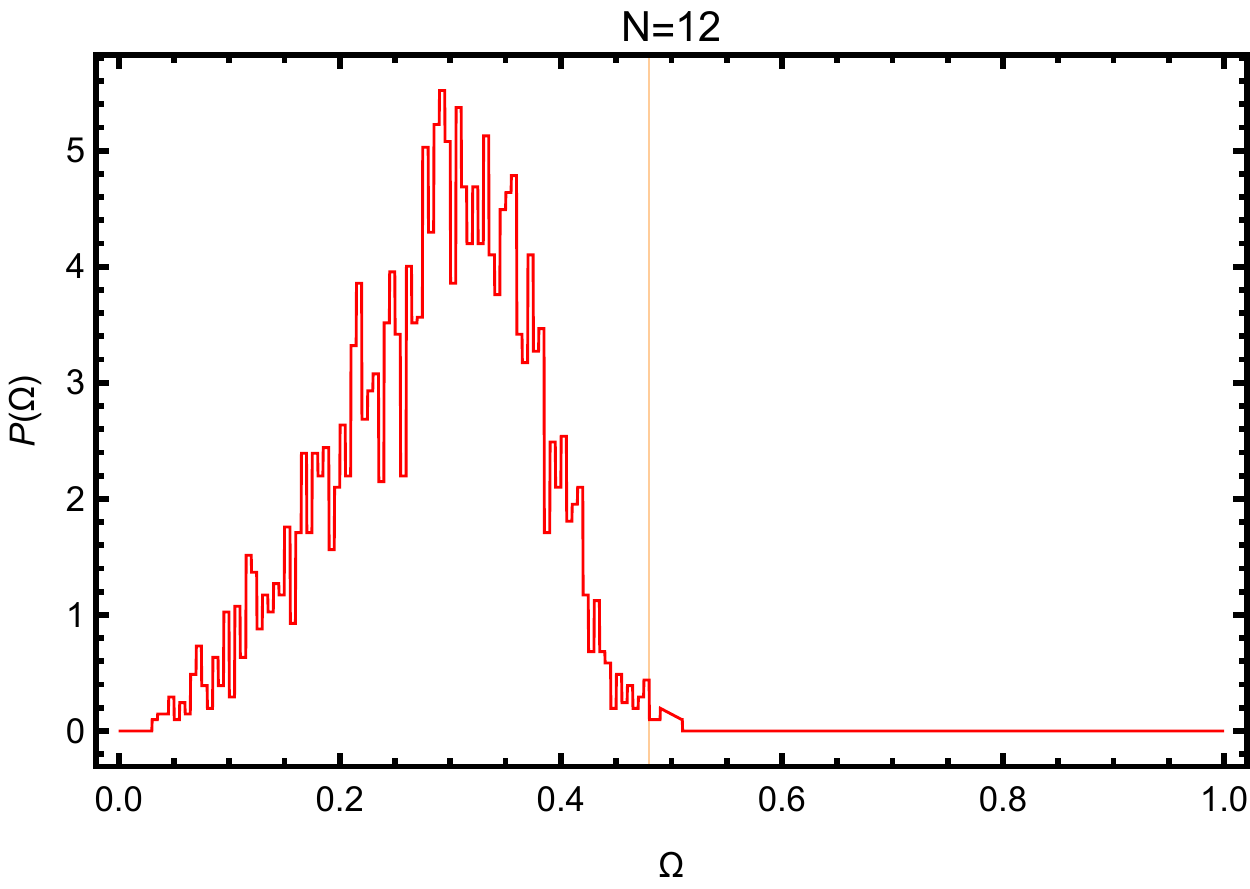}
  \caption{\footnotesize Eigenvalue gap statistics and IR diversities of the critical Ising model with $N = 12$ spins and open boundary conditions. The theory has two $\Z_2$ symmetries, the global spin flip and the reflection across the center of the chain. These symmetries commute and the eigenvalue statistics are computed in one sector of the $\Z_2 \times \Z_2$ symmetry. There was no special choice of polarization when degeneracies were present. The eigenvalue gaps follow a Poisson distribution, as befits an integrable system. The wide profile of the $\Omega$-distribution shows that the eigenstates are far from being ergodic. The localized states with low IR diversity can be interpreted as few-particle states with definite spatial momentum. States with $k$ free particles are maximally delocalized across $\binom N k$ basis vectors, so their diversity numbers are $\#(\Omega) \sim D \Omega$ with $0 \leq \Omega \leq  \frac1{\sqrt{\log D}} \approx 0.3$ for $D = 2^N = 4096$. The linearity, $P(\Omega) \propto \Omega$, is consistent with our result; the ``noise'' comes from unresolved degeneracies. As $D$ grows, the peak slowly shifts towards $\Omega = 0$.}\label{fig ising}
\end{figure}

We start from the transverse-field Ising model,
\bel{
  H = \sum_i \sigma^x_i \sigma^x_{i + 1} + h \sum_i \sigma^z_i.
}
This system is integrable at any $h$ as it is dual to free fermions \cite{Lieb:1961}. At the critical point its eigenvalue statistics are approximately Poisson, which is not the case at $h \rar 0$ or $h \rar \infty$ (for a discussion on when integrable models display Poisson statistics of eigenvalue gaps, see Ref.\ \citen{Poilblanc:1993, Barba:2008pc, Berry:1977}). The system never displays maximal ergodicity (see Fig.~\ref{fig ising}), and in particular its IR diversities clearly show a heavy tail that arises from free particle states.

More generic systems with Poisson statistics are various localized theories. To access these, we introduce quenched disorder. The simplest model of interest is the Edwards-Anderson model \cite{Edwards:1975}, which for a $d = 1$ chain has Hamiltonian
\bel{
  H = \sum_{i}  J_i \sigma^x_i \sigma^x_{i + 1},
}
with $J_i$ taken to be independent random variables. An exactly solvable variant is the Sherrington-Kirkpatrick model \cite{Sherrington:1975} where the sum runs over all pairs of spins,
\bel{
  H = \sum_{i < j} J_{ij} \sigma^x_i \sigma^x_j.
}
One can think of these models as of strongly disordered versions of the Ising model such that $h \rar 0$. The eigenstates are elementary: they are just all the usual momentum ($\sigma^x$) eigenstates. Nearby energy states are often very different from each other, however --- this is why relaxation via local spin flips is very slow in these models, and because of that these are called spin glasses \cite{Binder:1986}. Glassy models are thus trivially localized in their entire spectrum.

There exist many models which possess both an integrable and an ergodic regime. One example is the Sachdev-Ye model \cite{Sachdev:1992},
\bel{
  H = \sum_{i < j} J_{ij} \vec S_i\cdot \vec S_j,
}
where $\vec S$ are the spin operators of some representation of the $SU(M)$ group. This theory can be dualized to a model of fermions whose generalization, the SYK model, is of much current interest, and whose ergodicity properties are very different.

A related and extremely interesting set of models  displays many-body localization (MBL) \cite{Basko:2005}. These are interacting theories in which disorder can be made strong enough to make states localized \cite{Huse:2013,Serbyn:2013}. The canonical example is
\bel{
  H = \sum_i J \vec \sigma_i \cdot \vec \sigma_{i + 1} + \sum_i h_i \sigma^z_i
}
with transverse fields randomly chosen from $[-h, h]$; when $h$ is large enough, the eigenstates are localized \cite{Pal:2010}. Another simple example with an even richer phase structure is
\bel{
  H = \sum_i J_i \sigma_i^x \sigma^x_{i + 1} - \sum_i h_i \sigma_i^z + \trm{weak\ interactions}.
}
As the disorder in $J_i$ grows (i.e.~the larger its variance compared to the magnetic field variance), the more localized the theory will get. There are in fact several transitions that happen as this model approaches the Edwards-Anderson one; one is the standard ``order-disorder'' transition of the Ising model, and the other ones are reached as \emph{all} states become localized and as the $\Z_2$ symmetry of the first term in the Hamiltonian re-emerges in the spectrum, reverting back to the Edwards-Anderson model \cite{Huse:2013b}.

\begin{figure}
  \centering
  \includegraphics[width=0.48\textwidth]{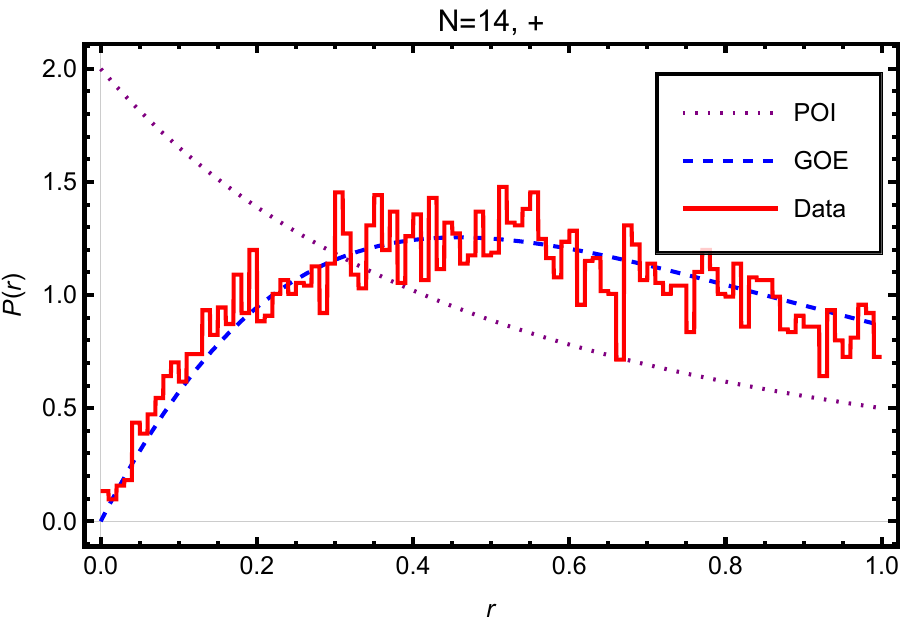} \includegraphics[width=0.4625\textwidth]{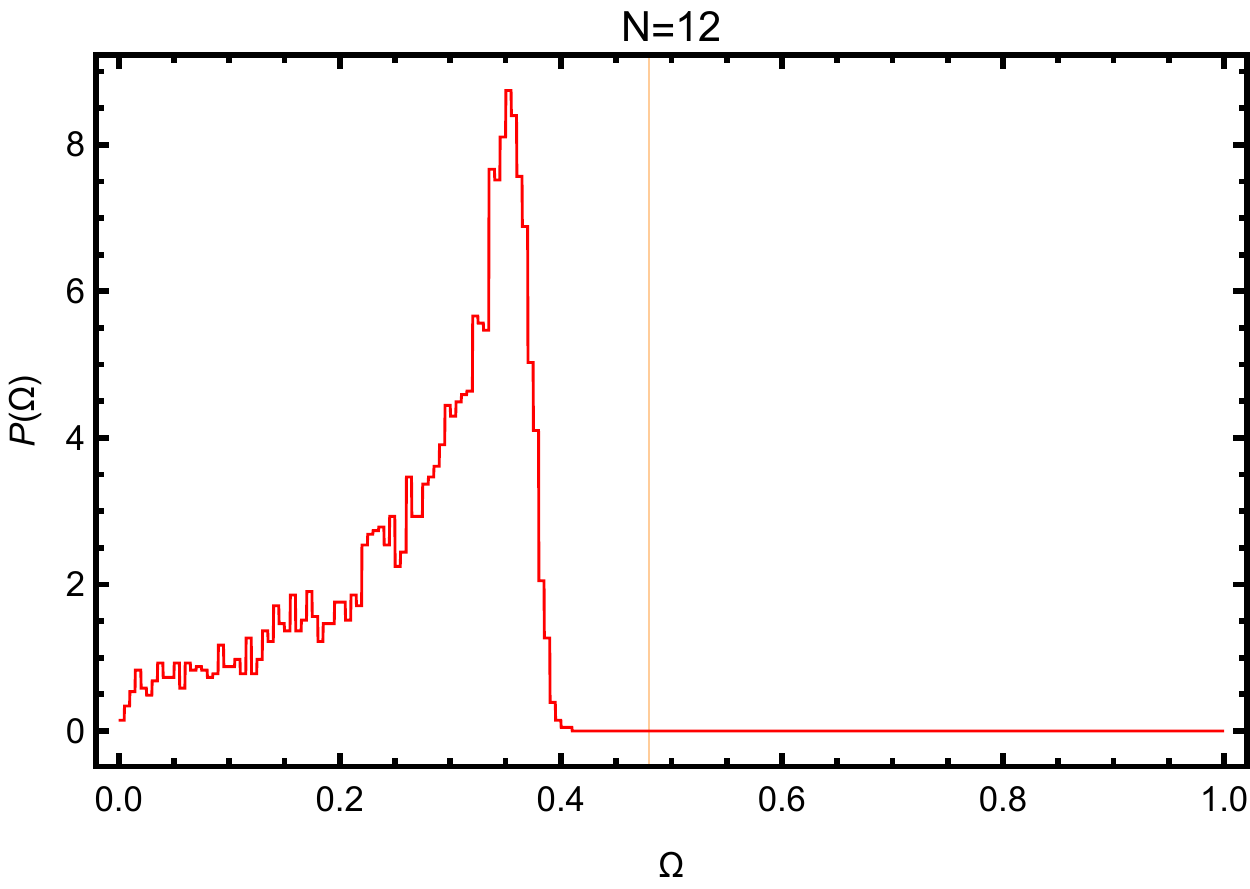}

  \caption{\footnotesize Eigenvalue gap statistics (for $N = 14$ spins) and IR diversities (for $N = 12$) of an interacting spin chain with Hamiltonian \eqref{eq spins}, with open boundary conditions and $g = \frac{5 + \sqrt 5}8$, $h = \frac{1 + \sqrt 5}4$. The system has a $\Z_2$ reflection symmetry and the eigenvalue statistics are shown only for reflection-symmetric eigenstates. The eigenvalue gaps follow the GOE surmise. There are no degenerate energy levels, but due to the reflection symmetry, no states can have IR diversity exceeding $\Omega = \frac12 + \frac1{2\sqrt D}$. The peak in the $\Omega$-distribution indicates that the theory is close to maximally ergodic within each superselection sector. The distribution is reminiscent of the one found in disordered random graphs on Fig.~\ref{fig large disorder RR}.}\label{fig spins}
\end{figure}

Generic spin systems without overly large disorder will be ergodic. Their ergodicity is often not complete, or it very slowly converges towards the random matrix one as $D$ is increased. A typical example is the chain with next-nearest-neighbor hoppings of $O(1)$ strength, e.g.~Ref~\citen{Santos:2010}. On Fig.~\ref{fig spins} we present results for
\bel{\label{eq spins}
  H = \sum_i \sigma_i^z \sigma_{i + 1}^z + g \sum_i\sigma_i^x + h \sum_i\sigma_i^z,
}
where we also see a heavy tail of localized states that may be due to (sufficiently long-lived) quasiparticles. Some spin models are even amenable to an analytic demonstration of ergodicity \cite{Keating:2014}. The class of ergodic models in 1d also contains one-dimensional fermionic models (see e.g.~Ref.~\citen{Jacquod:1997}) where experimental breakthroughs are also becoming possible \cite{Bordia:2015}.

We close this section with the belle de jour: the SYK system \cite{Kitaev:2015, Maldacena:2016hyu, Polchinski:2016xgd}.  This model is most simply presented in terms of $N \equiv 2N'$ Majorana fermions $\chi_i$,
\bel{
  H = \sum_{1\leq i < j < k < l\leq N}\!\!\!\!\! J_{ijkl} \ \chi_i \chi_j \chi_k \chi_l,
}
with random couplings $J_{ijkl}$ drawn from the Gaussian $\mathcal N(0, 1)$. Via a Jordan-Wigner transformation this theory can be represented as a spin system with Hamiltonian
\bel{\label{eq syk spins}
  H
  = \sum_{1\leq i < j < k < l\leq N'} \!\!\!\! J^{\mu\nu\rho\sigma}_{ijkl} \ \Sigma_{ij}^{\mu\nu} \Sigma_{kl}^{\rho\sigma} + \sum_{1\leq i < j \leq N'}\sum_{k < i \trm{\,or\,} k > j} \!\! J^{\mu\nu}_{ijk}\ \Sigma_{ij}^{\mu\nu} \sigma^z_k + \sum_{1 \leq i < j \leq N'} \!\! J_{ij}\ \sigma^z_i \sigma_j^z,
}
with string-like objects
\bel{
  \Sigma_{ij}^{\mu\nu}
  \equiv \sigma^\mu_i \sigma_{i + 1}^z \cdots \sigma_{j - 1}^z \sigma^\nu_j,\quad \mu, \nu \in \{x, y\}.
}
At first sight this model looks related to the 4-spin version of the Sherrington-Kirkpatrick model,
\bel{
  H = \sum J_{ijkl} \sigma^x_i \sigma^x_j \sigma_k^x \sigma_l^x,
}
but the crucial difference is that the operators summed in the Hamiltonian no longer commute due to the $\sigma^z$'s sandwiched between the $\sigma^x$'s and due to the potential $\sum_{ij} \sigma^z_i \sigma_j^z$.\footnote{A similar way to achieve this is to consider couplings between different orientations of the spins, with $H = \sum J_{ijkl}^{\mu\nu\rho\lambda} \sigma^\mu_i \sigma^\nu_j \sigma^\rho_k \sigma^\lambda_l$. This model was studied rigorously in Refs.~\citen{Erdos:2014, Schroeder:2015}.} In terms of scrambling and eigenvalue statistics, SYK is known to be ergodic \cite{Maldacena:2016hyu, Cotler:2016fpe}.\footnote{For variants of SYK with $q$-fermion interactions, it is expected that the model is fully ergodic when $N \gg q \gg \sqrt N$. Small but observable deviations from ergodicity occur when $q \ll \sqrt N$, and at $q = 2$ the system is integrable \cite{Maldacena:2016hyu}. These facts are invariant under $q \mapsto N - q$. The numerically accessible regime is $q^2 \sim N$ and therefore we expect almost complete ergodicity. We also note that the GOE/GUE/GSE alternation (Bott periodicity) reported on Fig.~\ref{fig   SYK evals} is only a feature of systems with $q\!\mod\!4 = 0$. For $q = 6$, for instance, the SYK model is always in the GUE universality class.} As seen on Figs.~\ref{fig   SYK evals} and \ref{fig   SYK ents}, our results confirm that SYK eigenstates are indeed maximally ergodic in each of the two superselection sectors labeled by the eigenvalues of the fermion number parity $\prod_{i = 1}^{N} \chi_i = \prod_{i = 1}^{N'} \sigma_i^z$. This holds even if the disorder is heavy-tailed, as in the case of Cauchy-distributed disorder. We will discuss the SYK model and its other variants in more detail in section \ref{sec syk}, after developing more tools that shed light on the origin of ergodicity in quantum theories.

\begin{figure}
  \centering
  \includegraphics[width=0.32\textwidth]{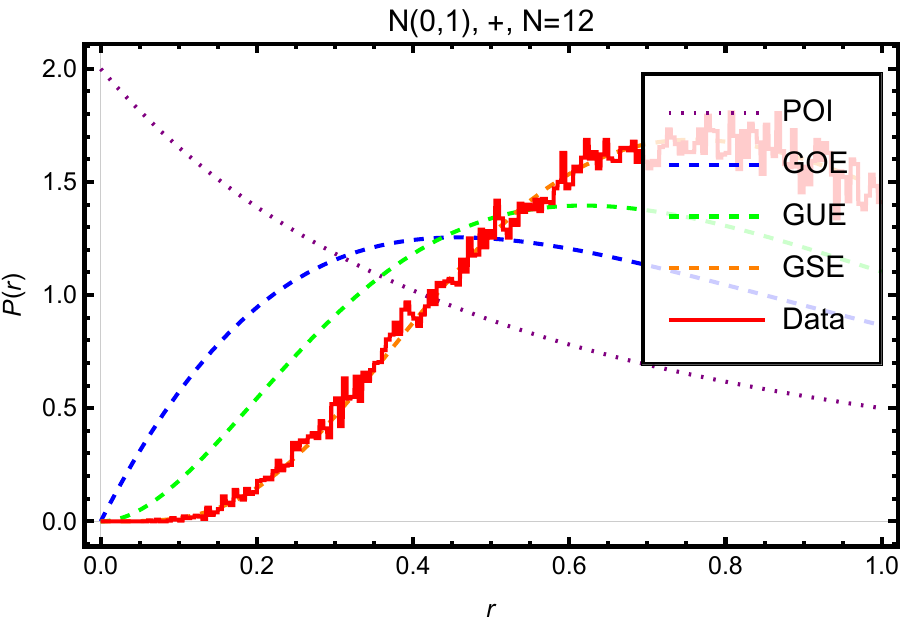} \includegraphics[width=0.32\textwidth]{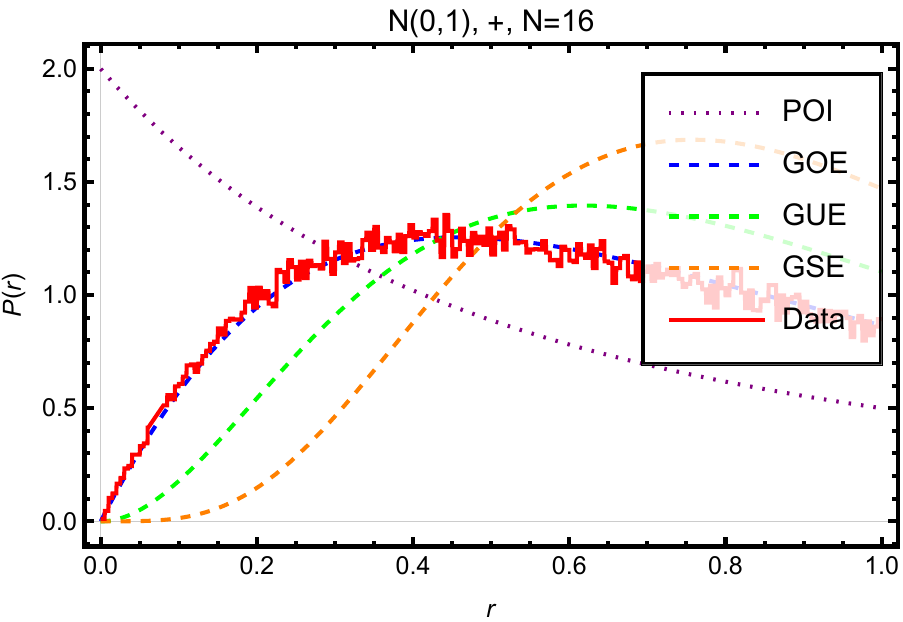} \includegraphics[width=0.32\textwidth]{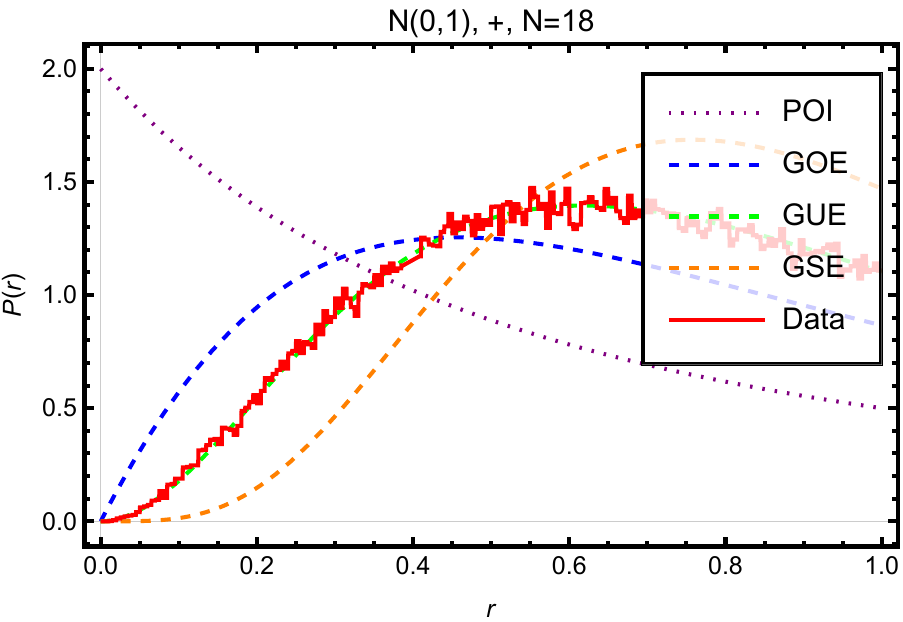}

  \caption{\footnotesize Eigenvalue statistics of the SYK model. The random matrix ensemble in question is GOE for $N \!\! \mod 8 = 0$, GUE for $N \!\!\mod 8 = \pm 2$, and GSE for $N \!\!\mod 8 = 4$ \cite{You:2016}. The model has a $\Z_2$ fermion number parity symmetry; here we present eigenvalue statistics for one of the symmetry sectors. The distributions are aggregates of $1,024,000/D$ instances of disorder, with Hilbert space dimension $D = 2^{N/2}$. }\label{fig SYK evals}
\end{figure}

\begin{figure}
  \centering
  \includegraphics[width=0.48\textwidth]{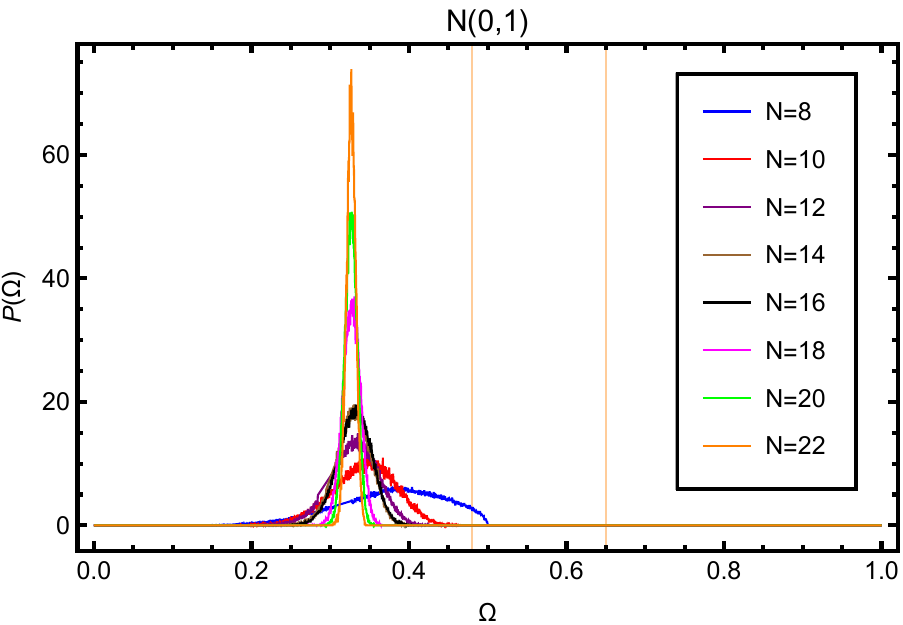}
  \includegraphics[width=0.48\textwidth]{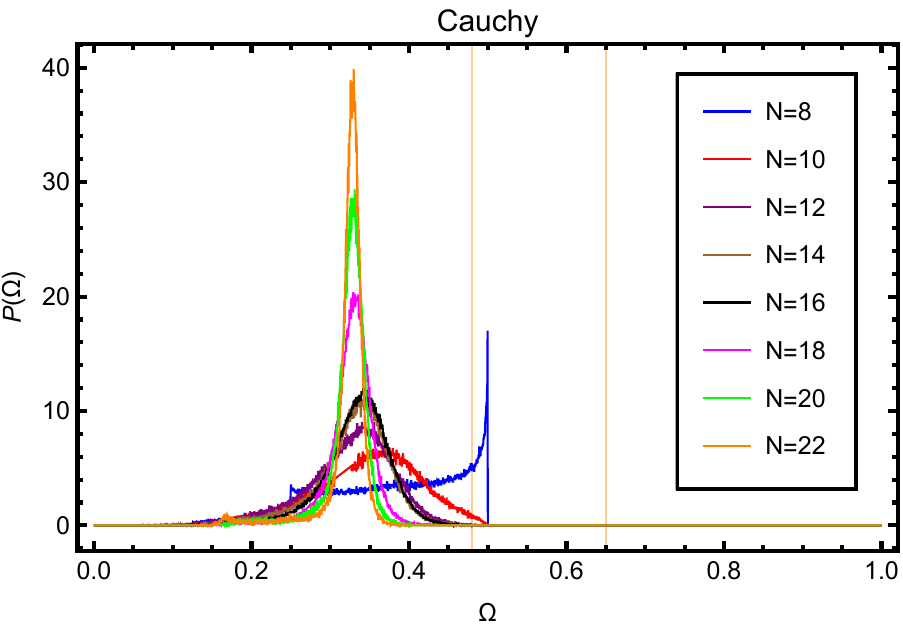}
  \caption{\footnotesize IR diversities for the SYK model with Gaussian and Cauchy disorder of zero mean. Results are aggregated over $1,024,000/D$ instances of disorder. In both cases the large $N$ behavior tends to full ergodicity on one half of the Hilbert space, with $\Omega \leq 1/2$ for each state. The IR diversities accumulate at $\Omega\_{GUE}/2 \approx 0.33$. The factor of $1/2$ is present because the Hamiltonian is block diagonal in the spin basis due to the global fermion number parity symmetry. We caution the reader that when the system is in the GUE or GSE universality class, each state is doubly degenerate, so a choice of polarization is implicit in the above plots.}\label{fig SYK ents}
\end{figure}

\section{Charting the landscape} \label{sec chart}

Having studiously reviewed the archetypical problems related to quantum ergodicity, we now aim to understand what makes them exhibit such universal behavior. Since quantum chaos is already a feature of theories with the minimal product structure in their Hilbert space, as reviewed in section \ref{subsec qm2}, our analysis will be framed in the language of this simple setup. We will view each theory, spin chains included, as a particle moving between $D$ sites; a closely related paradigm was exploited in Ref.\ \citen{Bauer:2013}. Our analysis will be most neatly carried out on the space of all free hopping Hamiltonians among these sites --- or, in other words, the space of unweighted adjacency matrices on graphs with $D$ vertices. We will argue that it is the symmetry structure of this \emph{state graph} that primarily determines the ergodicity of the theory, and that small modulations do not affect the ergodicity significantly.

\subsection{Localized operators} \label{subsec ops}

Diagnostics of quantum ergodicity are basis-dependent. As mentioned in section \ref{subsec qc}, our approach is to choose certain operators --- $U$ and $L$, or $U_i$ and $L_i$ --- with a highly regular spectrum, and to use their eigenvectors to formulate definitions of ergodicity. Here we will examine this convention in more detail.

A necessary but not sufficient condition for an operator to have such a useful eigenbasis is that it has uniformly spaced eigenvalues --- for instance, $e^{2\pi i n/D}$ or $n/D$ for $n = 1,\ldots, D$ --- because then we may hope to arrive at a picture where each eigenstate can be assigned a regularly shaped area in phase space that is labeled by the corresponding eigenvalue. This would then allow us to understand ergodicity as a characterization of the spread of a given energy eigenstate across these regular cells in phase space.

A regular spectrum is not enough to ensure this, however. If an operator $\O$ has eigenvalues $e^{2\pi i n/D}$, the same will hold for $R \O R^T$ with any orthogonal matrix $R$. Out of all operators with this spectrum, we need to find a convention that singles out a specific one and represents it as the diagonal matrix $\trm{diag}(e^{2\pi i n/D})$. This convention must be physically motivated: we must specify a reference state (or a reference minimal projection operator) as a localized state, i.e.~as a state that is the opposite of being ergodic. Then we can demand that this state be an eigenstate of an operator with a regular spectrum. Here is an example. If $\qvec{\b 1}$ is the given localized state, let $U$ be the operator with a regular spectrum $\{e^{2\pi i n/D}\}$ such that $U \qvec{\b 1} = \qvec{\b 1}$. Then, consider the conjugate operator $L$ defined so that it has the same spectrum as $U$, and such that $UL = e^{2\pi i/D} LU$. It follows that the states $L^n \qvec{\b 1} = \qvec{e^{2\pi i n / D}}$ now provide a basis in which $U$ is diagonal. Alternative conventions could have consisted of choosing any other eigenvalue to correspond to $\qvec{\b 1}$.

From now on, we always assume that the original localized state is known, that $U$ is a diagonal operator in the basis containing this state, and that $L$ is the corresponding conjugate operator that generates all other eigenstates of $U$ by acting on the original state. This is the most conservative (least model-specific) way to measure ergodicity that we are aware of.

There are other operators whose eigenbases can be used to measure ergodicity. We will call these operators \emph{fully localized}. Roughly, we expect that if $\O$ is fully  localized, so is $\O_R = R \O R^T$ whenever $R$ is sufficiently regular. By this we refer to matrices whose entries fall off exponentially with the distance from a diagonal, but also to permutation matrices and Hadamard matrices that transform position eigenstates into momentum ones. In particular, if $R$ is block-diagonal with $\sqrt D$ identical $\sqrt D \times \sqrt D$ blocks of Hadamard form gives an operator $U_R$ whose eigenbasis corresponds the usual coherent states. This basis is also known to be a good sensor of ergodicity, and the corresponding IR entropy is also called \emph{Wehrl entropy} \cite{Wehrl:1979, Lieb:1978}. Yet another fully localized basis is furnished by eigenstates of the harmonic oscillator, with the raising and lowering operators corresponding to the shifts $L$. In general, any basis of integrable Hamiltonians may work, and sometimes the basis of an underlying integrable theory (the ``mean-field basis'') is used to diagnose ergodicity \cite{Santos:2011}.

A word of caution is necessary here: the bases listed above can all be used to detect ergodicity, but it is not generally true that they will all lead to the same distribution of IR diversities $\Omega^{(n)}$, defined in \eqref{def Omega}. The cleanest analysis can be carried out when the Hamiltonian and eigenvectors are all real-valued in the chosen basis, in which case we always expect to find the GOE result $\Omega \approx 0.48$ for an ergodic eigenstate.

The notion of fully localized operators is also important because it provides a way to count symmetries relevant to the ergodicity of a given theory. For any Hamiltonian $H$, there are many operators that commute with it --- in particular, any eigenstate $\qvec n$ defines a projection operator $P_n = \qvec n \qvecconj n$ such that $[H, P_n] = 0$. However, $P_n$ will have one nonzero eigenvalue equal to unity, so it will not be fully localized. In fact, if a fully localized operator $Q$ commutes with $H$, then the theory is classical (or completely integrable), as each eigenvalue of $Q$ will label a single localized state whose time evolution will be independent of all other sectors.

To quantify this amount of ``usefulness'' in an operator that commutes with $H$, let a \emph{partially localized} operator be one whose eigenvalues are regularly spaced but which may still have degeneracies. For instance, if $D$ is even, $U^2$ has two eigenvalues $e^{2\pi i n/D}$ for each even $n \leq D$; for any $D$, a projection onto a pure state $P_n$ has one nonzero eigenvalue and $D - 1$ zero ones; and the identity operator has $D$ equal eigenvalues. The number of distinct eigenvalues is what we call the \emph{degree of localization}. The lower the degree of localization, the less useful the operator is for diagnosing ergodicity.

Operators with an unevenly spaced spectrum generically do not usefully diagnose ergodicity, and the definition of full/partial localization is not meant to apply to them. The one exception we anticipate are operators whose eigenvalue gap statistics is Poissonian. Reversing the Berry-Tabor conjecture \cite{Berry:1977}, we expect that a localized eigenbasis will be furnished by any operator with Poisson-distributed gaps in the spectrum and for which the reference localized state is an eigenstate. Moreover, we expect that such an operator can be constructed out of a fully localized operator $\O$ as $\sum_{n = 1}^D \zeta_n \O^n$ for some (typically randomly distributed) coefficients $\zeta_n$. Of course, similarity transformations by a ``regular'' matrix $R$ will, as above, preserve the localization of the operator.

In terms of spectral statistics, then, we conjecture that the distinction between ergodic and localized regimes is controlled, to a first approximation, by the maximal degree of localization found among all the symmetries of the Hamiltonian. If a fully localized operator is a symmetry, all energy eigenstates will be localized. This follows trivially from the definition of localized operators. The same holds true if $D = \prod_{i = 1}^N D_i$ and the Hamiltonian commutes with $N$ partially localized operators that correspond to motion along each direction $i$, as these can be simultaneously diagonalized to provide a fully localized eigenbasis. On the other hand, if only a few localized operators of degree one or two are symmetries, we a priori have no information how close the theory is to being maximally ergodic. This is where we must look for deeper structure in the Hamiltonian, as we describe next.

\subsection{Ergodicity in free hopping theories} \label{subsec free hopping}

We begin by studying theories whose Hamiltonians are unweighted adjacency matrices of graphs. Many single-particle examples from section \ref{sec review} fall into this category, including billiards and particles moving on unweighted random graphs. Such Hamiltonians represent \emph{free hopping} on graphs. The corresponding graphs are what we call state graphs.\cite{Bauer:2013}

Some spin systems can be also described as free hopping theories. Consider a theory of $N$ spins with all unit couplings and built out of only $\sigma^x$ operators, e.g.
\bel{\label{Ising free}
  H = \sum_i \sigma_i^x + \sum_{i < j} \sigma^x_i \sigma^x_j + \ldots
}
In the $\sigma^z$ basis, this is a free hopping theory on $2^N$ sites. These sites are remarkably simple to visualize: each position eigenstate can be assigned to a corner of an $N$-dimensional hypercube \cite{Bauer:2013, Huse:2014}. Each term in the Hamiltonian corresponds to a link connecting two vertices of this hypercube, and the result is a very structured, regular state graph with connectivity equal to the number of different basis operators in the Hamiltonian.

When are free hopping theories ergodic? Results reviewed in sections \ref{subsec qm1} and \ref{subsec qm2} show that hopping on random regular graphs is ergodic, that hopping on graphs describing the usual billiards is almost ergodic, and that hopping on metric graphs is essentially not ergodic at all. In addition, the spin system \eqref{Ising free} is clearly integrable, and in general a particle hopping freely on a regular grid-like graph is also integrable. These observations suggest that it is the amount of structure or symmetry in the graph that determines the ergodicity of its adjacency matrix.

From the point of view of trace formulae, this is not surprising. Recall from eq.~\eqref{eq trace} that the density of states and other eigenvalue statistics are controlled by sums over periodic orbits of all lengths. In a generic free hopping theory, the trace formula requires all trajectories to be summed over --- there is in general no semiclassical limit that picks out a small subset of classical orbits --- and the weight of each trajectory depends merely on its length. If the numbers of periodic trajectories of different lengths are essentially independent of each other, we should expect that their summation within the trace formula will give rise to a central limit theorem effect and lead to universality. This property was indeed exploited for random regular graphs in Ref.\ \citen{Oren:2010}.

We do not possess a perfectly satisfactory analytical tool that would connect ergodicity with a specific trace formula (but see Refs.~\citen{Kaplan:2001, Berkolaiko:2003} for some special cases, and the references in Ref.\ \citen{ORourke:2016} for other methods for accessing eigenvector statistics analytically). Being unable to make analytic progress, we turn to standard quantum-mechanical perturbation theory to illustrate our point heuristically. Consider a free, fully localized Hamiltonian $X$ with nondegenerate eigenvalues $x_n$. If its eigenstates are denoted by $\qvec{n}$, the eigenstates $\qvec{n(\lambda)}$ of a perturbed Hamiltonian $X + \lambda H$ can be written as
\bel{
  \qvec{n(\lambda)} = \sum_m \psi_{mn} (\lambda) \qvec{m},
}
with
\algnl{\notag
  \psi_{mn} (\lambda)
  &= \delta_{nm} + \lambda (1 - \delta_{nm}) \frac{H_{mn}}{x_n - x_m} - \frac{\lambda^2} 2 \delta_{nm} \sum_{l \neq n} \frac{|H_{ln}^2|}{(x_n - x_l)^2} + \\ \label{eq pert thy}
  & + \lambda^2 (1 - \delta_{nm})\left(\sum_{l \neq n} \frac{H_{ml} H_{ln}}{(x_n - x_l) (x_n - x_m)} - \frac{H_{mn} H_{nn}}{(x_n - x_m)^2}\right)  + O(\lambda^3).
}
The assumption that $X$ is fully localized means that there are no degeneracies, so ordinary perturbation theory should be valid. In general we must resum the entire perturbation series to get the correct eigenstates of the full Hamiltonian. The fact that the full Hamiltonian is finite and diagonalizable ensures that this series will converge.

This expansion can be obtained to arbitrary order as a solution to $(X + \lambda H)\qvec{n(\lambda)} = E_n(\lambda) \qvec{n(\lambda)}$ using a formal power series in $\lambda$. Writing
\bel{
  \psi_{mn}(\lambda) = \delta_{mn} + \sum_{k = 1}^\infty \lambda^k \psi_{mn}^{(k)}, \quad E_n(\lambda) = x_n + \sum_{k = 1}^\infty \lambda^{k} E_n^{(k)},
}
we find the recurrence relation
\bel{
  (x_m - x_n) \psi_{mn}^{(k + 1)} = \sum_{k' = 0}^k E_n^{(k' + 1)} \psi_{mn}^{(k - k')} - \sum_l H_{ml}\psi_{ln}^{(k)}.
}
The $m = n$ case of this equation determines $E_n^{(k + 1)}$ in terms of the lower-order terms, while $m \neq n$ determines $\psi_{mn}^{(k + 1)}$. The ``diagonal'' coefficients $\psi_{nn}^{(k)}$ are determined order-by-order in $\lambda$ by demanding that states have unit norm. This condition can be written as the series of equations
\bel{
  \sum_m \sum_{k' = 0}^k \left(\psi_{mn}^{(k')}\right)^* \psi_{mn}^{(k - k')} = 0, \quad k \geq 1.
}

We are interested in large values of $\lambda$, where we expect that eigenvectors of $X + \lambda H$ are approximately equal to those of $V$. The full perturbation series must be resummed to get the needed answer. The situation looks similar to the one that arises when using trace formulae, and the behavior of eigenvector coefficients will be determined by high orders in perturbation theory. At very large $k$, the terms contributing to $\psi^{(k)}_{mn}$ will be of the form
\bel{\label{traj}
  \sum_{l_1,\,\ldots,\,l_{k - 1} \neq n} \frac{H_{m l_1} H_{l_1 l_2} \cdots H_{l_{k - 1}n}}{(x_n - x_m) (x_n - x_{l_1}) \cdots (x_n - x_{l_{k - 1}})}.
}

Let us now assign each unperturbed state $\qvec n$ to a site, creating a state graph. The free Hamiltonian does not cause any hoppings between these states; links in the graph correspond to nonzero elements $H_{mn}$. Thus, the $k^{\trm{th}}$ order contribution to the $\psi_{mn} $ coefficient of the $n^{\trm{th}}$ eigenstate depends on the sum over all trajectories of length $k$ that connect sites $n$ and $m$ of the graph whose adjacency matrix is determined by $H$.

The case of free hopping theories is the simplest version of the above scenario. Here a Hermitian version of the position operator --- of the form $X \sim \sum_n c_n U^n$ --- can be chosen as the unperturbed Hamiltonian, and the hopping is taken to be the perturbation $H$ with strength $\lambda$ much greater than unity. The coefficients $c_n$ can be chosen to get an unperturbed Hamiltonian whose eigenvalues are $n/D$ for $n = 1, \ldots D$. Since $H_{nm}$ is either zero or one, $\psi^{(k)}_{mn} $ is determined by tracking the trajectories of length $k$ that connect sites $n$ and $m$. The eigenvalues $x_n$ in \eqref{eq pert thy} and \eqref{traj} are now actual \emph{position} labels running between 0 and 1 in $1/D$ increments. The contribution of each trajectory connecting $n$ and $m$ is simply the product of $k$ terms of the form $1/(x_l - x_n)$. If the graph described by adjacency matrix $H$ is highly irregular, we can expect the weights of trajectories to be distributed randomly and almost independently. The sum of contributions of trajectories of all lengths --- dominated by trajectories of large lengths, since there will be exponentially many of them --- can now reasonably be expected to look like a sum of almost i.i.d.~random variables. For a given $n$ and $m$, by the central limit theorem this sum should look like a Gaussian random variable. In other words, the coefficients $\psi_{mn}$, $m = 1,\ldots, D$ of any eigenstate $\qvec{n(\lambda)}$ can be expected to be Gaussian-distributed at large enough $D$.

This argument is at best a heuristic, as we have not proven that the sum over trajectories amounts to a sum over identically distributed random variables. Nevertheless, it is interesting to apply this idea to the problem of a particle hopping on a tree or on an otherwise expanding graph. The key feature of such graphs is that it is impossible to assign the position labels $x_n$ to sites such that trajectories of a given length between two fixed sites have very similar weights; as the $x_n$ grow linearly with $n$ while the graph expands exponentially (meaning that there are $\sim e^k$ sites at distance $\leq k$ away from a given site), for large enough $k$ the contributions from almost all trajectories of length $k$ will be uncorrelated products of $1/(x_n - x_l)$ factors where $x_n$ and $x_l$ are very far from each other in the spectrum of the free (``position'') Hamiltonian.

What happens in the converse case, when we study a particle hopping on a highly regular graph? Consider the motion on a regular, rectangular grid. Then we may assign the position labels $x_n$ to sites in a regular, lexicographic way. For instance, if the grid is $\Z_N \times \Z_N$ with $D = N^2$, we can assign to a site with coordinates $n = (x, y)$ the label $x_n = \frac1D[(x - 1)N + y]$. The regularity of the grid means that $x_n - x_l$ for two linked sites will only depend on $n - l$, so that the trajectories connecting $(x, y)$ and $(x', y')$ will have the same weights as those connecting $(x + a, y + b)$ and $(x' + a, y' + b)$. Thus the sum over trajectories in \eqref{eq pert thy} cannot be represented as a sum over almost independent random variables, and no central limit theorem is applicable.

While this picture is appealing, it is not at all obvious why going to high orders in perturbation theory will cause the contributions from individual trajectories to cancel out and give a finite answer at $\lambda \gg 1$ --- indeed, numerical checks on truncated versions of \eqref{eq pert thy} do not indicate a perfect cancellation. A better (albeit model-dependent) approach is to start from an integrable free hopping theory and then to perturb it by $\lambda \sim 1$ terms that implement the additional hoppings that ruin integrability. The heuristic analysis presented above should still hold and illustrate what we mean when talking about structure in a graph.  A detailed numerical study of this proposal is left for future work.

\subsection{Perturbations of free hopping theories} \label{subsec pert hopping}

We have seen that free hopping theories range from fully integrable to fully ergodic ones. There are two ways to perturb these theories: diagonally and off-diagonally. Both of them have been discussed in various places in section \ref{sec review}, and here we systematize this information. Note that in all the cases given here, the strength of the perturbation is expressed in units in which the original free hopping theory has a Hamiltonian with entries 0 and 1; this sets the scale of the theory.

\subsubsection{Diagonal perturbations}

We first consider the possibility of assigning weights to different positions. In quantum mechanics this corresponds to turning on a position-dependent potential $V(U)$, in spin chains this corresponds to turning on a background magnetic field and $\sigma^z \sigma^z \ldots$ interactions, and in quantum field theory this corresponds to adding any function of the fields (interaction terms included) to the Hamiltonian. All of these perturbations appear as diagonal terms in the Hamiltonian (in the appropriate position basis).

If we deform an integrable free hopping theory by a small diagonal perturbation (say, by dialing a little bit of disorder), its localized symmetries become broken and the model starts becoming ergodic. Moreover, if a fully ergodic regime occurs as the perturbation strength is increased, it must be smoothly connected to the integrable regime one started with, as we are working with a finite Hilbert space dimension $D$. As the results in sections \ref{subsec qm1} and \ref{subsec qm2} show, we may expect that perturbations of free hopping of size $1/D$ will start affecting the spectrum and delocalizing states.

On the one extreme in the space of all possible deformations lie the very simple perturbations, for instance adding smooth functions of the position to the Hamiltonian (corresponding to weighting vertices without breaking many symmetries). For example, we may add $\lambda \prod_i \sigma^z_i$ to the Hamiltonian \eqref{Ising free}. Doing this does not turn the system fully ergodic, even if the strength of the perturbation was $O(1)$. What does happen, however, is that localization in the $\sigma^x$ (momentum) basis partially crosses over to localization in the $\sigma^z$ (position) basis as $\lambda$ grows. A similar story happened in the simple quantum mechanics models in section \ref{subsec qm1}, where complete momentum space localization crossed over to complete position space localization for potentials like $V = \lambda (U + U^{-1})$. For such slowly varying, highly symmetric perturbations, the crossover to localization ends only when coupling hits the scale set by the Hilbert space size, $\lambda \sim D$.

On the other extreme are the fully disordered perturbations, with weights on graphs randomly distributed. Starting from a random regular graph, adding $O(1)$ diagonal disorder does not significantly alter ergodicity. However, starting from a nonergodic free hopping theory, adding a finite amount of disorder breaks the symmetries of the theory and renders it ergodic, as shown in studies of spin chains (e.g.~Refs.~\citen{Huse:2013b, Santos:2010}). Perturbations of $O(1)$ strength that are more slowly varying but still not fully symmetric render the theory partly ergodic, with some states localized, others fully ergodic, and yet others caught in the crossover between fully localized and fully ergodic. This is the typical scenario that happens in spin chains (and interacting quantum field theories in general), where ergodicity is often achieved through interaction terms that are not disordered but that vary sufficiently fast across the state graph.

The perturbation theory heuristic presented in the previous subsection once again allows us to understand this. If we start from free hopping on a very regular graph, one way to make each trajectory contributing to $\psi_{mn}$ essentially random is to turn the matrix elements $H_{l_i l_j}$ into sufficiently complicated (irregular) functions of $l_i$ and $l_j$. Doing so will wash out the regularity of the $x_n$'s and once again lead to a central limit theorem. Modulating the diagonal terms $H_{ll}$ can do the same thing, as terms such as $H_{nn}$ and its powers will also enter the formula \eqref{eq pert thy}.

As an example, consider a local spin chain with $N$ spins. The Hilbert space has dimension $D = 2^N$. Locality means that the Hamiltonian only has $O(N)$ terms; assuming only nearest- and next-nearest-neighbor interactions, there are $2N$ terms. The free hopping theory has $H = \sum_i \sigma_i^x \sigma_{i + 1}^x + \sigma_i^x \sigma^x_{i + 2}$, which corresponds to a particle living on sites of an $N$-dimensional hypercube, with each site (corresponding to $\sigma^z_i$ eigenstates) connected to $2N$ others by a single hopping term. This state graph is very regular and the theory is integrable. Now consider introducing a potential term $V = \sum h_i \sigma_i^z$ into the Hamiltonian, with $h_i \in [-1, 1]$. This term weights each spin configuration by a number between $-N$ and $N$. If the magnetic fields $h_i$ are distributed randomly, the weights of configurations connected by a single spin flip will be essentially independent random variables. Trajectories that contribute to the perturbative corrections of various states will be products of these numbers, and adding them together may be expected to give a Gaussian result. The same conclusion can be reached if $h_i$ is not random but still sufficiently quickly and irregularly varying with $i$, so that the weights of trajectories can still look effectively random. Any leftover regularity in the system, due e.g.~to particularly nice functions $h_i$, may be reflected by the partial localization of the spectrum. Note that typical quantum field theories encountered in the high-energy literature do not have any disorder or externally varying parameters, and instead rely only on local potential terms of the form $V = \sum_i f(\sigma_i^z, \sigma_{i + 1}^z)$ --- e.g.~$V \sim (\bfnabla \phi)^2 + \phi^4 + \ldots$ --- to enforce interactions; it is therefore not surprising that many field theories typically possess non-ergodic parts of the spectrum, often manifested as \hbox{(quasi-)particle} excitations.

As reviewed in sections \ref{subsec qm2} and \ref{subsec spin}, adding strong disorder to any theory typically results in Anderson or many-body localization. For a fixed dimension $D$, there exists a critical disorder strength below which no position space localization can be seen. In addition, above an upper critical disorder strength, all states are localized. (The nature of the free hopping theory and perturbation determine these critical strengths; we remind the reader that for the mildest perturbations and integrable hopping theories, these critical strengths scale as $\sim 1/D$ and $D$, respectively.) The reason why this happens is qualitatively clear from our analysis: a strong enough perturbation turns trajectory weights into random variables whose variance may grow quickly with $k$. There are at least two ways the central limit theorem could fail now. On the one hand, $\psi_{mn}$ may no longer be a sum of approximately identically distributed contributions, as trajectories with different lengths may be drawn from distributions with different variances. On the other hand, these variances may also diverge, meaning that the major contribution to $\psi_{mn}$ at large $D$ comes from random variables with variance $\gtrsim D$, and the central limit theorem does not apply in this case. A more detailed discussion of the localization transitions that happen as diagonal disorder is increased in Gaussian random matrices and random regular graphs is given in Ref.\ \citen{Kravtsov:2015}.

\subsubsection{Off-diagonal perturbations}

Off-diagonal perturbations involve momentum operators, and they move the free hopping Hamiltonian away from the unweighted adjacency matrix form by modifying entries of unit value. We need only analyze the case of perturbations that decrease the weight of links, since we may always rescale the Hamiltonian and ensure that the heighest-weight link has weight one and that all off-diagonal elements are within $[-1,1]$.\footnote{If the perturbation we are considering adds more hoppings by converting some entries $H_{ij}$ from zero to nonzero values, we may view that theory as a perturbation of the free hopping theory where the entries $H_{ij}$ were unity to begin with.}  What matters for ergodicity is merely the absolute value of the hopping terms, and in the following we will (for simplicity) assume that all terms in $H$ are positive. In the perturbative picture of section \ref{subsec free hopping}, the contributions to $\psi_{mn}$ already come with strongly varying signs; changing the signs of some terms in the potential is not expected to affect the central limit theorem, if it holds.

This case  is in many ways dual to the case of diagonal perturbations. In the latter case, a site is rendered inaccessible if the potential at the site is much greater (e.g.~by an $O(D)$ factor) than that of nearby sites --- it then acts as a $\delta$-function obstacle that supports a localized state on top of it. In the off-diagonal case, if a link between two sites outweighs all other links emanating from each of those sites by a huge factor (e.g.~$\sim D$), we find a state localized on the two sites, and all the other eigenstates will act as if the two sites on the state graph are fused into one. If the original graph had very low connectivity, increasing disorder will effectively remove many links and force eigenstates to localize in the remaining clusters of connected links. If the original graph connectivity was large enough, we still expect partial localization of the spectrum. These effects are clearly visible in random regular graphs with strong disorder (Fig.~\ref{fig large disorder   RR}).

Based on the examples studied, we once again expect there to be two critical perturbation strengths associated to each integrable free hopping theory. When typical link weights between the upper critical strength and unity, the eigenstates are localized in momentum space. For typical weights between the two critical strengths, the system is ergodic, and below the lower critical strength, the system is localized in position space. For very slowly varying perturbations, these critical strengths are expected to be $1 - 1/D$ and $1/D$. For more dramatic perturbations (such as Anderson disorder),  the critical strengths may both be $O(1)$ and the ergodic regime may be altogether absent.

\subsection{Ergodicity charts} \label{subsec charts}

\subsubsection{Local ergodicity structure}

\begin{figure}[tb!]
\begin{center}

\begin{tikzpicture}[scale = 2, axis/.style={very thick, ->, >=stealth'}]

  \fill[fill=blue!20]
    (0,0) rectangle (3,2.8);
  \fill[fill=blue!50]
    (0,0)  -- (2.5, 0) arc (0:90:2.5) -- (0,0);
  \fill[fill=blue!80]
    (0,0) -- (0.9, 0) arc (0:90:0.9) -- (0,0);

  \draw[axis] (0,0)  -- (3.1,0) node(xline)[right]
        {diagonal perturbation};
  \draw[axis] (0,0) -- (0,3.1) node(yline)[above] {off-diagonal perturbation};

  \draw[thick] (0.9, -0.05) -- (0.9, 0.05);
  \draw[thick] (2.5, -0.05) -- (2.5, 0.05);
  \draw[thick] (-0.05, 0.9) -- (0.05, 0.9);
  \draw[thick] (-0.05, 2.5) -- (0.05, 2.5);
  \draw[thick, dashed] (-0.05, 2.8) -- (3.05, 2.8);

  \draw (-0.05, 2.85) node[anchor = east] {$1$};
  \draw (-0.05, 2.5) node[anchor = east] {$\lambda\_{off-diag}\^{upper}$};
  \draw (-0.05, 0.9) node[anchor = east] {$\lambda\_{off-diag}\^{lower}$};
  \draw (0.9, -0.05) node[anchor = north] {$\lambda\_{diag}\^{lower}$};
  \draw (2.5, -0.05) node[anchor = north] {$\lambda\_{diag}\^{upper}$};

  \draw[white] (0.45, 0.1) node[anchor = south] {\textsc{\small localized}};
  \draw[white] (1, 1) node[anchor = south] {\textsc{\small mixed/ergodic}};
  \draw[white] (1.5, 2.6) node[anchor = west] {\textsc{\small localized}};

\end{tikzpicture}

\end{center}
\caption{\footnotesize A schematic representation of the ergodicity of theories around a typical integrable free hopping theory. In our convention, integrable free hopping theories must be localized in the momentum basis. Small perturbations do not affect this ergodicity (the darkest blue region). Medium perturbations, whether diagonal or off-diagonal, either turn part of the spectrum position-localized, or turn part of the spectrum ergodic (medium dark blue region). Strong perturbations always cause position space localization (light blue region).}
\label{fig landscape local}
\end{figure}
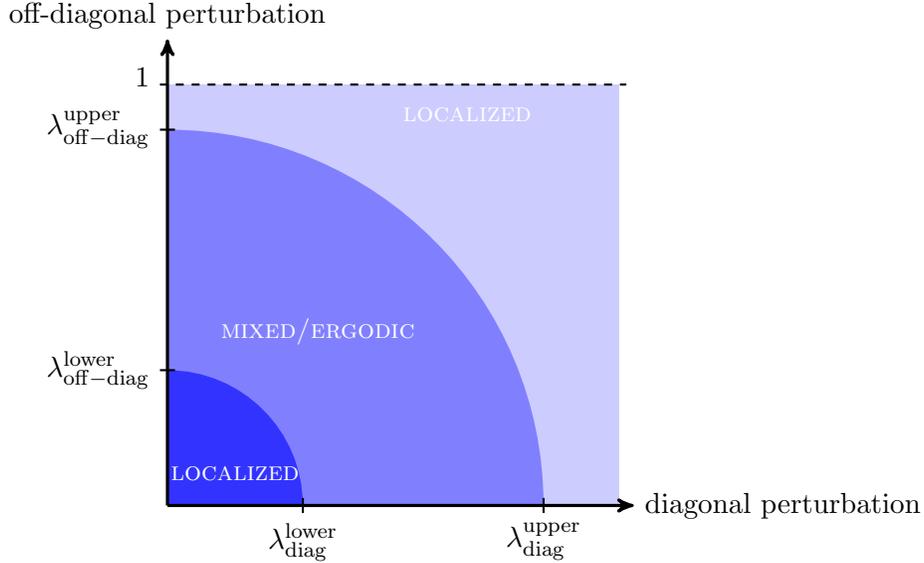

Let us now contemplate what happens when both diagonal and off-diagonal perturbations can be turned on. These perturbations define a ``bubble'' in theory space around each integrable free hopping theory. In our basis choice, integrability of free hopping theories corresponds to localization in momentum space, or equivalently to localization with respect to the Abelian algebra of hopping operators. The cross-section of the bubble is shown on Fig.~\ref{fig landscape local}.  The crossover couplings $\lambda\^{lower}$ and $\lambda\^{upper}$ are functions of $D$, of the integrable theory in question, and of the direction in theory space along which one is moving away from this theory. As discussed, smooth, slowly varying perturbations are expected to have extremely large distances between integrable regimes, with $\Delta \lambda \equiv \lambda\^{upper} - \lambda\^{lower} \sim D$. On the other hand, Aubry-Andr\'e Hamiltonians may have $\Delta \lambda \sim O(1)$, while Anderson-type disorder gives rise to a sharp transition between two localized regimes, with $\Delta \lambda \sim 1/D$.

Transitions between localized regimes with no intermediate ergodic regime happen in other models too. A good example to keep in mind is the Ising model in a constant transverse field, which is solvable for any field strength $h$. Along this line there exists a crossover from momentum ($\sigma^x$) to position ($\sigma^z$) space localization. At $h = 1$ there is a critical point (the Ising CFT) which is also localized in momentum space (as can be seen by dualizing to a free Majorana fermion).

Lines of integrability like the transverse field deformation or the $U + U^{-1}$ potential are atypical in the space of all theories. A typical deformation of a localized free hopping theory is the most disordered one possible. For instance, the majority of diagonal deformations have $D$ different parameters $\zeta_n$, one on each diagonal entry. The change in ergodicity along these typical directions is very different from the change along integrable lines. In the latter case, eigenstates delocalized one by one, each slowly spreading in a structured way that never leads to GOE ergodicity. The critical disorder strengths here demarkated the edges of this crossover. In the former case --- the typical one --- all states spread out chaotically as the result of the perturbation, and the crossover between full ergodicity and full localization can be expected to have length $\sim 1/D$ in theory space. The position of this transition interval is marked by the critical disorder strengths.

Existence of symmetries or lack of complete disorder in perturbations of integrable free hopping may stagger the crossover and make it happen piecemeal, spreading it over a macroscopic length in theory space when a large number of symmetries is present. In this case full ergodicity is not reached suddenly, and instead we can see intermediate regimes. This is typically the case in interacting quantum field theories without disorder; the translation-invariance of the couplings is \emph{effectively} a symmetry of the theory that slows down the onset of ergodicity by introducing more structure into the sum over trajectories in formulas like \eqref{eq pert thy}. Indeed, we see this effect at work in the interacting Ising model (Fig.~\ref{fig spins}), where full ergodicity is never reached without introducing disorder.

Finally, we remark that for each integrable free hopping theory that corresponds to particle motion in $d$ dimensions, there exists a $(d - 1)$-parameter class of off-diagonal perturbations that preserve full localization. For $d = 2$, for instance, the free hopping term is $H = L_1 + L_1^{-1} + L_2 + L_2^{-1}$, and the deformation in question is simply $H = L_1 + L_2^{-1} + \lambda(L_2 + L_2^{-1})$. For spin systems like \eqref{Ising free}, the effective dimension scales with $\log D$, and the perturbations in question are glassy Hamiltonians of the form $H = \sum_i J_i \sigma^x_i + \sum_{i > j} J_{ij} \sigma^x_i \sigma^x_j + \ldots$. These theories generically have Poisson eigenvalue statistics. In the space of all possible perturbations, this family still  takes up very little room, but it is worth remembering that in a typical problem the set of localized deformations of an integrable free hopping theory is at least a $(d - 1)$-dimensional submanifold. Note that a similar type of localization-preserving perturbation can be found in models where multiple integrable free hopping theories are superimposed. For instance, $H = L + L^{-1} + \lambda(L^2 + L^{-2})$ is a free hopping theory at $\lambda = 1$, but remains localized at any $\lambda < 1$.

\subsubsection{Global ergodicity structure}

The perturbations discussed so far span a natural subset in theory space: the space of all possible weightings of a fixed graph with $D$ sites. Each such graph corresponds to precisely one free hopping theory, and so far we have focused primarily on the perturbations around an integrable free hopping theory. The remainder of theory space is reached via perturbations that add extra hopping terms to the Hamiltonian. As seen with metric graphs, adding a few hopping terms to an integrable theory can already alter the eigenvalue spectrum of the theory, but it does not fundamentally change the ergodicity. Adding an $O(D)$ amount of hopping terms generically leads to a fully ergodic system (even if the hopping remains free), as long as this is done in an unstructured way. This is readily seen from the flat-space billiards studied in section \ref{subsec qm2}. Adding the hopping terms even more erratically can destroy all structure in the theory and lead to random graphs which are fully ergodic. We do not know if perturbations involving new hopping terms have a useful notion of critical perturbation strength.

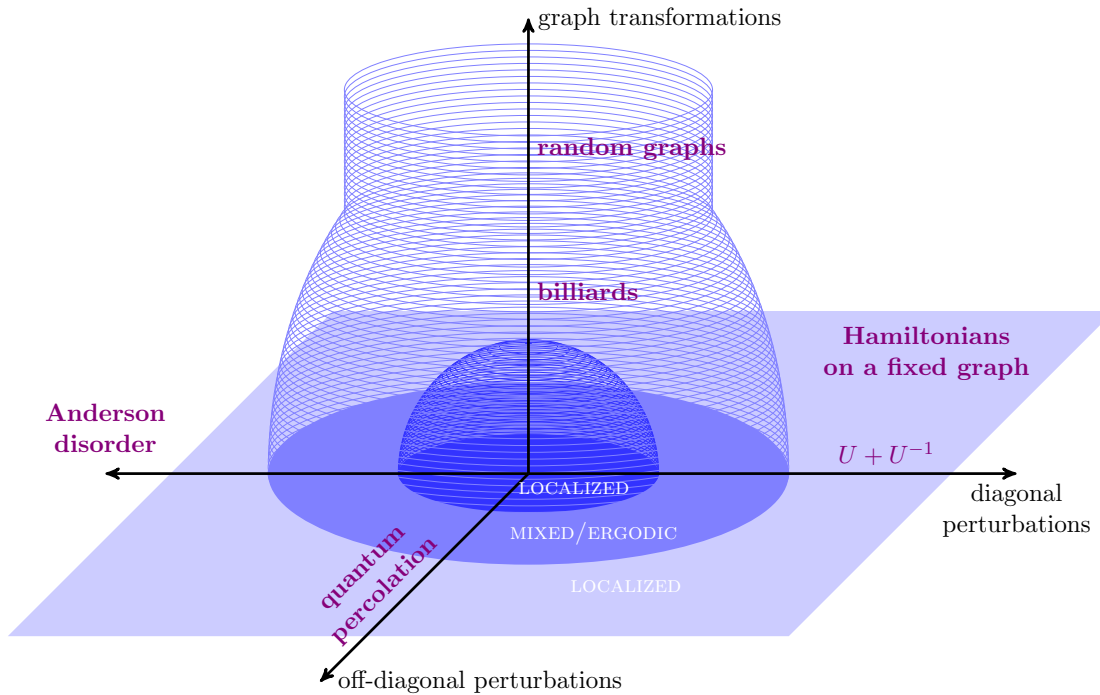
\begin{figure}
\begin{center}

\resizebox{.9\linewidth}{!}{
\begin{tikzpicture}[scale = 2,  axis/.style={very thick, ->, >=stealth'}]

  \fill[fill=blue!20]
    (-4, -1.25) -- ++(6, 0) -- ++(2.5, 2.5) -- ++(-6,0) -- cycle;
  \fill[fill=blue!50]
    (0,0) ellipse (2 and 0.7);
  \fill[fill=blue!80]
    (0,0) ellipse (1 and 0.3);

  \foreach \z in {0, 0.025, ..., 1} \draw[blue!80] (0, \z) ellipse ({sqrt(1 - \z^2)} and {0.25*sqrt(1- \z^2)});

  \foreach \z in {0, 0.05, ..., 2} \draw[blue!50] (0, \z) ellipse ({sqrt(4 - 0.5*\z^2)} and {0.25*sqrt(4- 0.5*\z^2)});

  \foreach \z in {2, 2.05, ..., 3} \draw[blue!50] (0, \z) ellipse ({sqrt(2)} and {0.25*sqrt(2)});


  \draw[axis] (0,0)  -- (3.75,0) node(xline)[below, align=center]
        {diagonal\\ perturbations};
  \draw[axis] (0,0)  -- (-3.25,0) node(xline)[right]
        {};
  \draw[axis] (0,0)  -- (-1.6, -1.6) node(yline)[right]
        {\ off-diagonal perturbations};
  \draw[axis] (0,0)  -- (0, 3.5) node(zline)[right]
        {graph transformations};

  \draw[purple!70!blue] (2.2, 1.2) node[anchor = north west, align=center] {\bf  Hamiltonians \\ \bf on a fixed graph};
  \draw[purple!70!blue] (2.75, 0) node[anchor = south] {$U + U^{-1}$};
  \draw[purple!70!blue] (-3.25, 0.1) node[anchor = south, align = center] {\bf Anderson\\ \bf disorder};
  \draw[purple!70!blue] (-1, -1) node[anchor = south, rotate = 45, align=center] {\bf quantum \\ \bf percolation};
  \draw[purple!70!blue] (0, 1.4) node[anchor = west] {\bf billiards};
  \draw[purple!70!blue] (0, 2.5) node[anchor = west] {\bf random graphs};

  \draw[white] (0.35, 0) node[anchor = north] {\textsc{\small localized}};
  \draw[white] (0.5, -0.3) node[anchor = north] {\textsc{\small mixed/ergodic}};
  \draw[white] (0.75, -0.75) node[anchor = north] {\textsc{\small localized}};

\end{tikzpicture}
}

\end{center}
\caption{\footnotesize A bigger slice of theory space around an integrable free hopping theory such as a free particle on a circle. There are many possible perturbations; we represent completely random potentials (``Anderson disorder'') and very smooth potentials (like $U + U^{-1}$) on the opposite sites of the space of diagonal perturbations. The shaded plane corresponds to all possible weightings of a given state graph; moving along the third axis changes the graph by adding or removing links. Usual billiard problems correspond to those deformations of integrable theories where links on the state graph were removed in a non-random way, resulting in a macroscopic obstacle in position space; they may still have some non-ergodic states (scars). Random graphs are even more extreme deformations of integrable theories, where links have been added and removed randomly, resulting in a theory that is typically fully ergodic.}
\label{fig landscape medium}
\end{figure}

Most free hopping theories are not integrable, and in fact they are often fully ergodic. Each of them can nevertheless be perturbed by both diagonal and off-diagonal disorder. When perturbing ergodic theories, only the upper critical disorder strength is meaningful, and it marks the transition from ergodic to localized. This critical strength depends strongly on the type of perturbation involved. For instance, if we perturb a random regular graph with Gaussian off-diagonal disorder, the disorder strength needs to scale with $D$ in order to localize the theory (see Fig.~\ref{fig large disorder   RR}). On the other hand, perturbing the same theory with \emph{any} finite Cauchy off-diagonal disorder causes full position-space localization, and the critical breadth of Cauchy disorder of mean one is found to scale as $1/D$. The reason for this behavior is that the scale of the theory is set by the largest random coupling. In the case of Cauchy disorder, if $d$ is the number of links attached to each site, there are $\sim d D$ random couplings drawn from a heavy-tailed distribution of breadth $\sigma$. A good heuristic for the typical maximal coupling is that it scales as $\sim \sigma d D$. (Compare this with the Gaussian maximal coupling $\sim \sigma \log(d D)$.) After rescaling to set this maximal coupling to unity, most couplings are now of order $\sim 1/d D$. Thus, Cauchy-disordering a random regular graph amounts to severely diluting the links of the graph. This leads to localization, just like in the percolation scenarios discussed in section \ref{subsec qm2}.

We can now envision how the whole theory space is classified by ergodicity. The theory space is the set of Hermitian $D \times D$ matrices, modulo permuting basis elements and an overall rescaling that we use to set the largest off-diagonal element to unity. After fixing an admissible basis, position-localized theories correspond to the subspace of diagonal matrices. Conversely, momentum-localized free hopping theories are a small part of the subspace of matrices with no diagonal elements. Each localized theory has a neighborhood of perturbations in which the ergodicity properties are unaffected, and outside this neighborhood there exists an intermediate regime where the theory is neither position- nor momentum-localized (see Fig.~\ref{fig landscape medium}). The theory in this regime is ergodic or displays nonuniversal behavior associated to crossovers, depending on the kind of the perturbation.

\section{Ergodicity in the SYK and Gurau-Witten models} \label{sec syk}

Where does the SYK model fit into the ergodicity landscape? This question can be answered by studying its Hamiltonian expressed in terms of Pauli spin operators, \eqref{eq syk spins}. The corresponding free hopping theory is that of the two- and four-spin Sherrington-Kirkpatrick model,
\bel{
  H = \sum_{1 \leq i < j \leq N'} \sigma^x_i \sigma^x_j + \sum_{1\leq i < j < k < l \leq N'} \sigma^x_i \sigma^x_j \sigma^x_k \sigma_l^x.
}
Recall that $N' = N/2$, where $N$ is the number of Majorana fermions. The corresponding graph consists of all plaquette and four-cube diagonals that can be drawn on an $N'$-dimensional hypercube. Each site has $\binom {N'}4 + \binom {N'}2 \sim N^4$ links emerging out of it.

The disorder that defines SYK is both diagonal and off-diagonal. The full SYK Hamiltonian has three types of terms: $\Sigma^2$, $\Sigma \sigma^z$, and $\sigma^z \sigma^z$ (cf.~\eqref{eq syk spins}). The model has a total of $\binom{N}4 \sim N^4$ different couplings, much fewer than the number of links in the corresponding graph, which is $\sim N^4 2^{N/2}$. This is very far from the standard off-diagonal disorder that we considered in the section \ref{subsec pert hopping}, and correspondingly we may expect that the SYK model is not as ergodic as a high-dimensional quantum percolation model. However, it turns out that even this comparatively small amount of disorder makes SYK maximally ergodic, as seen on Fig.~\ref{fig   SYK ents}. We recall, however, that the numerical results accessible to us all have $q^2 = 16 \sim N \leq 22$. All evidence indicates that SYK transitions from fully to partially ergodic as $N$ is dialed up across $q^2$ \cite{Cotler:2016fpe, Garcia-Garcia:2017pzl, Erdos:2014, Schroeder:2015}, and indeed the signatures of integrability and deviations from random matrix theories discussed in Refs.\ \citen{Maldacena:2016hyu, Polchinski:2016xgd, Cotler:2016fpe} at $N \gg q^2$ are not likely to be captured by our numerics. Conversely, we expect that our results may start deviating from random matrix behavior at sufficiently high $N$.

The fact that all couplings are drawn from the same distribution means that the rescaling that normalizes the maximal coupling in the hopping part will also normalize the diagonal disorder to be $O(1)$. Thus the SYK model will always be a strongly interacting theory with the position and momentum operators balanced, similarly to how the Ising model becomes critical when the position and momentum operators have equal strengths. Unlike random graphs where Cauchy disorder --- or even a sufficiently high-variance Gaussian disorder --- was sufficient to localize the theory, this is not possible in SYK.

\begin{figure}
  \centering
  \includegraphics[width=0.47 \textwidth]{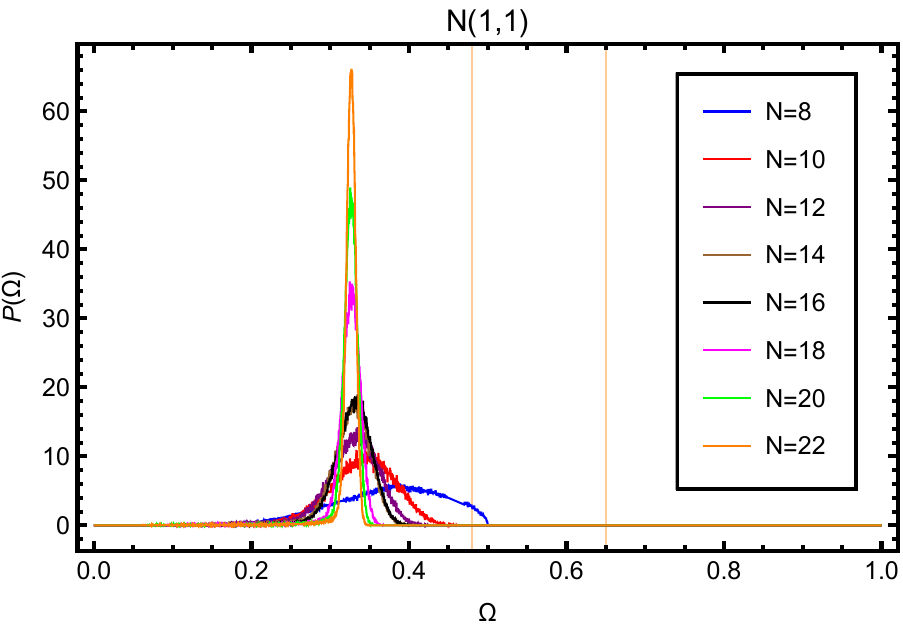}
  \includegraphics[width=0.47 \textwidth]{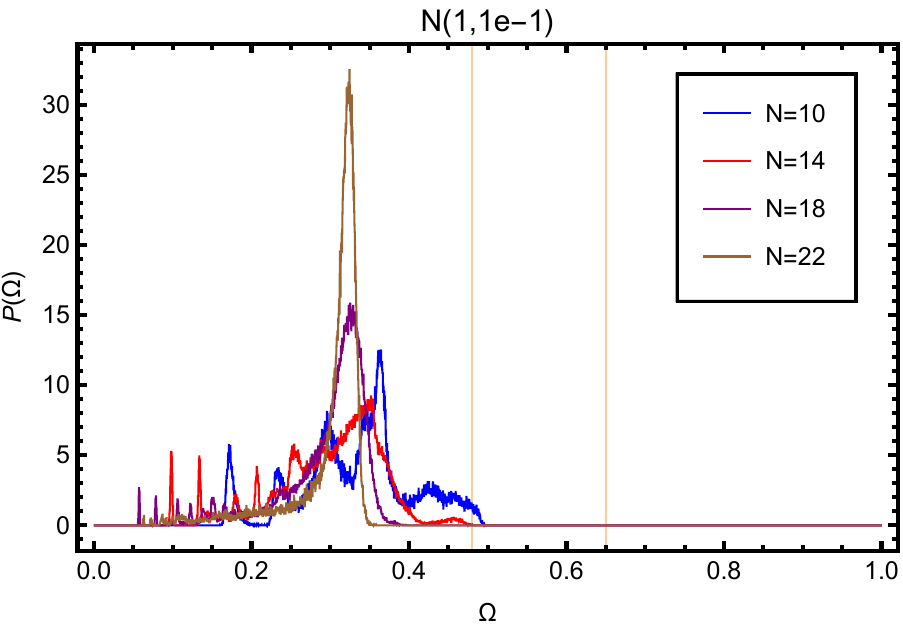}

  \includegraphics[width=0.47 \textwidth]{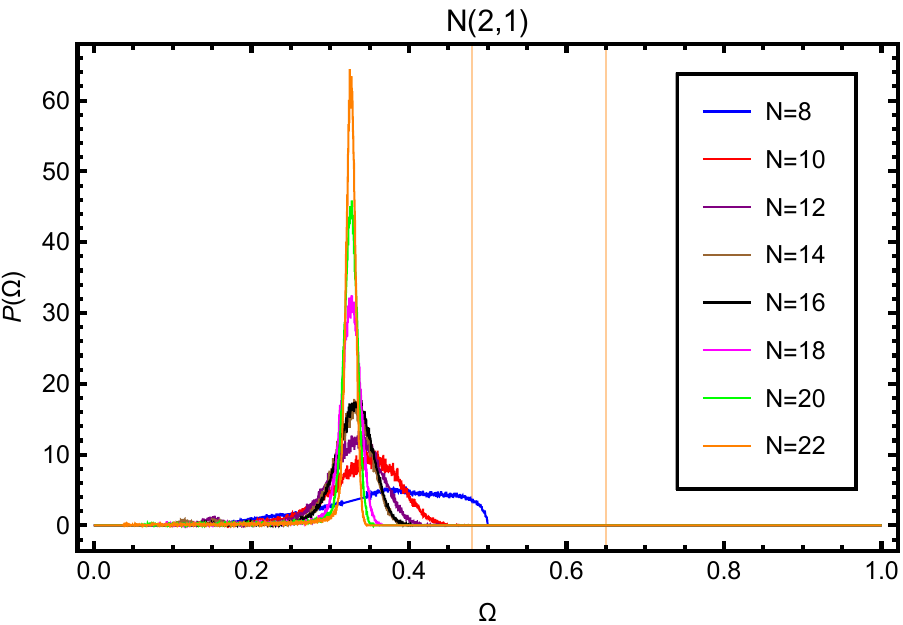}
  \includegraphics[width=0.47 \textwidth]{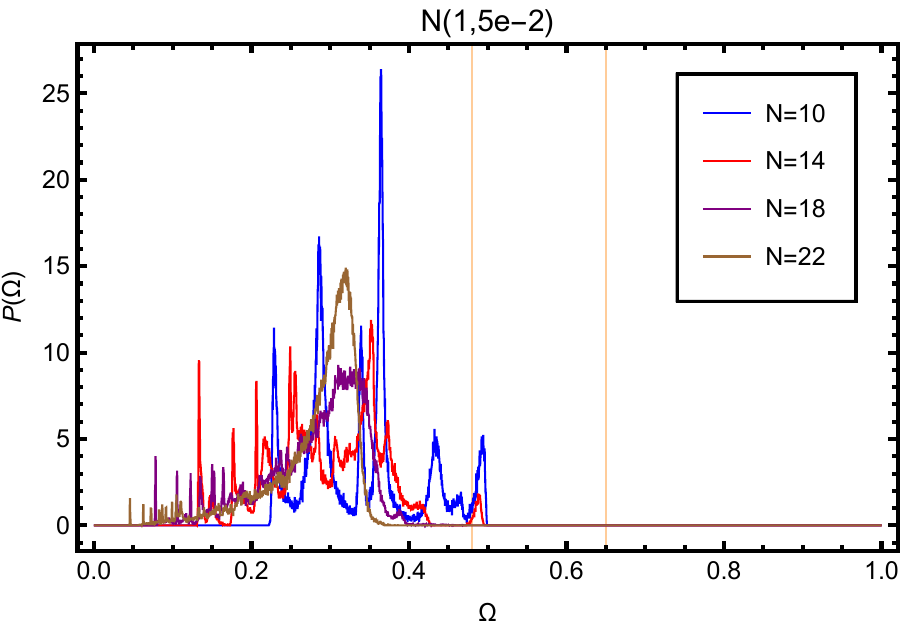}

  \includegraphics[width=0.47 \textwidth]{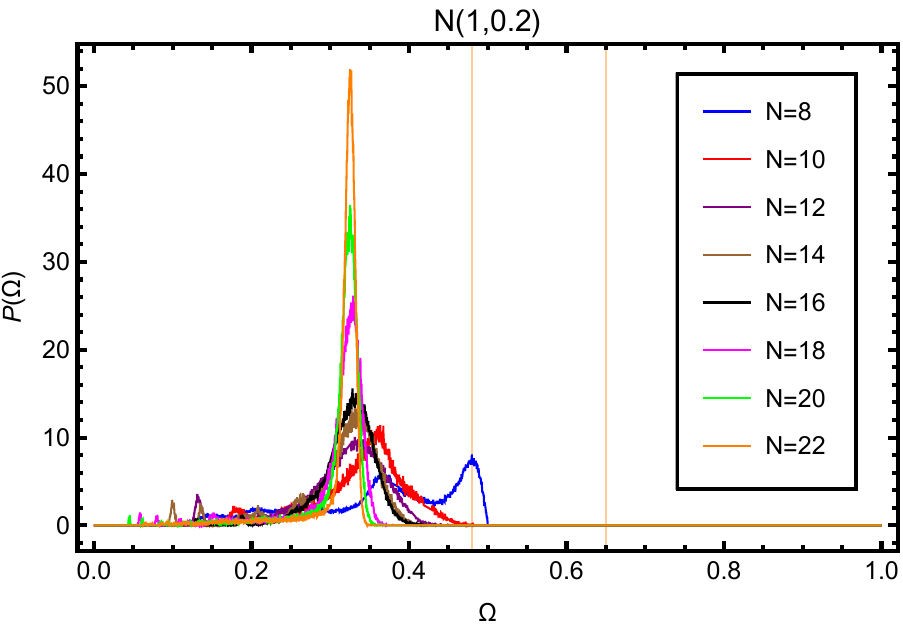}
  \includegraphics[width=0.47 \textwidth]{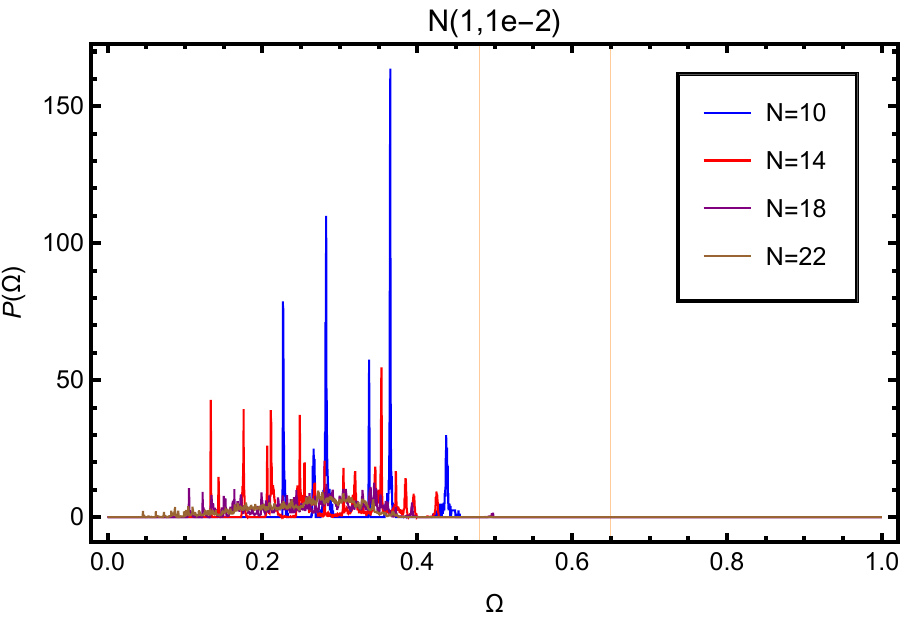}
  \caption{\footnotesize IR diversities of the SYK model when couplings are chosen from the Gaussian distribution $\mathcal N(1, J)$, with $J$ varying from $J = 1$ (upper left) to $J = 0.01$ (bottom right). Each plot shows an aggregate of $1,024,000/D$ instances of disorder. As $J$ decreases at fixed $N$, the $\Omega$ profile flattens, developing a heavy tail towards low IR diversities; ergodicity is diminished. The theory does not localize with large $N$ even at the smallest examined values of $J$.}\label{fig syk medium disorder}
\end{figure}

The one deviation from ergodicity that we can numerically access comes from tuning the disorder away (see Fig.~\ref{fig syk medium disorder}) by choosing random couplings from a Gaussian distribution like $\mathcal N(1, J)$ for small $J$. At fixed $N$, making all couplings effectively equal reduces ergodicity but does not fully localize the theory. For a fixed disorder strength $J$, increasing $N$ brings the $\Omega$-distribution closer to that of strongly disordered random regular graphs and ergodic spin chains without disorder (Figs.~\ref{fig large disorder RR} and \ref{fig spins}). These distributions have heavy tails that become less important in the large $N$ limit. In the case of the critical Ising model, we have argued that the heavy tails seen there are due to free particle states giving rise to a linear $\Omega$-distribution that cuts off at $\Omega \sim \frac1{\sqrt{N}}$ (Fig.~\ref{fig ising}). In the case of SYK, we know that stringy  states exist in the spectrum \cite{Maldacena:2016hyu}. It is tempting to speculate that the heavy tails seen at weak disorder come from these states, and that they disappear as $N \rar \infty$ because the spectrum is dominated by maximally ergodic states, the analogs of ``black hole microstates'' \cite{Cotler:2016fpe}.

We do not know whether this weak disorder limit becomes critical if $J$ scales with $N$ in a particular way. The fact that the $\Omega$-distribution at low $J$ gives the same distribution $P(\Omega)$ at all $N$, modulo a rescaling of bin widths, is intriguingly similar to the  behavior seen near the critical region of the Aubry-Andr\'e model, Fig.~\ref{fig AA transition}, where conformal quantum mechanics is known to exist. Studying this question in greater numerical detail is currently beyond our reach.

Additional perspective can be gained by studying the Gurau-Witten model \cite{Bonzom:2011zz, Witten:2016iux}, another Majorana model whose large $N$ thermodynamics largely matches that of the SYK model (as both are governed by the same diagrams in a particular temperature regime). The GW model has $\mathcal D + 1$ fields $\psi^a$, $a = 0, 1, \ldots, \mathcal D$, each containing $N^\mathcal D$ Majorana fermions. The Hamiltonian is, up to prefactors,
\bel{
  H = \sum_{i_{01},\, i_{02},\,\ldots,\ i_{\mathcal D - 1, \mathcal D} = 1}^N \psi^0_{i_{01}\ldots i_{0\mathcal D}} \psi^1_{i_{01}i_{12}\ldots i_{1\mathcal D}} \ldots \psi^\mathcal D_{i_{0\mathcal D}i_{1\mathcal D}\ldots i_{\mathcal D-1,\mathcal D}}.
}

When $\mathcal D$ is odd, this model is very simply dualized to a spin system. Let $I, J, \ldots$ be multi-indices composed of $\mathcal D$ numbers between $1$ and $N$, and let us impose the lexicographical ordering (e.g. $NN\ldots N > 11\ldots 1N > 11\ldots 11$) on these multi-indices. Then we dualize via
\bel{
  \psi^{2a}_I \equiv X^a_I  \prod_{J < I} Z^a_J \prod_{b < a} \prod_{\trm{all\ }J} Z^b_J, \quad
  \psi^{2a + 1}_I \equiv  Y^a_I \prod_{J < I} Z^a_J \prod_{b < a} \prod_{\trm{all\ }J} Z^b_J.
}
The operators $X^a, Y^a, Z^a$ are copies of the usual Pauli matrices, with $a = 0, \ldots, \frac{\mathcal D - 1}2$. The GW model thus dualizes to $\frac{\mathcal D + 1}2$ coupled spin chains of $N^\mathcal D$ spins each, for a total Hilbert space dimension $D = 2^{(\D + 1) N^\D / 2}$.

Note that
\bel{
  \psi_I^{2a} \psi_J^{2a + 1} = X_I^a \prod_{K < I} Z_K^a\  Y_I^a \prod_{L < J} Z_L^a
}
is always an operator on a single spin chain, and it contains at most two momentum (spin-flip) operators. The GW Hamiltonian thus consists of $N^{\mathcal D(\mathcal D + 1)/2}$ terms, and each flips at most two spins in each of the $\frac{\mathcal D + 1}2$ chains. In this sense the GW model is very similar to the SYK model without disorder. We expect that their ergodic properties (Fig.~\ref{fig syk medium disorder}) will be very similar.

In terms of state graphs, the ones corresponding to Gurau-Witten and SYK at low disorder will be diagonally perturbed free hopping theories, as was the graph of the ergodic spin chain without disorder. These graphs will have a lot of symmetries and we expect some deviation from random matrix behavior in all of these models. The graph connectivities $d(N)$ differ in all these cases, and this will be reflected in different details of the tail behavior in the $\Omega$-distributions.  This is in qualitative agreement with our results (Figs.~\ref{fig   spins} and \ref{fig syk medium disorder}) and with numerical studies of the Gurau-Witten model, which have found that already the eigenvalue diagnostics of chaos fail to agree with predictions from random matrix theory, even though numerics are only viable for the modest values of $\D = 3$ and $N = 2$  \cite{Krishnan:2016bvg}. Of course, we caution that more numeric and analytic work is needed to understand whether these similarities hold below the surface.

Finally, recall that the Feynman diagrams that compute the free energy in a large window of temperatures agree between the Gurau-Witten model and SYK with the usual $\mathcal N(0, 1)$ disorder.\cite{Witten:2016iux} This leads us to conjecture that the deviation from maximal ergodicity in the $q^2 \ll N$ regime of the usual SYK model will be analogous to the deviation from maximal ergodicity found in the weakly disordered SYK model, as the latter is expected to be in the same region of the ergodicity landscape as the Gurau-Witten model based on our graph structure considerations. Beyond spin systems, this regime of deviating from ergodicity is also exhibited in strongly disordered random regular graphs, Fig.~\ref{fig large disorder RR}. It is natural, although precipitous, to interpret all of these deviations as signatures of quasi-particles or ``quasi-strings/branes'' that correspond to relatively localized eigenstates within the spectrum.\footnote{See Refs.\ \citen{Shenker:2014cwa, Gu:2016oyy} for related discussions on how the presence of stringy states reduces the ergodicity as measured by Lyapunov exponents.}  It would be extremely interesting to understand these issues in greater detail.

\section{Outlook}

The goal of this paper was to understand how quantum ergodicity varies across the landscape of all quantum theories with a given Hilbert space dimension. We have delineated the rough outlines of each important region of this landscape, but have not attempted to understand the fate of the landscape in the continuum/thermodynamic limit. In particular, we have not provided a criterion that distinguishes which crossovers become phase transitions in this limit. Understanding this would provide valuable insight into the structure of the ergodicity landscape.

On a related note, we reported glimpses of universal behavior that is neither localized nor ergodic. This critical regime is found only in a handful of theories --- Aubry-Andr\'e models and perhaps weakly disordered SYK --- and its relation to what is often called ``conformal quantum mechanics'' is quite unclear. While conformal symmetry in general does not strongly affect ergodicity (even in 2D, there are conformal theories that are ergodic and have BTZ black holes in their gravity duals; see also Ref.~\citen{Turiaci:2016cvo}), the Aubry-Andr\'e kind of criticality appears to bear a remarkable ergodic signature that deserves further scrutiny.

A question related to our ergodicity landscape program concerns classifying the ways complexity can grow under Hamiltonian evolution. For a fixed (ensemble of) $H$, what percentage of the space of $SU(D)$ matrices is covered by the set $\{e^{iHt}\}$ as the time $t$ grows? Alternatively, how well does a long-time evolution by a given Hamiltonian approximate an evolution by a Haar-random unitary? These questions have recently been addressed in Ref.~\citen{Roberts:2016hpo}, and its main diagnostic tool --- the frame potential --- may well serve the same landscaping purpose as our IR entropies and diversities. The related question of developing a complexity-based metric on the space of all theories was discussed in Ref.~\citen{Brown:2017jil} and references therein. Yet another question related to studying the space of all unitary evolution operators pertains to the spectral universality of periodically driven systems \cite{DAlessio:2014}.

Finally, a topic that we did not address at all concerns the distinction between Anderson- and many-body-localization in the position-localized theories. These regimes can be distinguished, for instance, by the dynamics of subsystem entanglement following a quench \cite{Huse:2014}, and it seems that our diagnostics are blind to such nuances. It could be illuminating to understand the connection between our measures of ergodicity and dynamical processes such as thermalization.

\section*{Acknowledgments}

It is a pleasure to acknowledge useful discussions with Pranjal Bordia, Daniel Brattan, Hrant Gharibyan, Steve Shenker, Guifre Vidal, Sho Yaida, Horng-Tzer Yau, and Beni Yoshida. We also thank the audiences who heard preliminary versions of this research at Harvard, MIT, Brandeis, TIFR Mumbai, and the 2016 Indian Strings Meeting in Pune. Research at Perimeter Institute is supported by the Government of Canada through Industry Canada and by the Province of Ontario through the Ministry of Economic Development \& Innovation.


\begin{thebibliography}{120}

\bibitem{Radicevic:2016kpf}
  \DJ.~Radi\v cevi\'c,
  ``Quantum Mechanics in the Infrared,''
  arXiv:1608.07275 [hep-th].


\bibitem{Shenker:2013pqa}
  S.~H.~Shenker and D.~Stanford,
  ``Black holes and the butterfly effect,''
  JHEP {\bf 1403}, 067 (2014),
  arXiv:1306.0622 [hep-th].

\bibitem{Maldacena:2015waa}
  J.~Maldacena, S.~H.~Shenker and D.~Stanford,
  ``A bound on chaos,''
  JHEP {\bf 1608}, 106 (2016),
  arXiv:1503.01409 [hep-th].


\bibitem{Kitaev:2015}
  A.~Kitaev,
  ``A simple model of quantum holography,''
   KITP strings seminar and Entanglement program (Feb 12, Apr 7, and May 27, 2015) \href{http://online.kitp.ucsb.edu/online/entangled15/} {http://online.kitp.ucsb.edu/online/entangled15/}

\bibitem{Polchinski:2016xgd}
  J.~Polchinski and V.~Rosenhaus,
  ``The Spectrum in the Sachdev-Ye-Kitaev Model,''
  JHEP {\bf 1604}, 001 (2016),
  arXiv:1601.06768 [hep-th].


\bibitem{Maldacena:2016hyu}
  J.~Maldacena and D.~Stanford,
  ``Remarks on the Sachdev-Ye-Kitaev model,''
  Phys.\ Rev.\ D {\bf 94}, no. 10, 106002 (2016),
  arXiv:1604.07818 [hep-th].


\bibitem{Gutzwiller:1990}
  M.~Gutzwiller,
  ``Chaos in classical and quantum mechanics,''
  Springer (1990).

\bibitem{Wigner:1951}
  E.~P.~Wigner and P.~A.~M.~Dirac,
  Proceedings of the Cambridge Philosophical Society {\bf 47}, 790 (1951).

\bibitem{Dyson:1970tza}
  F.~J.~Dyson,
  ``Correlations between the eigenvalues of a random matrix,''
  Commun.\ Math.\ Phys.\  {\bf 19}, no. 3, 235 (1970).



\bibitem{Bohigas:1983er}
  O.~Bohigas, M.~J.~Giannoni and C.~Schmit,
  ``Characterization of chaotic quantum spectra and universality of level fluctuation laws,''
  Phys.\ Rev.\ Lett.\  {\bf 52}, 1 (1984).

\bibitem{Berry:1977}
  M.~V.~Berry and M.~Tabor,
  ``Level clustering in the regular spectrum,''
  Proc.\ R.\ Soc.\ Lond.\ A {\bf 356}, 375 (1977).


\bibitem{Shnirelman:1974}
  A.~I.~Shnirelman,
  ``Ergodic properties of eigenfunctions,''
  Usp.\ mat.\ nauk {\bf 29} (1974).

\bibitem{CdV:1985}
  Y.~Colin de Verdi\`ere,
  ``Ergodicit\'e et fonctions propres du laplacien,''
  Commun.~Math.~Phys.~{\bf 102} (1985).

\bibitem{Zelditch:1987}
  S.~Zelditch,
  ``Uniform distribution of eigenfunctions on compact hyperbolic surfaces,''
  Duke Math.\ Journal, {\bf 55}, 919 (1987).

\bibitem{Muller:2005}
  S.~M\"uller, S.~Heusler, P.~Braun, F.~Haake, A.~Altland,
  ``Periodic-orbit theory of universality in quantum chaos,''
  Phys.~Rev.~E {\bf 72}, 4 (2005).

\bibitem{Muller:2009}
  S.~M\"uller, S.~Heusler,  A.~Altland, P.~Braun, F.~Haake,
  ``Periodic-orbit theory of universal level correlations in quantum chaos,''
  New J.~Phys.~{\bf 11}, 103025 (2009).

\bibitem{Erdos:2011}
  L.~Erd{\H o}s,
  ``Universality of Wigner random matrices: a survey of recent results,''
   Russian Mathematical Surveys, {\bf 66}, 507 (2011),
   arXiv:1004.0861 [math-ph].

\bibitem{Altland:1997}
  A.\ Altland and M.\ R.\ Zirnbauer,
  ``Nonstandard symmetry classes in mesoscopic normal-superconducting hybrid structures,''
  Phys.~Rev.~B, {\bf 55}, 1142 (1997).

\bibitem{Gnutzmann:2003}
  S.~Gnutzmann and B.~Seif,
  ``Universal spectral statistics in Wigner-Dyson, chiral and Andreev star graphs I: construction and numerical results,''
  Phys.~Rev.~E {\bf 69} (2003),
  arXiv:nlin/0309049 [nlin.CD].

\bibitem{Kachiga:1986}
  K.~Machiga and M.~Fujita,
  ``Quantum energy spectra and one-dimensional quasiperiodic systems,''
  Phys.\ Rev.\ B {\bf 34}, 7367 (1986).

\bibitem{Geisel:1991}
  T.\ Geisel, R.\ Ketzmerick, and G.\ Petschel,
  ``New class of level statistics in quantum systems with unbounded diffusion,''
  Phys.\ Rev.\ Lett.\ {\bf 66}, 1651 (1991).

\bibitem{Kravtsov:1997}
  V.~E.~Kravtsov and K.~A.~Muttalib,
  ``New Class of Random Matrix Ensembles with Multifractal Eigenvectors,''
  Phys.~Rev.~Lett.~{\bf 79} (1997),
  arXiv:cond-mat/9703167.

\bibitem{Mirlin:2000}
  A.~D.~Mirlin and F.~Evers,
  ``Multifractality and critical fluctuations at the Anderson transition,''
  Phys. Rev. B {\bf 62}, 7920 (2000),
  arXiv:cond-mat/0003332 [cond-mat.mes-hall].

\bibitem{Kaplan:2009kr}
  D.~B.~Kaplan, J.~W.~Lee, D.~T.~Son and M.~A.~Stephanov,
  ``Conformality Lost,''
  Phys.\ Rev.\ D {\bf 80}, 125005 (2009),
  arXiv:0905.4752 [hep-th].

\bibitem{Chirikov:1988}
  B.~V.~Chirikov, F.\ M.\ Izrailev, and D.\ L.\ Shepelyansky.
  ``Quantum chaos: localization vs.\ ergodicity.'' Physica D: Nonlinear Phenomena {\bf 33}, 1 (1988).

\bibitem{Izrailev:1990}
  F.~Izrailev,
  ``Simple models of quantum chaos: spectrum and eigenfunctions,''
  Physics Reports, {\bf 196}, 5 (1990).

\bibitem{Santos:2010}
  L.~Santos, M.~Rigol,
  ``Localization and the effects of symmetries in the thermalization properties of one-dimensional quantum systems,''
  Phys.~Rev.~E {\bf 82}, 031130 (2010),
  arXiv:1006.0729 [cond-mat.stat-mech].


\bibitem{Santos:2011}
  L.\ F.\ Santos, F.\ Borgonovi, and F.\ M.\ Izrailev,
  ``Onset of chaos and relaxation in isolated systems of interacting spins-1/2: energy shell approach,''
  Phys.\ Rev.\ E {\bf 85}, 036209 (2012),
  arXiv:1201.0186 [cond-mat.stat-mech].



\bibitem{Kaplan:2001}
  L.~Kaplan,
  ``Eigenstate Structure in Graphs and Disordered Lattices,''
  Phys.\ Rev.\ E {\bf 64}, 036225 (2001),
  arXiv:nlin/0101048v1.

\bibitem{Berkolaiko:2003}
  G.\ Berkolaiko, J.\ P.\ Keating, and B.\ Winn,
  ``No quantum ergodicity for star graphs,''
  Commun. Math. Phys. \textbf{250}, 259 (2004),
  arXiv:math-ph/0308005v1.

\bibitem{Ohya:2004}
  M.~Ohya and D.~Petz,
  ``Quantum Entropy and its Use,''
  Springer, 2004.

\bibitem{Balachandran:2013}
  A.~P.~Balachandran, T.~R.~Govindarajan, A.~R.~de Queiroz, A.~F.~Reyes-Lega,
  ``Algebraic approach to entanglement and entropy,''\
  Phys.\ Rev.\ A\ {\bf 88}, 022301 (2013),
  arXiv:1301.1300 [mat-ph].

\bibitem{Polyakov:1974gs}
  A.~M.~Polyakov,
  ``Nonhamiltonian approach to conformal quantum field theory,''
  Zh.\ Eksp.\ Teor.\ Fiz.\  {\bf 66}, 23 (1974).

\bibitem{Kawahigashi}
  Y.~Kawahigashi,
  ``Conformal Field Theory and Operator Algebras,''
  in V.\ Sidoravi\v cius, ``New Trends in Mathematical Physics,''
  Springer (2009).

\bibitem{Jones}
  V.\ Jones, V.~S.~Sunder,
  ``Introduction to Subfactors,''
  Cambridge UP (1997).

\bibitem{Deutsch:1991}
  J.~Deutsch,
  ``Quantum statistical mechanics in a closed system,''
  Phys.\ Rev.\ A {\bf 43}, 2046 (1991).

\bibitem{Srednicki:1994}
  M.~Srednicki,
  ``Chaos and Quantum Thermalization,'' Phys.~Rev.~E. {\bf 50}, 888 (1994)
  arXiv:cond-mat/9403051v2.


\bibitem{Berry:1977b}
  M.~V.~Berry,
  ``Regular and irregular semiclassical wave functions,''
  J.~Phys.~A {\bf 10}, 2083-91 (1977).

\bibitem{Brody:1981}
  T.~A.~Brody et al,
  ``Random-matrix physics: spectrum and strength fluctuations,''
  Rev.~Mod.~Phys.~{\bf 53} 385 (1981).

\bibitem{Tao:2010}
  T.~Tao, V.~Vu,
  ``Random matrices: Universality of local eigenvalue statistics up to the edge,''
  Comm.~Math.~Phys, {\bf 298} 549 (2010),
  arXiv:0908.1982v4 [math.PR].

\bibitem{ORourke:2016}
  S.~O'Rourke, V.~Vu, K.~Wang,
  ``Eigenvectors of random matrices: A survey,''
  arXiv:1601.03678v3 [math.PR].

\bibitem{Zelditch:2005}
  S.~Zelditch,
  ``Quantum ergodicity and mixing of eigenfunctions,''
  [arXiv:math-ph/0503026].

\bibitem{Nonnenmacher:2010}
  S.~Nonnemacher,
  ``Anatomy of quantum chaotic eigenstates,''
  arXiv:1005.5598 [math.DS].

\bibitem{Oganesyan:2006}
  V.~Oganesyan and D.~A.~Huse,
  ``Localization of interacting fermions at high temperature,''
  Phys.\ Rev.\ B {\bf 75}, 155111 (2007),
  arXiv:cond-mat/0610854 [cond-mat.str-el].

\bibitem{Atas:2012}
  Y.\ Y.\ Atas, E.\ Bogomolny, O.\ Giraud, and G.\ Roux,
  ``The distribution of the ratio of consecutive level spacings in random matrix ensembles,''
  Phys.\ Rev.\ Lett.\ {\bf 110}, 084101 (2013),
  arXiv:1212.5611 [math-ph].

\bibitem{Anderson:1958vr}
  P.~W.~Anderson,
  ``Absence of Diffusion in Certain Random Lattices,''
  Phys.\ Rev.\  {\bf 109}, 1492 (1958).

\bibitem{Stolz:2011}
  G.~Stolz,
  ``An Introduction to the Mathematics of Anderson Localization,''
  arXiv:1104.2317 [math-ph].


\bibitem{Aubry:1980}
  S.~Aubry and G.~Andr\'e,
  ``Analyticity breaking and Anderson localization in incommensurate lattices,''
  Ann.~Israel Phys.~Soc.~{\bf 3}, 133 (1980).

\bibitem{Wilkinson:1984}
  M.~Wilkinson,
  ``Critical Properties of Electron Eigenstates in Incommensurate Systems,''
  Proc.~R.~Soc.~A {\bf 391}, 1801 (1984).


\bibitem{deAlfaro:1976vlx}
  V.~de Alfaro, S.~Fubini and G.~Furlan,
  ``Conformal Invariance in Quantum Mechanics,''
  Nuovo Cim.\ A {\bf 34}, 569 (1976).

\bibitem{Chamon:2011xk}
  C.~Chamon, R.~Jackiw, S.~Y.~Pi and L.~Santos,
  ``Conformal quantum mechanics as the CFT$_1$ dual to AdS$_2$,''
  Phys.\ Lett.\ B {\bf 701}, 503 (2011),
  arXiv:1106.0726 [hep-th].


\bibitem{Kottos:1997}
  T.~Kottos and U.~Smilansky,
  ``Quantum Chaos on Graphs,''
  Phys.~Rev.~Lett.~{\bf 79}, 24 (1997).

\bibitem{GangOfFour}
  E.~Abrahams, P.~W.~Anderson, D.~C.~Licciardello, and T.~V.~Ramakrishnan,
  ``Scaling Theory of Localization: Absence of Quantum Diffusion in Two Dimensions,''
  Phys.~Rev.~Lett.~{\bf 42}, 10 (1979).

\bibitem{deGennes:1959}
  P.~G.~de Gennes, P.~Lafore, and J.~P.~Millot,
  ``Amas accidentels dans les solutions solides d\'esordonn\'ees,''
  J.~Phys.~Chem.~Solids {\bf 11}, 105 (1959).

\bibitem{Shapir:1982}
  Y.~Shapir, A.~Aharony, and A.~Brooks Harris,
  ``Localization and Quantum Percolation,''
  Phys.~Rev.~Lett.~{\bf 49}, 7 (1982).

\bibitem{Chayes:1986}
  J.~T.~Chayes, L.~Chayes, J.~R.~Franz, J.~P.~Sethna, and S.~A.~Trugman,
  ``On the density of states for the quantum percolation problem,''
  J.~Phys.~A: Math.~Gen.~{\bf 19} (1986).

\bibitem{Sinai:1963}
  Ya.~G.~Sinai,
  ``On the foundations of the ergodic hypothesis for a dynamical system of statistical mechanics,''
  Doklady Akademii Nauk SSSR, {\bf 153}, 6 (1963).

\bibitem{Bunimovich:1970}
  L.~A.~Bunimovich,
  ''On the Ergodic Properties of Nowhere Dispersing Billiards,''
   Commun.~Math.\ Phys.\ {\bf 65}, 3 (1970).

\bibitem{Berry:1981}
  M.~V.~Berry,
  ``Quantizing a classically ergodic system: Sinai's billiard and the KKR method,''
  Ann.~Phys.~{\bf 131}, 1 (1981).

\bibitem{Heller:1984}
  E.~J.~Heller,
  ``Bound-state eigenfunctions of classically chaotic Hamiltonian systems: scars of periodic orbits,''
  Phys.\ Rev.\ Lett.\ {\bf 53} (1984).

\bibitem{Hassell:2008}
  A.~Hassell and L.~Hillairet,
  ``Ergodic billiards that are not quantum unique ergodic,''
  [arXiv:0807.0666].

\bibitem{Zelditch:1992}
  S.~Zelditch,
  ``Quantum Ergodicity on the Sphere,''
  Commun.\ Math.\ Phys.\ {\bf 146} (1992).

\bibitem{Brooks:2015}
  S.~Brooks, E.~Le Masson, and E.~Lindenstrauss,
  ``Quantum Ergodicity and Averaging Operators on the Sphere,''
  arXiv:1505.03887 [math.SP].

\bibitem{Rudnick:1994}
  Z.~Rudnick and P.~Sarnak,
  ``The behaviour of eigenstates of arithmetic hyperbolic manifolds,''
  Commun.~Math.~Phys.~{\bf 161} (1994).

\bibitem{Bourgain:2003}
  J.~Bourgain and E.~Lindenstrauss,
  ``Entropy of Quantum Limits,''
  Commun.~Math.~Phys.~{\bf 233} (2003).

\bibitem{Anantharaman:2008}
  N.~Anantharaman,
  ``Entropy and the localization of eigenfunctions,''
  Ann.~Math.~{\bf 168} (2008).

\bibitem{Bogomolny:2003}
  E.~Bogomolny,
  ``Quantum and Arithmetical Chaos,''
  arXiv:nlin/0312061.

\bibitem{Bogomolny:2003b}
  E.~Bogomolny and C.~Schmit,
  ``Multiplicities of Periodic Orbit Lengths for Non-Arithmetic Models,''
  arXiv:nlin/0312057.

\bibitem{Gnutzmann:2006}
  S.~Gnutzmann and U.~Smilansky,
  ``Quantum Graphs: Applications to Quantum Chaos and Universal Spectral Statistics,''
  arXiv:nlin/0605028.

\bibitem{Gutzwiller:1971}
  M.~C.~Gutzwiller,
  ``Periodic Orbits and Classical Quantization Conditions,''
  J.~Math.~Phys.~{\bf 12}, 343 (1971).

\bibitem{Berry:1989}
  M.~V.~Berry,
  ``Some quantum-to-classical asymptotics,'' in Les Houches Lecture Series LII (1989), eds.~M.-J.~Giannoni, A.~Voros and J.~Zinn-Justin, North-Holland, Amsterdam, 251-304.

\bibitem{Bauerschmidt:2016}
  R.~Bauerschmidt, J.~Huang, H.-T.~Yau,
  ``Local Kesten--McKay law for random regular graphs,''
  arXiv:1609.09052 [math.PR].

\bibitem{Efetov:1997}
  K.~Efetov,
  ``Supersymmetry in Disorder and Chaos,''
  Cambridge University Press (1997).

\bibitem{Tao:2011}
  T.~Tao and V.~Vu,
  ``Random matrices: The Four Moment Theorem for Wigner ensembles,''
  arXiv:1112.1976 [math.PR].

\bibitem{Erdos:2016}
  L.~Erd\H os and H.-T.~Yau,
  ``Dynamical Approach to Random Matrix Theory,'' preliminary version available at \href{http://www.math.harvard.edu/~htyau/Random-Matrix-Aug-2016.pdf} {http://www.math.harvard.edu/~htyau/Random-Matrix-Aug-2016.pdf}
  (retrieved in December 2016)

\bibitem{Kravtsov:2015}
   V.\ E.\ Kravtsov, I.\ M.\ Khaymovich, E.\ Cuevas, and M.\ Amini,
   ``A random matrix model with localization and ergodic transitions,''
   New Journal of Physics {\bf 17}, 122002 (2015),
    	arXiv:1508.01714 [cond-mat.dis-nn].

\bibitem{Soshnikov:2006}
   A.\ Soshnikov,
   ``Poisson statistics for the largest eigenvalues in random matrix ensembles,''
   in ``Mathematical physics of quantum mechanics,'' vol.\ 690 of Lecture Notes in Physics, Springer (2006).

\bibitem{Lieb:1961}
  E.~Lieb, T.~Schultz, and D.~Mattis,
  ``Two soluble models of an antiferromagnetic chain,''
  Ann.~Phys.~{\bf 16}, 3 (1961).

\bibitem{Poilblanc:1993}
   D.\ Poilblanc, T.\ Ziman, J.\ Bellissard, F.\ Mila, and G.\ Montambaux,
   ``Poisson vs. GOE Statistics in Integrable and Non-Integrable Quantum Hamiltonians,''
   Europhys.\ Lett.\ {\bf 22}, 537 (1993).

\bibitem{Barba:2008pc}
  J.~C.~Barba, F.~Finkel, A.~Gonzalez-Lopez and M.~A.~Rodriguez,
  ``The Berry-Tabor conjecture for spin chains of Haldane-Shastry type,''
  Europhys.\ Lett.\  {\bf 83}, 27005 (2008),
  arXiv:0804.3685 [hep-th].

\bibitem{Edwards:1975}
  S.~F.~Edwards and P.~W.~Anderson,
  ``Theory of spin glasses,''
  J.~Phys.~F, {\bf 5}, 5 (1975).

\bibitem{Sherrington:1975}
  D.~Sherrington and S.~Kirkpatrick,
  ``Solvable model of a spin-glass,''
  Phys.\ Rev.\ Lett.\ {\bf 35}, 26 (1975).

\bibitem{Binder:1986}
  K.~Binder and A.~P.~Young,
  ``Spin glasses: Experimental facts, theoretical concepts, and open questions,''
  Rev.~Mod.~Phys.~{\bf 58}, 4 (1986).

\bibitem{Sachdev:1992}
  S.~Sachdev and J.~Ye,
  ``Gapless Spin-Fluid Ground State in a Random Quantum Heisenberg Magnet,''
  Phys.~Rev.~Lett.~{\bf 70}, 3339 (1992).

\bibitem{Basko:2005}
  D.\ M.\ Basko, I.\ L.\ Aleiner, and B.\ L.\ Altshuler,
  ``Metal-insulator transition in a weakly interacting many-electron system with localized single-particle states,''
  Ann.~Phys.~{\bf 321}, 1126 (2006),
  arXiv:cond-mat/0506617.

\bibitem{Serbyn:2013}
  M.~Serbyn, Z.~Papi\'c, D.~A.~Abanin,
  ``Local Conservation Laws and the Structure of the Many-Body Localized States,''
  Phys.~Rev.~Lett.~{\bf 111}, 17201 (2013).


\bibitem{Huse:2013}
  D.~A.~Huse and V.~Oganesyan,
  ``A phenomenology of certain many-body-localized systems,''
  Phys.~Rev.~B {\bf 90}, 174202 (2013),
  arXiv:1305.4915v1 [cond-mat.dis-nn].


\bibitem{Pal:2010}
  D.~A.~Huse and A.~Pal,
  ``The many-body localization phase transition,''
  Phys.\ Rev.\ B {\bf 82}, 174411 (2010),
  arXiv:1010.1992 [cond-mat.dis-nn].

\bibitem{Huse:2013b}
  D.~A.\ Huse, R.\ Nandkishore, V.\ Oganesyan, A.~Pal, and S.\ L.\ Sondhi,
  ``Localization-protected quantum order,''
  Phys.~Rev.~B {\bf 88}, 014206 (2013).

\bibitem{Keating:2014}
  J.\ P.\ Keating, N.\ Linden, and H.\ J.\ Wells,
  ``Random matrices and quantum spin chains,''
  arXiv:1403.1114 [math-ph].

\bibitem{Jacquod:1997}
  Ph.\ Jacquod  and D.\ L.\ Shepelyansky,
  ``Emergence of Quantum Chaos in Finite Interacting Fermi Systems,''
  Phys.~Rev.~Lett.~{\bf 79}, 10 (1997).


\bibitem{Bordia:2015}
  P.~Bordia \emph{et al},
  ``Coupling Identical 1D Many-Body Localized Systems,''
  Phys.\ Rev.\ Lett.\ {\bf 116}, 140401 (2016),
  arXiv:1509.00478 [cond-mat.quant-gas].

\bibitem{Erdos:2014}
  L.\ Erd\H{o}s and D.\ Schr\"oder,
  ``Phase Transition in the Density of States of Quantum Spin Glasses,''
  D.\ Math.\ Phys.\ Anal.\ Geom.\ {\bf 17}: 9164 (2014),
  arXiv:1407.1552 [math-ph].

\bibitem{Schroeder:2015}
  D.~Schr\"oder,
  ``Phase transition in the density of states of quantum spin glasses,''
  University of Munich, Master Thesis (2015).


\bibitem{Cotler:2016fpe}
  J.~S.~Cotler {\it et al},
  ``Black Holes and Random Matrices,''
  arXiv:1611.04650 [hep-th].

\bibitem{You:2016}
  Y.-Z.~You, A.\ W.\ W.\ Ludwig, and C.\ Xu,
  ``Sachdev-Ye-Kitaev Model and Thermalization on the Boundary of Many-Body Localized Fermionic Symmetry Protected Topological States,''
  arXiv:1602.06964 [cond-mat.str-el].

\bibitem{Bauer:2013}
  B.~Bauer and C.~Nayak,
  ``Area laws in a many-body localized state and its implications for topological order,''
  J.\ Stat.\ Mech.\ P09005 (2013),
  arXiv:1306.5753 [cond-mat.dis-nn].

\bibitem{Wehrl:1979}
  A.~Wehrl,
  ``On the relation between classical and quantum-mechanical entropy,''
  Rept.~Math.~Phys.~{\bf 16}, 3 (1979).

\bibitem{Lieb:1978}
  E.~Lieb,
  ``Proof of an entropy conjecture of Wehrl,''
  Comm.~Math.~Phys.~{\bf 62}, 1 (1978).

\bibitem{Huse:2014}
  R.~Nandkishore and D.~A.~Huse,
  ``Many-body localization and thermalization in quantum statistical mechanics,''
  arXiv:1404.0686.

\bibitem{Oren:2010}
  I.~Oren and U.~Smilansky,
  ``Trace formulas and spectral statistics for discrete Laplacians on regular graphs (II),''
  J.\ Phys.\ A: Math.\ Theor.\ {\bf 43} (2010).

\bibitem{Garcia-Garcia:2017pzl}
  A.~M.~Garc\'ia-Garc\'ia and J.~J.~M.~Verbaarschot,
  ``Analytical Spectral Density of the Sachdev-Ye-Kitaev Model at finite N,''
  arXiv:1701.06593 [hep-th].

\bibitem{Bonzom:2011zz}
  V.~Bonzom, R.~Gurau, A.~Riello and V.~Rivasseau,
  ``Critical behavior of colored tensor models in the large N limit,''
  Nucl.\ Phys.\ B {\bf 853}, 174 (2011),
  arXiv:1105.3122 [hep-th].

\bibitem{Witten:2016iux}
  E.~Witten,
  ``An SYK-Like Model Without Disorder,''
  arXiv:1610.09758 [hep-th].

\bibitem{Krishnan:2016bvg}
  C.~Krishnan, S.~Sanyal and P.~N.~Bala Subramanian,
  ``Quantum Chaos and Holographic Tensor Models,''
  arXiv:1612.06330 [hep-th].

\bibitem{Shenker:2014cwa}
  S.~H.~Shenker and D.~Stanford,
  ``Stringy effects in scrambling,''
  JHEP {\bf 1505}, 132 (2015),
  arXiv:1412.6087 [hep-th].

\bibitem{Gu:2016oyy}
  Y.~Gu, X.~L.~Qi and D.~Stanford,
  ``Local criticality, diffusion and chaos in generalized Sachdev-Ye-Kitaev models,''
  arXiv:1609.07832 [hep-th].

\bibitem{Turiaci:2016cvo}
  G.~Turiaci and H.~Verlinde,
  ``On CFT and Quantum Chaos,''
  JHEP {\bf 1612}, 110 (2016),
  arXiv:1603.03020 [hep-th].

\bibitem{Roberts:2016hpo}
  D.~A.~Roberts and B.~Yoshida,
  ``Chaos and complexity by design,''
  arXiv:1610.04903 [quant-ph].

\bibitem{Brown:2017jil}
  A.~R.~Brown and L.~Susskind,
  ``The Second Law of Quantum Complexity,''
  arXiv:1701.01107 [hep-th].


\bibitem{DAlessio:2014}
  L.~D'Alessio and M.\ Rigol,
  ``Long-time Behavior of Isolated Periodically Driven Interacting Lattice Systems,''
  Phys.\ Rev.\ X {\bf 4}, 041048 (2014).
\end{thebibliography}
\end{document}